\documentclass[11pt]{article}
 \usepackage{jheppub}
\pdfoutput=1
\usepackage{amsmath,amssymb}
\usepackage{epsfig}
\usepackage{hyperref}
\usepackage{color}
\usepackage{slashed}
\usepackage{placeins}
\usepackage{hyperref}

\usepackage{verbatim}
\usepackage{subfigure}
\usepackage{acronym}
\usepackage{multirow}
 \usepackage{epsfig,multicol,bbm}
 \usepackage{float}

\usepackage{amsfonts}
\usepackage{graphicx, rotating}
\usepackage{epstopdf}
\usepackage{latexsym}
\usepackage{rotating}

\usepackage[normalem]{ulem}

\usepackage[font={small}]{caption}   

\definecolor{myred}{rgb}{0.7, 0, 0}
\definecolor{myblue}{rgb}{0, 0, 0.7}
\definecolor{mygreen}{rgb}{0.04, 0.7, 0.5}
\definecolor{mygray}{rgb}{0.1, 0.1, 0.1}

\hypersetup{colorlinks,citecolor=myred,linkcolor=myblue,urlcolor=myblue,linktocpage=true}

 \def\be   {\begin{equation}}   \def\ee   {\end{equation}}
 \def\ba   {\begin{array}}      \def\ea   {\end{array}}
 \def\bea  {\begin{eqnarray}}   \def\eea  {\end{eqnarray}}
 \def\bean {\begin{eqnarray*}}  \def\eean {\end{eqnarray*}}
 
 \def\bry{\begin{array}}
 \def\ery{\end{array}}

\numberwithin{equation}{section}

\begin{document}

\begin{flushright}
\footnotesize
DESY-22-209 \\
\end{flushright}
\color{black}

\title{
Status of Electroweak Baryogenesis \\ in Minimal Composite Higgs
}
\date{\today}

\author[a,b]{Sebastian Bruggisser,}

\affiliation[a]{Institut f\"{u}r Theoretische Physik, Universit\"{a}t Heidelberg, D-69120 Heidelberg, Germany}

\affiliation[b]{Department of Physics and Astronomy, Uppsala University, 75120 Uppsala, Sweden}

\emailAdd{bruggisser@thphys.uni-heidelberg.de}

\author[c]{Benedict von Harling,}

\affiliation[c]{Institut de F\'isica d'Altes Energies (IFAE), The Barcelona Institute of Science and Technology, Campus UAB, 08193 Bellaterra (Barcelona), Spain}

\emailAdd{benedictvh@gmail.com}

\author[d]{Oleksii Matsedonskyi,}

\affiliation[d]{DAMTP, University of Cambridge, Wilberforce Road, Cambridge, CB3 0WA, United Kingdom}

\emailAdd{alexey.mtsd@gmail.com}

\author[e,f]{G\'eraldine Servant}

\affiliation[e]{Deutsches Elektronen-Synchrotron DESY, Notkestr. 85, D-22607 Hamburg, Germany}
\affiliation[f]{II. Institute of Theoretical Physics, Universit\"{a}t Hamburg D-22761, Germany}

\emailAdd{geraldine.servant@desy.de}

\vskip 5pt

\abstract{
We present an update on the status of electroweak baryogenesis in minimal composite Higgs models.
The particularity of this framework is that
 the electroweak phase transition can proceed simultaneously with the confinement phase transition of the new strong dynamics that produces the composite Higgs. The latter transition is controlled by the dilaton -- the pseudo-Goldstone boson of an approximate scale invariance of the composite sector. Since it naturally is first-order, the electroweak phase transition becomes first-order too.
Another appealing aspect is that 
the necessary additional source of CP violation can arise  from the variation of the quark Yukawa couplings during the phase transition, which is built-in naturally in this scenario. These two features address the shortcomings of electroweak baryogenesis in the Standard Model.
We confront this scenario with the latest experimental bounds derived from collider searches for new resonances and measurements of the Higgs couplings and electric dipole moments. All these constraints provide (or will be able to provide in the near future) important bounds on the considered scenario, with the most stringent ones coming from LHC searches for new resonances which constrain the dilaton mass and couplings. We identify the viable region of parameter space which satisfies all the constraints, and is characterized by a dilaton mass in the $300-500$~GeV range and a Higgs decay constant $f \lesssim 1.1\,$TeV.  We discuss its future tests.
}

\maketitle

\newpage

%
%


\section{Introduction}\label{sec:incon}

The observed imbalance between the amount of matter and antimatter, although being extremely important for the understanding of the evolution of the Universe, has no confirmed explanation. While the standard model (SM) appears to fail in generating the observed baryon asymmetry,
a number of promising beyond-the-SM scenarios have appeared (see e.g.~Refs.~\cite{Affleck:1984fy,Fukugita:1986hr,Shaposhnikov:1987tw,Cohen:1990it}). The asymmetry generation could {\it a priori} have occurred at any temperature between reheating after inflation and big bang nucleosynthesis (and could have involved particles of a corresponding range of mass scales). 
It is particularly interesting to analyse scenarios where the new-physics scale is sufficiently low to allow the mechanism of baryogenesis to be tested, even if only indirectly, in controlled laboratory conditions, such as particle accelerators. One such a scenario is electroweak baryogenesis (EWBG)~\cite{Shaposhnikov:1987tw,Cohen:1990it} (see e.g.~Refs.~\cite{Enomoto:2022rrl,Azatov:2022tii,Harigaya:2022ptp,Ellis:2022lft,Servant:2018xcs,vonHarling:2016vhf,Baldes:2016rqn,Cline:2021dkf,Postma:2022dbr,Cline:2021iff,Cline:2020jre,Kainulainen:2021oqs,Ellis:2019flb} for recent work and Refs.~\cite{Krauss:1999ng,Garcia-Bellido:1999xos,Konstandin:2011ds,Servant:2014bla,Hall:2019ank,Carena:2022qpf,Flores:2022oef} for its variations), which is the most studied baryogenesis option in $(B-L)$-preserving theories. It relies on non-perturbative processes at high temperature in the standard electroweak theory, the sphalerons, which violate the $(B+L)$-symmetry. The defining feature of this mechanism is that the asymmetry is produced during an electroweak (EW) phase transition that is first-order. Its out-of-equilibrium dynamics, where regions with broken and unbroken electroweak symmetry are spacially separated, allows to use the electroweak sphalerons which are active in the unbroken phase to create an excess of baryons if new additional sources of CP-violation are present in the vicinity of the bubble walls. 

Although the SM predicts the EW phase transition to be a crossover, various TeV-scale extensions can make the transition first-order. 
One well-motivated example of such an extension
is the composite Higgs scenario. Designed to solve the Higgs mass naturalness problem, it assumes that the Higgs is a composite state of some new confining dynamics and a pseudo-Nambu-Goldstone boson (PNGB) of some global flavour symmetry of the confining sector~\cite{Panico:2015jxa}. As is well-known from QCD, such a confining sector typically produces a large set of composite resonances. Some of these fields can be PNGBs which can undergo phase transitions together with the Higgs field, and thereby make the EW phase transition first-order~\cite{Espinosa:2011eu,Chala:2016ykx,Xie:2020bkl,DeCurtis:2019rxl}. Another possibility is that the compositeness-induced changes of the Higgs boson self-interactions can result in a first-order EW phase transition~\cite{Grojean:2004xa,Bodeker:2004ws,Delaunay:2007wb,Grinstein:2008qi}. Finally, the phase transition can be first-order if it happens simultaneously with the confinement phase transition of the new strong dynamics which is itself naturally first-order~\cite{Bruggisser:2018mus,Bruggisser:2018mrt}. In fact, cooling down from high temperatures in the early universe, the strong sector is expected to pass from the deconfined phase to the confined one. The deconfined phase contains no composite Higgs and the EW symmetry is unbroken. 
The composite Higgs is produced during confinement and if the phase transition happens at sufficiently low temperatures, the Higgs can simultaneously obtain a vacuum expectation value (VEV), breaking the EW symmetry.\footnote{We will not discuss the possible domain wall problem pointed out in~\cite{DiLuzio:2019wsw}, and assume that the degeneracy of inequivalent vacua is broken by some explicit symmetry-breaking source as proposed in~\cite{DiLuzio:2019wsw}.}

\begin{figure}
\begin{center}
\includegraphics[scale=0.55]{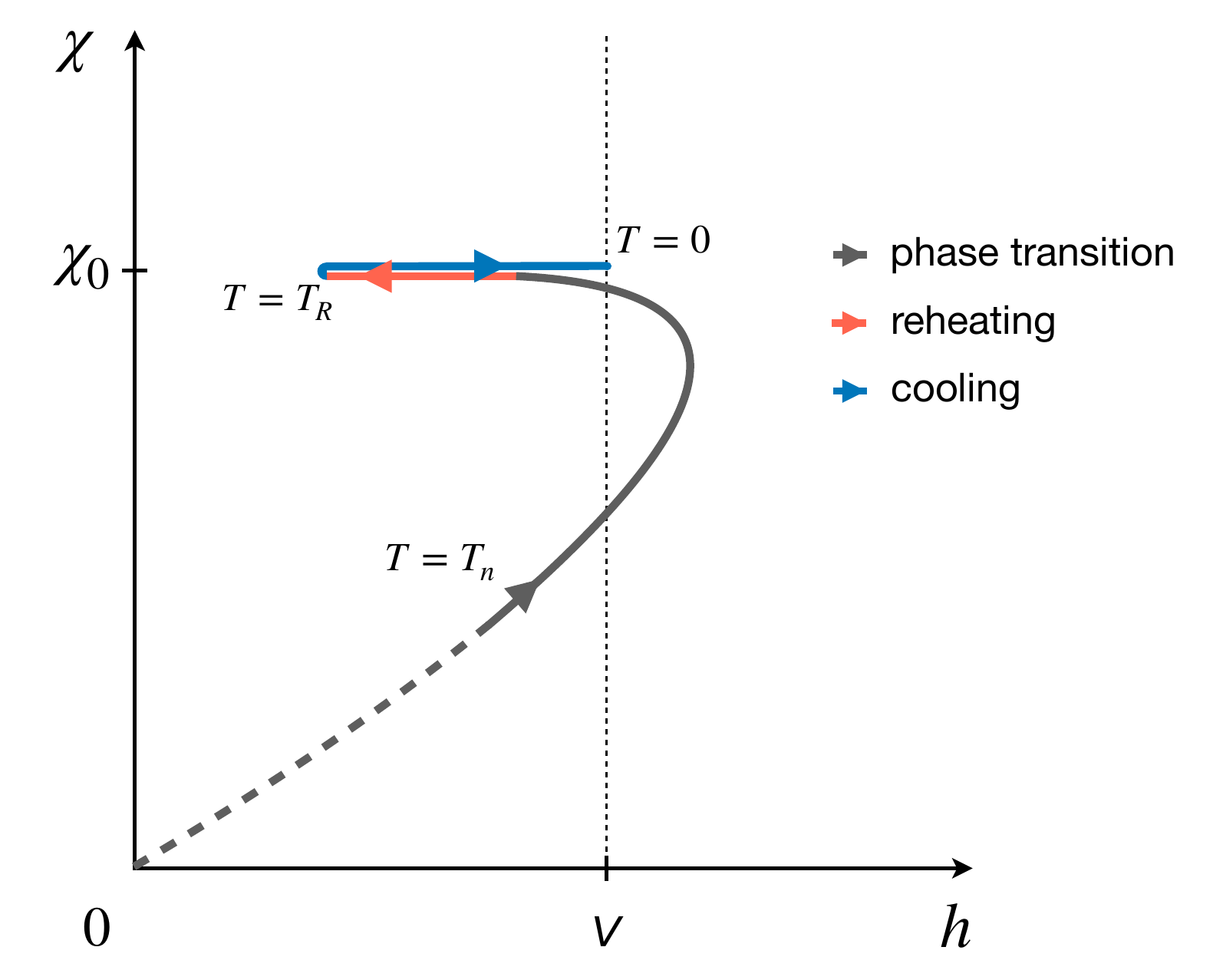}
\end{center}
\caption{{\it 
Schematic evolution of the Higgs $h$ and the dilaton $\chi$. 
At the nucleation temperature $T_n$, the dilaton tunnels from the metastable minimum at $\chi=0$ to some intermediate value (grey dashed line) and subsequently rolls towards the minimum of the potential (grey straight line).
The Higgs potential is detuned during the phase transition and hence the Higgs VEV can be larger than in today's minimum, see the end of Section~\ref{sec:Vh}.  The phase transition is followed by reheating, which tends to decrease the Higgs VEV (red line) compared to late times (blue line). The reheating and phase transition temperatures have to be below approximately $130\, \,$GeV to prevent $h$ falling below $T$ which would reactivate the EW sphalerons and wash out the baryon asymmetry. 
}}
\label{fig:sketch}
\end{figure}

In this work, we will concentrate on such a scenario with an EW phase transition induced by confinement, updating the earlier analyses of Refs.~\cite{Bruggisser:2018mus,Bruggisser:2018mrt}. 
One of the main ingredients of this model is an approximate conformal invariance of the new strongly-interacting sector, which is broken spontaneously by the confinement and thereby dynamically generates the mass scales in the theory. This can give rise to a relatively light PNGB of conformal invariance -- the dilaton $\chi$. 
We call this scenario ``Minimal Composite Higgs'' as the only extra light scalar beyond the Higgs is the dilaton, whose existence
does not depend on the details of the global symmetry-breaking pattern that delivers the Higgs as a PNGB.
The lightness of the dilaton allows to consider the confinement phase transition as a transition from a metastable minimum at the origin $\chi=0$ of the dilaton potential to a global minimum at $\chi=\chi_0$.  We thus study phase transitions during which the Higgs and dilaton simultaneously obtain VEVs. This is depicted schematically in Fig.~\ref{fig:sketch}. The confinement phase transition is first-order as a result of the large thermal barrier which is (qualitatively) generated by states of the strong sector becoming massive when the dilaton acquires a VEV, combined with the shallowness of the $T=0$ dilaton potential at large number of degrees of freedom.

For successful EWBG, a first-order phase transition is necessary, but this is not sufficient. 
Models of electroweak baryogenesis must rely on  new sources of CP violation (CPV) in the interactions of the plasma with the bubble wall, beyond what is present in the SM. This often results in sizeable contributions to the electron and neutron electric dipole moments (EDM)  which are notoriously tightly constrained by experiments, with the electron EDM currently giving the most stringent bounds~\cite{ACME:2018yjb}.
One appealing aspect of composite Higgs models is the 
way  the SM Yukawa couplings are generated. This goes under the name of partial compositeness, where elementary particles with SM quantum numbers couple to operators from the composite sector \cite{Kaplan:1991dc,Agashe:2004rs}. 
Importantly, the strongly-coupled nature of the underlying theory allows to have operators whose energy scaling can differ significantly from that of a free theory. In particular, the operators which give rise to the SM Yukawa couplings   
can experience significant running starting from some high UV scale, which can lead to the observed hierarchical fermion mass structure~\cite{Kaplan:1991dc,Agashe:2004rs}. Since the constituents of the strong sector confine at the dilaton VEV, this running stops there. The variation of the dilaton VEV during the phase transition can therefore lead to the variation of the quark Yukawa couplings. The latter can then source CP violation needed to produce the baryon 
asymmetry~\cite{Baldes:2016gaf,vonHarling:2016vhf,Bruggisser:2017lhc}.
 We will consider two qualitatively different benchmark models, where the CPV is induced by the running of either the top or charm quark Yukawa coupling. 
 
 With the EW phase transition being first-order and the existence of new CP-violating sources, we thus have in principle all ingredients  for successful EWBG.
 A crucial requirement on the model comes from an upper bound on the reheating temperature after the (potentially supercooled) phase transition. Indeed, too high a temperature would push the ratio of Higgs VEV and temperature $h/T$ below 1, leading to EW sphalerons being active again and washing out any baryon asymmetry generated during the phase transition. As we will demonstrate, this bound leads to an upper bound on the dilaton mass. This allows to effectively test this model using collider searches for new resonances. These bounds are actually stronger than the bounds  from the  electron EDM, which are known  to be generally very constraining on any model of EWBG.

In our analysis we will account for 
\begin{itemize}
\item a precise estimate of the sphaleron washout effects, 
\item 
 the latest improvements in the collider bounds on the dilaton~\cite{Bruggisser:2022ofg}, 
 \item
 the updated bounds from  the electron EDM, 
 \item the constraints on deviations of the Higgs couplings from the SM. 
 \end{itemize}
 The collider bounds are especially efficient in constraining the model parameter space, yet they leave a window to successfully realize EWBG. 
The main focus in this work will be on analysing the combined confinement and electroweak phase transition, with less emphasis on the CP-violating sources which are also necessary to generate the baryon asymmetry (see Refs.~\cite{Bruggisser:2018mus,Bruggisser:2018mrt} for an extensive discussion on this point).

The paper is organized as follows.
 In Section~\ref{sec:model}, we define the effective scalar potential of the model.  Thermal corrections to the potential and the dynamics of the phase transition are discussed in Section~\ref{sec:temp}. In Section \ref{sec:EWBG}, we recap the key 
features of EWBG in the minimal composite Higgs framework. 
We present the results of our numerical computations in Section~\ref{sec:NumericalResults}, showing the parameter space where EWBG can be realized and the predicted gravitational-wave signals.  In Section~\ref{sec:edm}, the constraints from the electron EDM measurements are analysed. We discuss our findings in Section~\ref{sec:conc}. Appendices contain a comparison between the 4D and 5D approaches to the analysis of the phase transition and details of the one-loop corrections to the potential.

\section{The Higgs-dilaton potential}
\label{sec:model}

Instead of solving the full UV theory of new strong interactions we will limit ourselves to the more tractable problem of analysing the dynamics of the lightest states in the theory, laying below some mass scale $m_*$. These light states include all the SM particles (although with altered properties compared to the SM) and the dilaton. The presence of the light dilaton is justified by assuming an approximate conformal symmetry in the new strong sector~\cite{cpr1,cpr2,cpr3,Coradeschi:2013gda,Bellazzini:2013fga,Chacko:2012sy,Megias:2014iwa,Megias:2016jcw,Pomarol:2019aae}. In this case the confinement phase transition of the new strong dynamics can be modelled as a phase transition of the dilaton field $\chi$ from the phase with $\chi=0$ to $\chi\simeq \chi_0$, where $\chi_0 \propto m_*$ sets the mass scale of the confined theory. The value of the Higgs field, being a composite object, has to also be related to the value of the dilaton field, vanishing for $\chi=0$ and taking some potentially non-zero value for $\chi=\chi_0$. The confinement phase transition therefore triggers the electroweak phase transition.

In order to quantitatively describe the coupled dynamics of the composite Higgs and the dilaton we will employ a 4D description based on a large-$N$ expansion, dimensional analysis, conformal invariance and the approximate shift symmetry of the composite Higgs~\cite{Bruggisser:2018mus,Bruggisser:2018mrt}.  We will discuss its ingredients below. In addition, we devote Appendix~\ref{sec:45D} to the discussion of the sources of model-dependence in our approach and compare it to 5D scenarios.

\subsection{Higgs effective potential}
\label{sec:Vh}

The Higgs boson is assumed to be a Goldstone boson associated with the spontaneous global-symmetry breaking $SO(5) \to SO(4)$ in the new strongly-interacting sector, which happens as this new sector confines.  The potential for the Higgs is generated via loops involving $SO(5)$-breaking interactions between the elementary fermions (such as the top quark) and the new strongly-interacting sector.  The SM electroweak gauge group is embedded in a subgroup of $SO(5)$ and a $U(1)_X$ factor. 
It is convenient to break down the Higgs potential into several parts as follows:

\subsubsection*{Tuned part}

After integrating out the heavy composite resonances with mass $\sim m_*$, the zero-temperature Higgs potential reads~\cite{Panico:2015jxa}
\be\label{eq:vCHtree}
V_h^0 = \alpha_0 \sin^2 h/f + \beta_0 \sin^4 h/f,
\ee
where $f$ is the breaking scale of the $SO(5)/SO(4)$ symmetry, also referred to as the Goldstone decay constant. The parameter $\xi \equiv (246\text{ GeV}/f)^2$ controls deviations of various observables from the SM predictions~\cite{Grojean:2013qca} and is therefore constrained from above. In this work we adopt the value $\xi=0.1$ or $f\simeq 800$~GeV.
The coefficients $\alpha_0$ and $\beta_0$ depend on the shift-symmetry-breaking parameters of the theory, such as elementary-composite mixings and SM gauge couplings. They have to be chosen to reproduce the observed Higgs mass and the Higgs VEV $v_{\text{CH}}$ satisfying 
\be
f \sin (v_{\text{CH}}/f) = v_{\text{SM}} = 246\, \,\text{GeV}.
\ee
Note that we prefer to keep the one-loop contribution of the top quark to the Higgs potential separate from~Eq.~(\ref{eq:vCHtree}) and will account for it together with other one-loop corrections from light degrees of freedom.

We call this part of the scalar potential tuned, as the coefficients $\alpha_0$ and $\beta_0$ have to be tuned down with respect to their generic values in order to reproduce the desired Higgs mass and $v_{\text{CH}} \ll f$~\cite{Matsedonskyi:2012ym,Redi:2012ha,Panico:2012uw}. The desire to minimize this tuning is the reason why we prefer to keep $f$ fixed around the minimal experimentally allowed value.

\subsubsection*{Accounting for varying $f$}
\label{sec:varf}

Our goal here is to consider the dynamics of the confinement phase transition, therefore we have to promote the compositeness scale $f$ to a dynamical variable proportional to the VEV of the dilaton $\chi$.
By assumption, the only source of mass in the theory is the dilaton VEV and therefore the dimension-4 coefficients $\alpha_0$ and $\beta_0$ have to scale $ \propto \chi^4$. This is reflected in the potential
\be\label{eq:vCH1}
V_h[h,\chi] = (\chi/\chi_0)^4 V_h^0,
\ee
where $\chi_0 \propto f$ is the dilaton VEV today.
In order to account for the scaling of $f$ with $\chi$ in the trigonometric functions of $h/f$ in the potential, we write the kinetic terms of the Higgs and the dilaton as 
\be
\label{eq:kinchih}
{\cal L}_{\text{kin}} = \frac 1 2 (\partial_\mu \chi)^2  + \frac 1 2 \frac{\chi^2}{\chi_0^2}(\partial_\mu h)^2.
\ee
Compared to simply substituting $\chi$ instead of $f$ in the potential (\ref{eq:vCHtree}), this choice ensures the invariance under the symmetry of the theory $h\to h +2 \pi f$.

For the following, when considering the dilaton interactions, we will need to specify the exact relation between $f$ and $\chi_0$.  We will assume that the strongly-coupled sector behaves as an $SU(N)$ confining theory, similar to QCD. Since the Higgs transforms non-trivially under the global $SO(5)$ symmetry of the strong sector, it is expected to be an analogue of the QCD mesons, and hence the value of $f$ is related to the strength of an analogue of the quark-antiquark condensate. Unlike the Higgs, the state controlling the confinement phase transition -- the dilaton -- can be composed of $SO(5)$-neutral constituents and hence can in principle behave as a glueball or a meson. The analyses based on the AdS/CFT correspondence prefer the former option, but we will consider both possibilities to make the discussion more general. In the limit of a large number of colors $N$, the interactions of mesons and glueballs are expected to have the parametric size~\cite{Witten:1979kh}
\be\label{eq:largencoupl}
g_{\text{mes}} \, \approx \, 4 \pi/ \sqrt{N}\, , \quad \quad 
g_{\text{glue}} \, \approx \, 4 \pi/ N.
\ee
The factors $4 \pi$ are chosen to reproduce strong coupling in the limit $N\rightarrow1$. The masses of mesons and glueballs, on the other hand, do not scale with $N$. Dimensional analysis then tells us that their VEVs scale as
\be
\begin{aligned}
\langle \text{meson} \rangle  & \propto  \frac{m_{\text{mes}}}{g_{\text{mes}}} \propto \sqrt{N}, \quad \; \;
\langle \text{glueball} \rangle  \propto \frac{m_{\text{glue}}}{g_{\text{glue}}} \propto N \quad \; \; \\
& \Rightarrow \quad \;\frac{\langle \text{meson} \rangle}{\langle \text{glueball} \rangle} \propto \frac{g_{\text{glue}}}{g_{\text{mes}}} \propto 1/\sqrt N .
\end{aligned}
\ee

The relations~(\ref{eq:largencoupl}) are expected to hold up to order-a-few factors. We introduce coefficients $c_k^{(h)}, c_k^{(\chi)}$ to account for this freedom and consider the following couplings associated with respectively the (meson) Higgs and the (glueball or meson) dilaton:
\bea
g_* &=&c_k^{(h)} \frac {4\pi} {\sqrt N} \\ 
g_\chi &=& c_k^{(\chi)} \frac {4\pi} {N} \; \; (\text{glueball}) \quad \text{or} \quad  c_k^{(\chi)} \frac {4\pi} {\sqrt N} \; \; (\text{meson}). \label{eq:gchi}
\eea
We then fix the relation between the dilaton VEV and the Higgs decay constant as
\be\label{eq:chioff}
\chi_0=(g_*/g_\chi) f = \frac{c_k^{(h)}}{c_k^{(\chi)}} f \times
\begin{cases}
\sqrt N & \text{for glueball dilaton}\\
1 & \text{for meson dilaton.}
\end{cases}
\ee
We refer the reader to Refs.~\cite{Bruggisser:2022ofg,Bruggisser:2018mrt} for further discussion of this relation. Notice in particular that, if $f$ is kept fixed,  the interactions of a glueball dilaton (which are controlled by $1/\chi_0$) become more and more suppressed at large $N$. Given the presence of order-one parameters which can alter the relations discussed in this section, we will consider an {\it effective number of colors} $N$ which can also take non-integer values.

\subsubsection*{Accounting for varying Yukawa couplings}

Finally, we account for the fact that $\alpha_0$ and $\beta_0$ can scale differently from $\propto \chi^4$ in the presence of explicit breaking of the conformal symmetry by quantum effects, in particular by the running of the Yukawa couplings. 
As discussed in the introduction, we consider two concrete scenarios for this running, both motivated by the need to have an additional source of CPV for successful electroweak baryogenesis. In the first scenario, we assume that the top quark Yukawa coupling changes significantly between $\chi=\chi_0$ and $\chi=0$, while the Yukawa couplings of the other quarks remain small and can be neglected. In the second scenario, the charm Yukawa grows from its small value today at $\chi=\chi_0$ to a large value at $\chi=0$. The top Yukawa, on the other hand, is taken to be approximately constant. As was shown in~\cite{Bruggisser:2018mus,Bruggisser:2018mrt} both possibilities can effectively produce the baryon asymmetry.

Let us begin with the first scenario. 
We will assume the usual partial compositeness mechanism for the generation of the top quark Yukawa coupling. The latter then is the result of the mixing between the elementary top quark $t_{L,R}$ and operators ${\cal O}_{tL,R}$ from the conformal field theory (CFT)
\be
\label{eq:topop}
y_{tL,R}(\mu) \,  \bar t_{L,R} {\cal O}_{tL,R}\, .
\ee
The dependence on the renormalization scale $\mu$ is induced by strongly-interacting degrees of freedom of the CFT above the compositeness scale, and can be significant.
The CFT operators can excite composite fermionic states (which we assign a mass $m_* = g_* f  = g_{\chi} \chi_0$). The term in Eq.~\eqref{eq:topop} then leads to mass mixing of these states with the top quark, schematically given by
\be\label{eq:toppartners}
y_{tL}(\mu) \, f \sin (h / f) \bar t_{L} T_{R}  \, + \, y_{tR}(\mu) \, f \, \bar t_{R} T_{L} \, + \, m_\star \bar T T.
\ee 
Evaluating the couplings at the condensation scale, $\mu=\chi$, we can find the value of the top quark Yukawa coupling. The latter can act as a source of CPV for electroweak baryogenesis if it has 
a varying complex phase. To achieve this, we will assume that the Yukawa coupling has the form
\be\label{eq:topquk}
{\cal L}_{\rm top} = -\frac{\lambda_t}{\sqrt 2} f \sin(h/f) \bar q_L t_R,\quad \; \; \lambda_t = y_{tL} (y_{tR}^{(1)}+y_{tR}^{(2)})/g_*,
\ee
where $y_{tR}^{(1)}$ and $y_{tR}^{(2)}$ have a non-zero relative complex phase.
This can be obtained if $t_R$ couples to two different CFT operators, with two different mixings $y_{tR}^{(1)}, y_{tR}^{(2)}$. As we mentioned earlier, we will not specify the physics at the scale $m_*$ and will use directly the expression~(\ref{eq:topquk}) for the top Yukawa.    
One choice allowing for a varying top Yukawa phase during the EW phase transition (i.e.~while $\chi$ is changing from 0 to $\chi_0$) is to assume constant $y_{tL}$ and $y_{tR}^{(2)}$, and to take $y_{tR}^{(1)} \equiv y_t$ varying according to the RG equation~\cite{Contino:2010rs}
\be\label{eq:yukrun}
\frac{\partial y_t}{\partial \log \chi} = \gamma_y y_t+ c_y y_t^3 /g_*^2.
\ee

Note that the heavy fermionic states at the scale $m_*$, which we have integrated out, also contribute to threshold corrections of the scalar potential which are sensitive to the varying coupling $y_t$.  
In order to account for this we make the following substitution in Eq.~(\ref{eq:vCHtree})
\be
\label{eq:genalphabeta}
\begin{aligned}
\alpha_0 \, \to &\,\, \, \alpha_0 + (\alpha[\chi]-\alpha[\chi_0]), \\
\beta_0 \, \to & \, \, \, \beta_0 + (\beta[\chi]-\beta[\chi_0]), 
\end{aligned}
\ee
where 
\be
\label{eq:ndaalphabeta}
\begin{aligned}
\alpha[\chi] \, = & \, \,  c_\alpha \frac{3 y_t^2[\chi]}{16 \pi^2} g_*^2 f^4, \\
\beta[\chi] \, = & \, \, c_\beta  \frac{3 y_t^2[\chi]}{16 \pi^2} g_*^2 f^4
\end{aligned}
\ee
are the parametric estimates of the one-loop contribution of fermionic top partners in Eq.~(\ref{eq:toppartners}) to the scalar potential~\cite{Bruggisser:2018mus,Bruggisser:2018mrt}.  As was noted in Refs.~\cite{Bruggisser:2018mus,Bruggisser:2018mrt}, the substitution~(\ref{eq:genalphabeta}) detunes the Higgs potential when $\chi$ is away from $\chi_0$. As a result, the Higgs VEV takes its detuned value during the phase transition (for intermediate $\chi$ values). This detuned value can be either 0 or $\sim \chi$, depending on the coefficients $c_{\alpha,\beta}$ which are free order-one parameters in our description. This, in turn, can respectively suppress or enhance the produced baryon asymmetry.

The discussion for the scenario with a varying charm quark Yukawa coupling is analogous. For most of the expressions in Eqs.~\eqref{eq:topop} -- \eqref{eq:ndaalphabeta} we just need to replace the corresponding quantities for the charm. 
One difference with respect to the top-induced CPV is that for the charm one can obtain large CPV even if it is coupled to only one CFT operator~\cite{Bruggisser:2018mrt}. Instead of Eq.~\eqref{eq:topquk}, the charm quark Yukawa coupling then reads
\be 
\lambda_c = \frac{y_{cL} \, y_{cR}}{g_*} \, .
\ee

\subsection{Dilaton potential}\label{sec:Vchi}

We assume that some unspecified strongly-coupled and approximately conformal sector generates a dilaton potential of the form~\cite{cpr1,cpr2,cpr3,Coradeschi:2013gda,Bellazzini:2013fga,Chacko:2012sy,Megias:2014iwa,Megias:2016jcw}
\be
\label{eq:vchi}
V_\chi = c_\chi g_\chi^2 \chi^4 - \epsilon(\chi) \chi^4.
\ee
The first term is scale-invariant and the second one breaks conformal invariance due to the running of $\epsilon$ with the dilaton VEV according to the RG equation
\be\label{eq:epsrunning}
\frac{\partial \epsilon}{\partial \log \mu} = \gamma_\epsilon \epsilon - c_\epsilon  \epsilon^2 /g_\chi^2.
\ee
This running is arranged such that the second term in Eq.~(\ref{eq:vchi}) is much smaller than the first term at large $\chi \gg \chi_0$ and then grows towards smaller $\chi$. At $\chi=\chi_0$ the two contributions in Eq.~(\ref{eq:vchi}) equilibrate each other, and thereby produce a minimum of the dilaton potential.\footnote{For simplicity, we neglect the term proportional to $c_{\chi y}$ of Ref.~\cite{Bruggisser:2018mrt} since it only has a minor effect on the phase transition properties.}

We will set $g_\chi$ to be either the glueball or the meson coupling of Eq.~(\ref{eq:largencoupl}), as was motivated in Sec.~\ref{sec:varf}. Furthermore, we will fix the coefficient $c_\epsilon$ to a constant value chosen to minimize $\epsilon$ at $\chi \ll \chi_0$ in order to avoid that it grows above $g_\chi^2$ which would make the perturbative description inadequate. 
The scaling dimension $\gamma_\epsilon$ in Eq.~(\ref{eq:epsrunning}) is then chosen to reproduce the desired dilaton mass -- a free parameter in our setup. Finally, the integration constant of Eq.~(\ref{eq:epsrunning}), $\epsilon(\chi_0)$, is determined by the requirement that the minimum of the dilaton potential be at $\chi=\chi_0$. 

The remaining free parameter of Eq.~(\ref{eq:vchi}) is $c_\chi$, which is expected to be of order one. If $c_\chi \approx 1$, $\epsilon(\chi)$ would need to be of order $c_\chi g_\chi^2\simeq g_\chi^2$ at $\chi=\chi_0$ to produce a minimum, and would be even larger than that at $\chi<\chi_0$. Given that our description is based on an expansion in $\epsilon/g_\chi^2$, we have to require $c_\chi<1$  to avoid a loss of perturbativity.

\subsection{Higgs-dilaton mixing}

As we will show in Section~\ref{sec:NumericalResults}, the size of mass mixing between the Higgs and the dilaton has important consequences for the phenomenological tests of the considered scenario. We will now present a simple analytic estimate of this mixing which will be useful in the following. 

The scale-invariant part of the composite Higgs potential~(\ref{eq:vCH1}) does not lead to any mass mixing between the Higgs and the dilaton~\cite{Chacko:2012sy,Bruggisser:2018mrt}. Indeed, in this case the mass mixing is given by $\partial_\chi \partial_h (\chi^4 V_h^0) = (\partial_\chi \chi^4) (\partial_h  V_h^0)$ which vanishes at the minimum of the potential where $\partial_h  V_h^0=0$. 
However, the running top or charm quark mixings, which we use as CPV sources, contribute to the Higgs-dilaton mixing through the scale variation of $\alpha$ and $\beta$~(\ref{eq:genalphabeta}) in the potential~(\ref{eq:vCH1}). For the top quark-induced CPV the mixing angle $\theta$ can be estimated as
\be\label{eq:mixangle}
\sin \theta \simeq \frac{\partial_\chi \partial_h V_h[h,\chi]}{m_\chi^2}  \simeq \gamma_y \frac{3  c_\alpha y_t^2}{4\pi^2} \frac{g_* g_\chi v f}{m_\chi^2} 
=\gamma_y \frac{3  c_\alpha y_t^2}{4 \pi^2} \frac{m_*^2}{m_\chi^2} \frac v f \frac{g_\chi}{g_*}.
\ee
The expression for the charm case is analogous, although the resulting mixing angle $\theta$ is generically expected to be much lower due to the smaller Yukawa coupling.

\section{Dilaton phase transition}
\label{sec:temp}

We will next discuss the thermal corrections to the potential and the phase transition of the dilaton. Our discussion is in many aspects similar to that in~\cite{Creminelli:2001th,Randall:2006py,Nardini:2007me,Konstandin:2010cd,vonHarling:2017yew,Dillon:2017ctw} for the phase transition in 5D dual models. 
Let us first consider the region of the potential for large dilaton VEVs with $ \chi\gtrsim T/g_\chi$, where $T$ is the temperature. The confinement scale $ \sim g_\chi \chi$ is larger than the temperature in this region and we can thus use the description of the strong sector in the confined phase. Furthermore, most of the confined states have masses $\sim g_\chi \chi \gtrsim T$ or larger and we can neglect their thermal corrections to the potential. We only have to include thermal corrections from the light states, namely the SM particles including the Higgs, and the dilaton. The free energy then reads
\be
\label{eq:feft}
F[\chi \gtrsim T/g_\chi]\,  \simeq \, V_h \, + \, V_\chi \, + \, \Delta V_{T, \text{light}} ,
\ee
where $V_h$ and $V_\chi$ are given in Eqs.~\eqref{eq:vCH1} and \eqref{eq:vchi} and 
\begin{equation}
\label{eq:v1Tloop}
	\Delta V_{T, \text{light}} \, = \,\sum_{\substack{\text{light}\\ \text{bosons}}}\frac{n T^4}{2\pi^2}J_b\left[\frac{m^2}{T^2}\right] \, - \, \sum_{\substack{\text{light}\\ \text{fermions}}}\frac{n  T^4}{2\pi^2}J_f\left[\frac{m^2}{T^2}\right] .
\end{equation}
The masses $m$ depend on $\chi$ and/or $h$.  The sums run over all SM bosons, fermions and the dilaton; $n$ is the number of {\it d.o.f.}~for each particle species. Furthermore, the functions $J_b$ and $J_f$ are given by
\begin{equation}
J_b[x] \, = \int_0^\infty dk~k^2 \log\left[1-e^{-\sqrt{k^2+x}}\right] \quad \text{and} \quad J_f[x]\, =\int_0^\infty dk~k^2 \log\left[1+e^{-\sqrt{k^2+x}}\right] .
\end{equation}

\begin{figure}
\begin{center}
\includegraphics[scale=0.7]{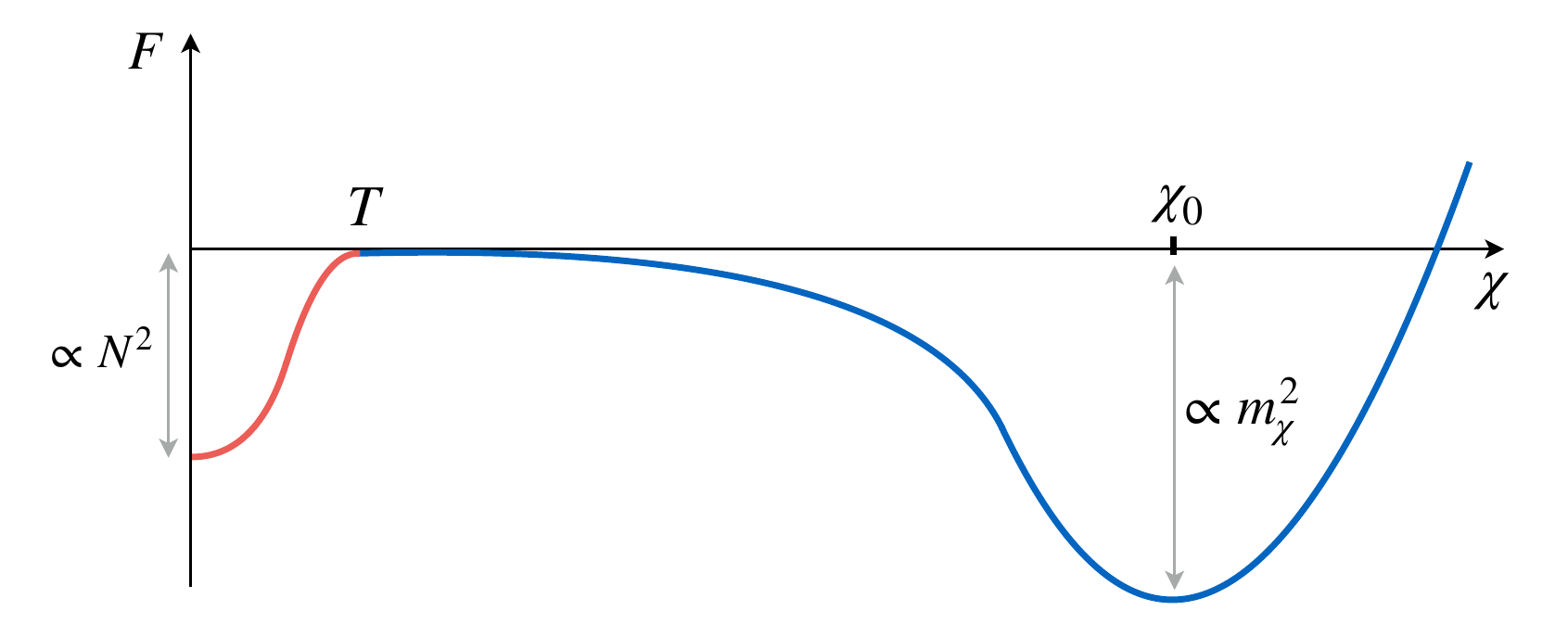}
\end{center}
\caption{{\it The dilaton potential, with the region dominated by the thermal effects (red) and the region dominated by the zero-temperature potential (blue).}}
\label{fig:chipotential}
\end{figure}

For $\chi=0$, on the other hand, the strong sector is in the deconfined and (nearly) conformal phase. Using dimensional analysis and large-$N$ counting, the free energy in this phase is then given by
\be\label{eq:fcft}
F[\chi=0]\, \simeq \, - \, c  N^2 T^4\, - \,  \frac{\pi^2 g_{\rm SM}}{90} T^4\, ,
\ee
where the constant $c$ is a function of the number of \emph{d.o.f.} per color in the strong sector and $g_{\rm SM}\simeq 100$ is the total number of \emph{d.o.f.}~of the SM. For definiteness, we will use $c = \pi^2 /8$ as arises in $\mathcal{N}=4$ $SU(N)$ super-Yang-Mills  (including a factor $3/4$ due to strong coupling which can be calculated from the AdS dual \cite{Gubser:1996de}). 

In the intermediate region $0\lesssim \chi \lesssim T/g_\chi$, the temperature is larger than the confinement scale  
and we would need knowledge of the UV description of the strong sector in terms of the deconfined constituent fields in order to properly account for the thermal corrections.\footnote{A proper calculation of the thermal corrections in this regime is difficult already due to the fact that the conformal sector is expected to be at strong coupling. }  In absence of this knowledge, we will model a smooth transition of the free energy between Eq.~\eqref{eq:feft} for $\chi  \gtrsim T/g_\chi$ and Eq.~\eqref{eq:fcft} for $\chi \sim 0$. As discussed in more detail in Ref.~\cite{Bruggisser:2018mrt}, we expect that our results do not depend sensitively on the precise way in which this modelling is performed. For concreteness, we follow Ref.~\cite{Randall:2006py}, and include the thermal corrections from $\sim N^2$ states with mass $g_\chi \chi$. The free energy then reads
\be
\label{eq:feft2}
F\,  \simeq \, V_h \, + \, V_\chi \, + \, \Delta V_{T, \text{light}} \, + \, \Delta V_{T, \text{trans.}},
\ee
where $\Delta V_{T,\text{trans.}}$ is given by an expression analogous to Eq.~\eqref{eq:v1Tloop} but with the sums running over the states with mass  $g_\chi \chi$. Since the difference between the thermal corrections of bosons and fermions is small, we assume that they are bosonic. Furthermore, we choose the number of these states such that 
\be
\label{eq:fixf}
\sum_{\substack{\text{states with}\\ {\text{mass} \, g_\chi \chi}}} \hspace{-.2cm} n \, = \, \frac{45 N^2}{4} \, .
\ee
This ensures that Eq.~\eqref{eq:feft2} reproduces Eq.~\eqref{eq:fcft} for $\chi =0$.\footnote{Note that $V_h$ and $V_\chi$ are normalized such that they vanish for $\chi=0$.} On the other hand, for $\chi \gtrsim T /g_\chi$ the thermal corrections from the states with mass  $g_\chi \chi$ are negligible and the free energy matches Eq.~\eqref{eq:feft} with only light states included. A sketch of the free energy is shown in Fig.~\ref{fig:chipotential}. 

Let us next discuss the dynamics of the phase transition of the dilaton. At high temperatures, we expect that the thermal potential has only one minimum, at $\chi=0$ and with free energy given in Eq.~\eqref{eq:fcft}. Going to lower temperatures, eventually a second minimum appears near the one of the zero-temperature potential at $\chi_0= g_\star f/g_\chi$. Neglecting the Higgs-dependent part and the thermal corrections in Eq.~(\ref{eq:feft}), the free energy in this minimum is given by
\be
\label{eq:vchimin}
F\, = \, V_{\chi}^{\text{min}} \, \simeq \,   {\gamma_\epsilon \over 4} c_\chi g_\chi^2 \chi_0^4 \, = \,  {\gamma_\epsilon \over 4} c_\chi \frac{g_\star^4}{g_\chi^2} f^4\, .
\ee 
The phase transition becomes energetically possible at the critical temperature $T_c$ where the free energies in both minima are equal. 
Using Eqs.~\eqref{eq:fcft} and \eqref{eq:vchimin} and neglecting again the thermal corrections from light particles, we find
\be
\label{eq:CriticalTemperature}
T_c \, \simeq \, 2 \left(\frac{g_\star^2}{4 \pi g_\chi N}\right)^{1/2} (2 \gamma_\epsilon c_\chi)^{1/4}f \, = \, 2 (2 \gamma_\epsilon c_\chi)^{1/4}f \times 
\begin{cases}
N^{-3/4}\,, \quad  &\text{for}\,\, g_\chi  =  4\pi/\sqrt{N} \\
N^{-1/2}\,, \quad  & \text{for}\,\, g_\chi  =  4\pi/N \, .
\end{cases}
\ee
The critical temperature is thus suppressed by the number of colors $N$. Furthermore, going from the minimum near $\chi_0= g_\star f/g_\chi$ of the thermal potential in Eq.~\eqref{eq:feft} at $T=T_c$ towards smaller values of $\chi$, the free energy grows. This shows that the two minima of the free energy are separated by a barrier and the phase transition is therefore first-order. The phase transition then proceeds by the nucleation of bubbles of broken phase which form inside the plasma of unbroken phase and which subsequently expand.  Initially, the nucleation rate of these bubbles is too small to compete with the expansion of the universe. Subsequently, the size of the barrier decreases with the temperature and the bubble nucleation rate grows. The phase transition then eventually completes at some temperature $T_n< T_c$ to which we refer as the nucleation temperature.

\section{Electroweak baryogenesis in composite Higgs models}
\label{sec:EWBG}

As we have discussed in Section~\ref{sec:temp}, the confinement phase transition involving the dilaton is naturally first-order. If this phase transition happens at a temperature $T_n \lesssim 130 \,$GeV, the EW phase transition takes place simultaneously and thereby becomes first-order too. This satisfies one requirement for successful EWBG. Another requirement is a source of CPV (beyond CPV from the CKM matrix which is not large enough). As we have discussed in Section~\ref{sec:Vh}, the size of the Yukawa couplings can naturally depend on the dilaton VEV in composite Higgs models. Since the dilaton VEV changes when moving from the inside of the bubbles during the phase transition towards the outside, the Yukawa couplings vary across the bubble walls too. As was pointed out in \cite{Bruggisser:2017lhc}, this can give rise to CPV near the bubble walls.
The CP-violating source term is given by 
\be\label{eq:cpv}
S_{\rm CPV} \propto \text{Im}[V^\dagger m_q^{\dagger \prime \prime} m_q V],
\ee 
where V is the unitary matrix diagonalizing the product of the quark mass matrices $m_q^{\dagger}m_q$, and the primes denote spacial derivatives along the direction of the bubble wall velocity.
The size of $S_{\rm CPV}$ is thus sensitive to the squared mass of the quarks during the phase transition whose varying Yukawa couplings provide CPV.  
The Higgs and dilaton tunnel along a trajectory in the two-field potential with approximately constant $h/f$ in the normalization\footnote{For a canonically normalized Higgs, its value is proportional to the dilaton value.} of Eq.~\eqref{eq:kinchih} which can be understood as the angle of  the trajectory (see \cite{Bruggisser:2018mrt} for more details). The top mass in turn is proportional to $\sin h/ f$ according to Eq.~\eqref{eq:topquk}. The size of $S_{\rm CPV}$ therefore depends on $\sin h/ f$ and we will be interested in maximizing the latter. 
We will consider two main cases, where the CPV is induced either by the top or the charm varying Yukawa coupling. 

The baryon asymmetry is primarily generated at the nucleation temperature.  After the phase transition is completed, the energy stored in the bubble walls is converted into thermal energy which increases the temperature of the plasma. The resulting reheat temperature $T_R$ is given by
\be
\frac{\pi^2 g_c}{30} T_R^4 \, \simeq \, \Delta V \, + \, \frac{3 \pi^2 N^2}{8} T_n^4 \, + \, \frac{\pi^2 g_c}{30} T_n^4 \, ,
\label{eq:TR}
\ee
where $g_c$ is the number of relativistic degrees of freedom in the confined phase (i.e.~the SM plus dilaton) and $\Delta V$ is the difference between the potential energies in the false and true vacuum. This reheating can reduce the baryon asymmetry, which was produced during the EW phase transition, by two effects. One is washout due to sphalerons. 
At high temperatures, thermal corrections move the minimum of the potential in the Higgs direction towards smaller Higgs VEVs. Denoting the Higgs VEV in the minimum at temperature $T$ by $h_{\rm min}[T]$, sphalerons become active again if $h_{\rm min}[T]/T \lesssim 1$ and begin to wash out the baryon asymmetry.
To quantify this washout, we  calculate the reduction factor $\omega_{\rm sph}$ of the baryon asymmetry due to sphalerons by integrating the corresponding Boltzmann equation from $T=T_R$ to $T=0$ (or some small enough temperature where sphalerons are guaranteed to be inactive). This gives \cite{Kamada:2016cnb}
\be
\omega_{\rm sph} \, = \, \exp \left[ \int^{0}_{T_R} d T \, \frac{111}{34} \frac{1}{H[T]} \exp \left(-\frac{2 C}{\alpha_W} \frac{m_W[T]}{T} \right)  \right] ,
\label{eq:wash}
\ee
where $H[T]$ is the Hubble rate at temperature $T$, $C\approx 1.9$ is a numerical constant and $\alpha_W = g_W^2/4\pi$ with $g_W$ being the $SU(2)$ gauge coupling. Here we have assumed the SM value for the sphaleron energy, which is expected to be very close to the one obtained in the composite Higgs set-up~\cite{Grojean:2004xa,Spannowsky:2016ile}.
Furthermore, 
\be
m_W[T] \, = \, \frac{g_W}{2} \frac{g_\chi}{g_*} \chi \sin\left[\frac{h_{\rm min}[T]}{f}\right]
\ee
is the mass of the $SU(2)$ gauge boson resulting from the Higgs VEV at temperature $T$.

Note that the reheat temperature is related to the dilaton mass. The latter can be approximated as
\be\label{eq:dilatonmass}
m_\chi^2 \,=  \, -4 \gamma_\epsilon c_\chi  m_*^2 \,  = \, 16 \,  \frac{g_\chi^2}{g_*^2} \frac{\Delta V}{f^2}.
\ee
Using Eq.~(\ref{eq:TR}) to obtain $\Delta V$ as a function of $T_R$, and accounting for the fact that the term $\propto T_n^4$ is typically subdominant compared to $\Delta V$ we then obtain
\be\label{eq:dilatonmass1}
m_\chi^2 \simeq \frac{8\pi^2 g_c}{15}\frac{g_\chi^2}{g_*^2} \frac{T_R^4}{f^2}.
\ee
This implies that an upper bound on the reheat temperature restricts the dilaton mass. 

After the confinement phase transition the dilaton rests around $\chi=\chi_0$, and the deformations of the composite Higgs potential with respect to the SM Higgs are suppressed by the large scale $f$. Hence the constraint from the sphaleron freeze-out condition $h/T<1$ implies almost the same bound on the reheat temperature as in the SM, $T_R<130 \;$GeV \cite{DOnofrio:2014rug}, as we have confirmed numerically. This leads to the bound on the dilaton mass
\be\label{eq:chimassbound}
m_\chi \lesssim \frac{g_\chi}{g_*} \, 500 \,\text{GeV}.
\ee
In the case of a glueball-like dilaton we have $g_\chi/g_* \propto 1/\sqrt N$ and the bound (\ref{eq:chimassbound}) becomes especially tight.  On the other hand, this bound can be relaxed by a factor $c_{k}^{(\chi)}/c_{k}^{(h)}$ if the latter is greater than one, see Eq.~(\ref{eq:chioff}).

The second effect that diminishes the baryon asymmetry is entropy injection due to the increase of the plasma temperature from $T_n$ to $T_R$ after the phase transition. This reduces the baryon asymmetry by an extra factor $(T_n/T_R)^3$.
Therefore, the total washout factor of the baryon asymmetry after reheating is given by
\be
\omega_{\rm tot}=  \omega_{\rm sph} \times \left(\frac{T_n}{T_R}\right)^3 .
\label{eq:totwash}
\ee
We will be interested in maximizing this quantity.

\section{Numerical results for the phase transition}
\label{sec:NumericalResults}

In Section~\ref{sec:model}, we have introduced the Higgs-dilaton potential and its thermal corrections. In addition, the potential receives loop corrections at zero-temperature which we discuss in Appendix~\ref{sec:V1L}. We next present numerical results for some concrete scenarios. In order to obtain a new CPV source for electroweak baryogenesis, we assume that the Yukawa coupling of either the charm or top changes significantly with the dilaton VEV. We have performed numerical scans for both cases. As we discuss below, however, for the charm we find no region in parameter space with reasonably small washout of the baryon asymmetry after the electroweak phase transition and where LHC constraints are fulfilled. In the following, we therefore only present numerical results for the top. Technical details about how the relevant quantities of the phase transition in the two-field potential for the Higgs and dilaton were calculated can be found in \cite{Bruggisser:2018mrt}.

\begin{figure}[t]
\centering
\includegraphics[width=7.5cm]{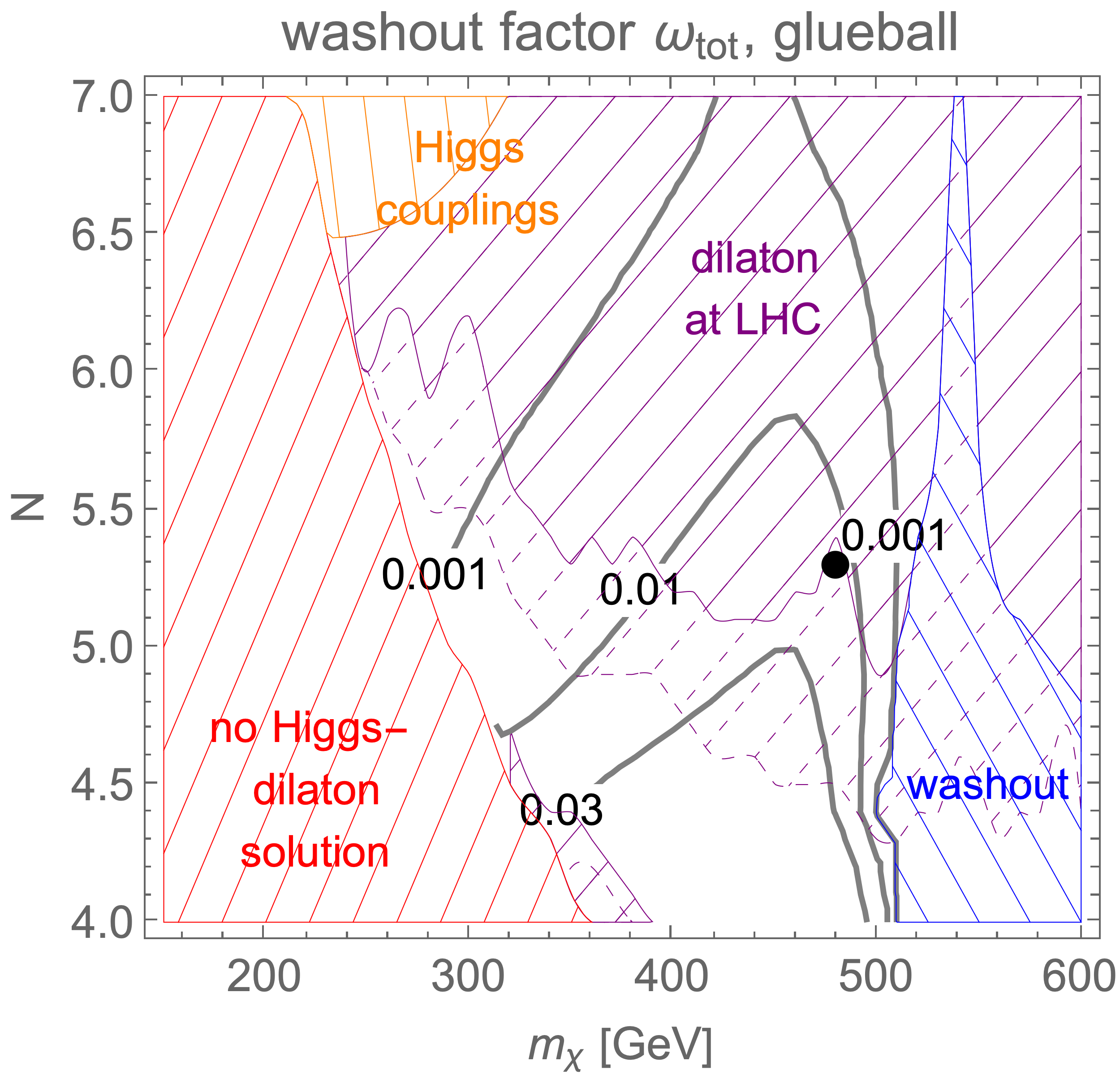}
\hspace{0.3cm}
\includegraphics[width=7.5cm]{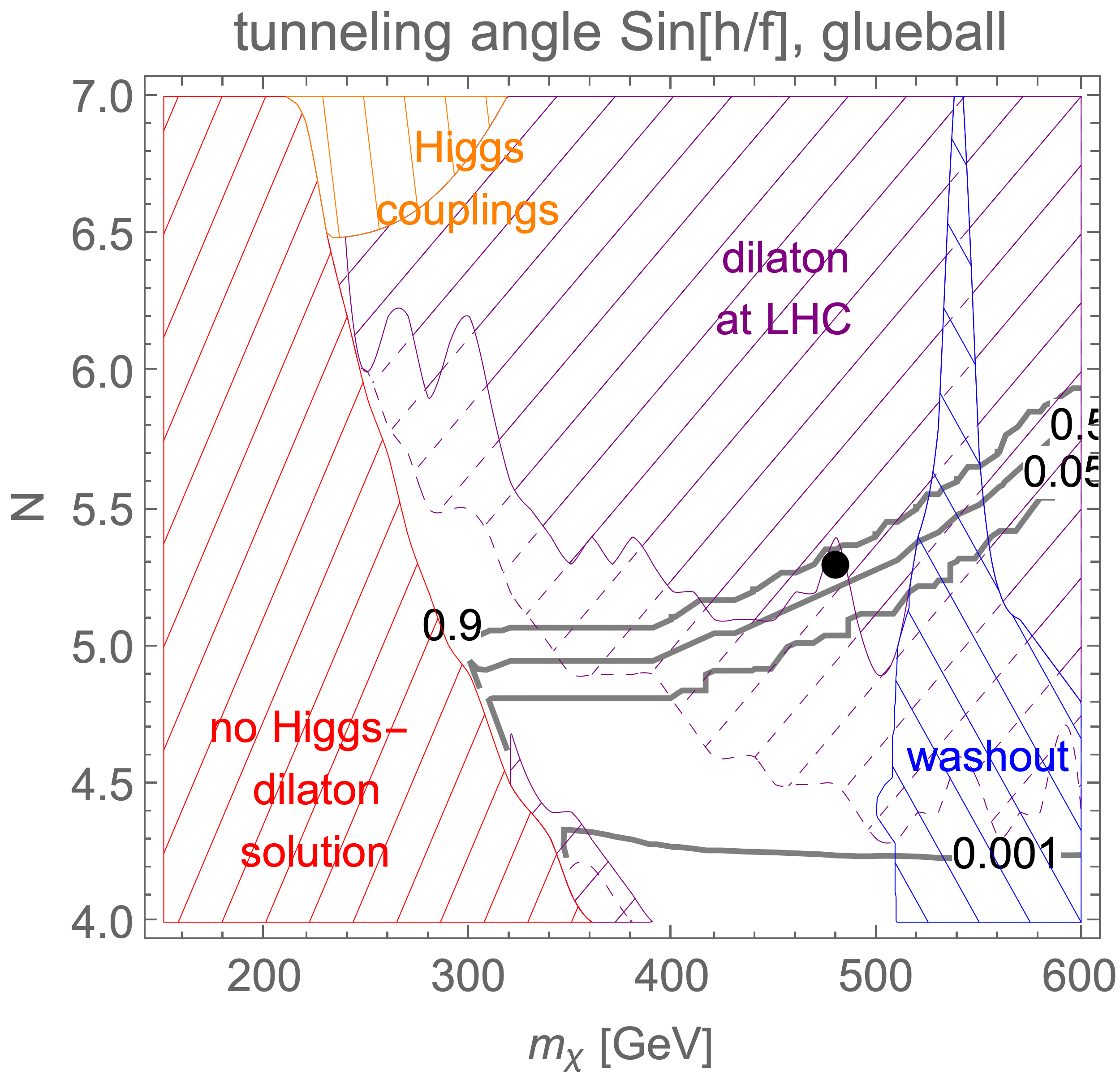} \\ 
\vspace{0.2cm}
\includegraphics[width=7.5cm]{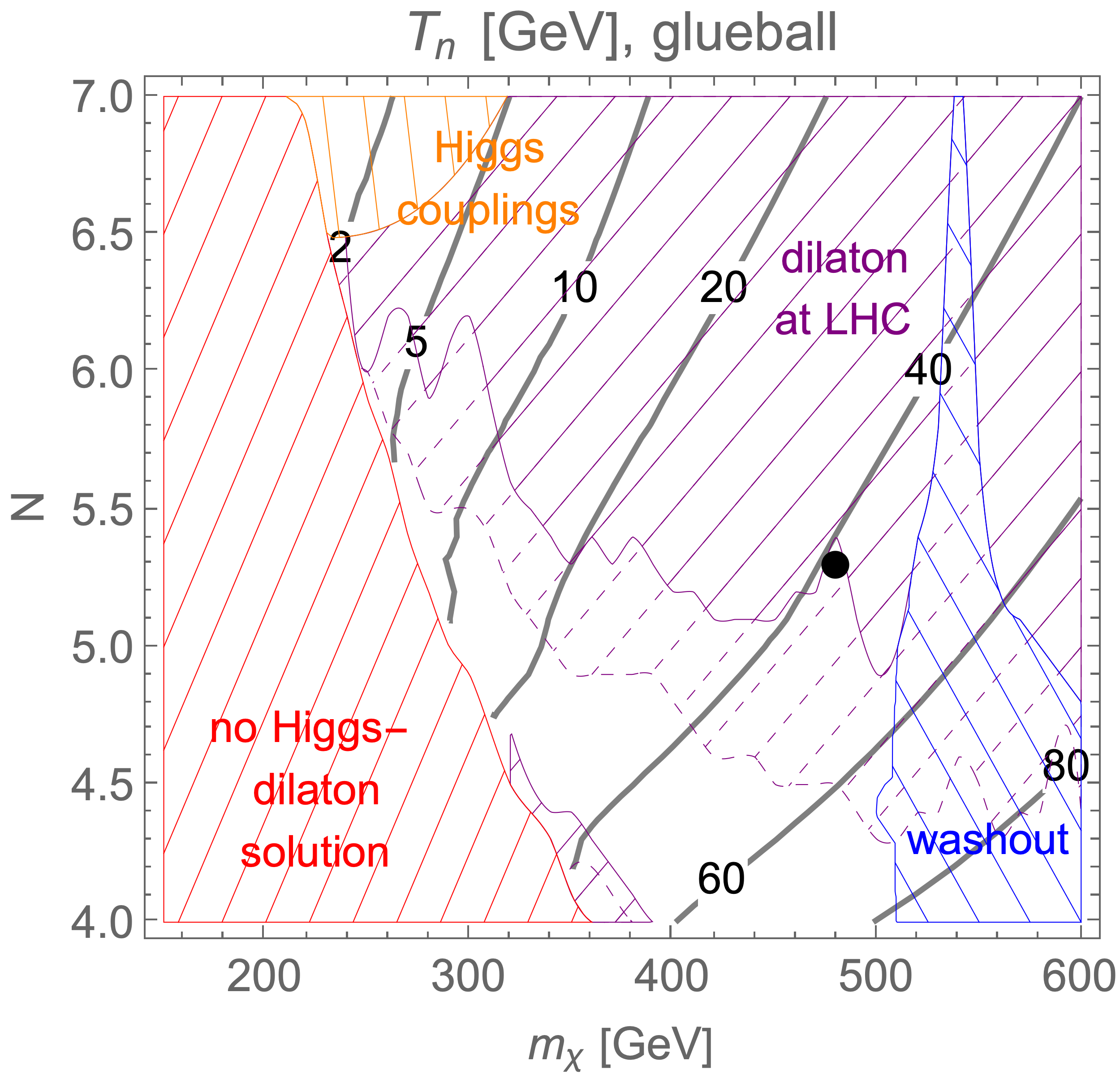}
\hspace{0.3cm}
\includegraphics[width=7.5cm]{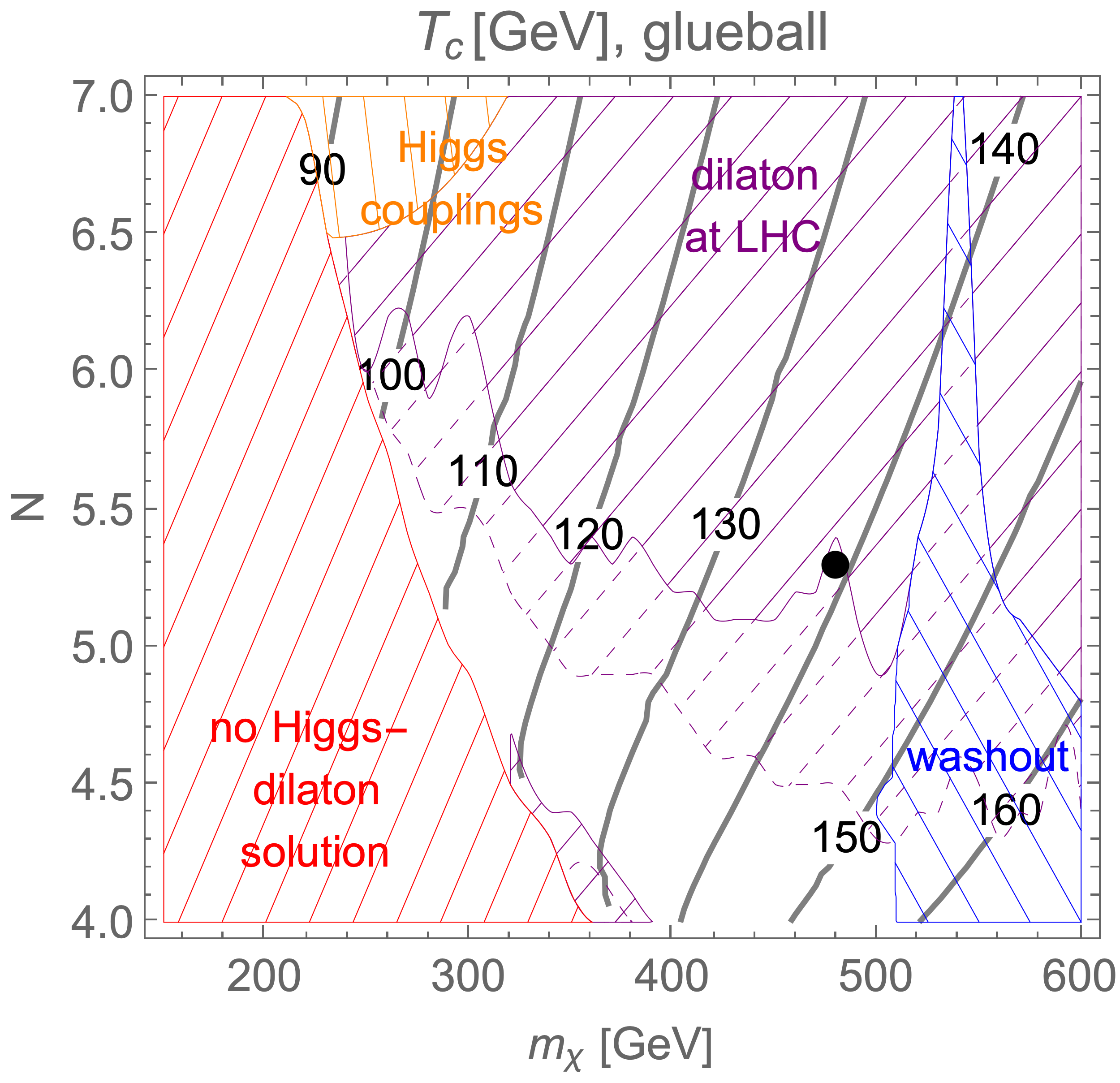}
\caption{\small \it{Results for a glueball dilaton and with varying top Yukawa. The parameters that we have used are given in Table~\ref{tab:bench}. {\bf Upper left panel}: the total washout factor $\omega_{\rm tot}$ of the baryon asymmetry due to sphalerons and entropy injection. {\bf Upper right panel}: the (sine of) the tunneling angle $\sin h/f$, which is important for the amount of CPV during the phase transition.  {\bf Lower left panel}: the nucleation temperature $T_n$ (in GeV). {\bf Lower right panel}: the critical temperature $T_c$ (in GeV). In the red hashed region, there is no consistent solution to the zero-temperature Higgs-dilaton potential. The orange hashed region is excluded because the Higgs couplings deviate too much from the SM. Furthermore, the purple hashed region with straight (dashed) lines is not allowed by LHC searches assuming $c_{gg}=0$ ($c_{gg}=0.1$). In the blue hashed region, the washout factor $\omega_{\rm sph}$ from sphalerons is below $10^{-2}$. The dot at $m_{\chi}=480\,\text{GeV}, N=5.3$ marks the point with the largest product of $\omega_{\rm tot}$ in the upper left panel and $sin[h/f]^2$ from the upper right panel, while satisfying all constraints for $c_{gg}=0$. }}
\label{fig:MNplotsTopGlueball}
\end{figure}

As discussed in Sec.~\ref{sec:Vh}, we assume that the right-handed top mixing has two contributions (see~Eq.~\eqref{eq:topquk}), one of which runs significantly between $ \chi = \chi_0$ and $ \chi = 0$. We choose the parameters which determine this running as in Table~\ref{tab:bench}. This choice ensures a large source of CPV during electroweak baryogenesis. Furthermore, we set several other parameters relevant for the Higgs-dilaton potential to the values given in Table~\ref{tab:bench}. This leaves $m_\chi$, $N$, $f$, $c^{(\chi)}_k$ undetermined. We have performed two numerical scans for the top quark scenario. In the first one, we fix $f=800\;$GeV. This minimizes the residual fine-tuning of the Higgs potential, while ensuring that the $v/f$-corrections to the Higgs couplings are small enough to satisfy constraints. We then set $c^{(\chi)}_k=2$ ($c^{(\chi)}_k=1$) for the glueball-like (meson-like) dilaton, and scan over $m_\chi, N$. We show contour lines in the $m_\chi, N$-plane of several quantities describing the phase transition in Figs.~\ref{fig:MNplotsTopGlueball}  and \ref{fig:MNplotsAlphaBeta} for a glueball-like dilaton and in Figs.~\ref{fig:MNplotsTopMeson} and \ref{fig:MNplotsAlphaBeta} for one that is meson-like. For the second scan, we in addition vary $f, c^{(\chi)}_k$ to better understand the dependence on these parameters. For each choice of $f, c^{(\chi)}_k$, we perform a scan over $m_\chi \subset [300 , 600] \, \text{GeV}$ and $N \subset [4,7]$ and demand that several constraints, to be discussed below, are fulfilled. The resulting allowed regions in the $f, c^{(\chi)}_k$-plane are shown in Fig.~\ref{fig:ckfplots}. In the following, we will discuss the results in more detail.

\begin{table}[t]
\centering
\begin{tabular}{|c|c|c|c|c|c| c| c| c| c| c| c| c|}
\hline
$c^{(h)}_k$ & $c_\epsilon$ & $c_\chi$   & $c_\alpha$ & $c_\beta$  & $\gamma_y$ & $c_y$ & $y_t[\chi_0]$  \\
\hline
1&  0.5 & 0.2 & -0.3 & 0.3  & -0.3 & 1.875 & $0.6 \sqrt{\lambda_t g_\star}$  \\
\hline 
\end{tabular}
\caption{\it \small Parameters chosen for the numerical studies. Here $\lambda_t$ is the top Yukawa and $y_t$ is one of two contributions to the right-handed top mixing, cf.~Eq.~\eqref{eq:topquk}. The remaining parameters, not given in the table, which determine the Higgs-dilaton potential are $m_\chi$, $N$, $f$, $c^{(\chi)}_k$. In the first scan, we fix $f=800\,$GeV and $c^{(\chi)}_k=2$ ($c^{(\chi)}_k=1$) for a glueball-like (meson-like) dilaton, and vary $m_\chi, N$. For the second scan, we in addition vary $f, c^{(\chi)}_k$.}
\label{tab:bench}
\end{table}

\subsection{Constraints on the parameter space}

Let us begin the discussion with the analysis of the relevant constraints. We have hashed the resulting excluded regions in the $m_\chi, N$-plane in Figs.~\ref{fig:MNplotsTopGlueball}, \ref{fig:MNplotsTopMeson} and \ref{fig:MNplotsAlphaBeta} and have also applied these constraints in Fig.~\ref{fig:ckfplots}. In the red hashed regions, there is no solution for the parameters $\alpha_0$, $\beta_0$, $\gamma_\epsilon$ and $\epsilon(\chi_0)$ which determine the zero-temperature potential leading to the required minima and masses for the Higgs and dilaton. In the orange and purple hashed regions, on the other hand, LHC constraints on respectively the Higgs couplings and the dilaton are not fulfilled. These constraints are discussed in more detail below. Furthermore, we have hashed regions in blue, where the washout factor $\omega_{\rm sph}$ due to sphalerons is smaller than $10^{-2}$. Note that this washout factor changes rapidly between 1 and 0 when crossing from the white region into the blue region. Demanding that the washout factor be larger than a somewhat different value will therefore not change the blue region significantly. Notice also that our numerical results for the blue hashed region confirm the bound that was derived in Eq.~\eqref{eq:chimassbound}.

\subsection{Properties of the phase transition}

\begin{figure}[t]
\centering
\vspace{.5cm}
\includegraphics[width=7.5cm]{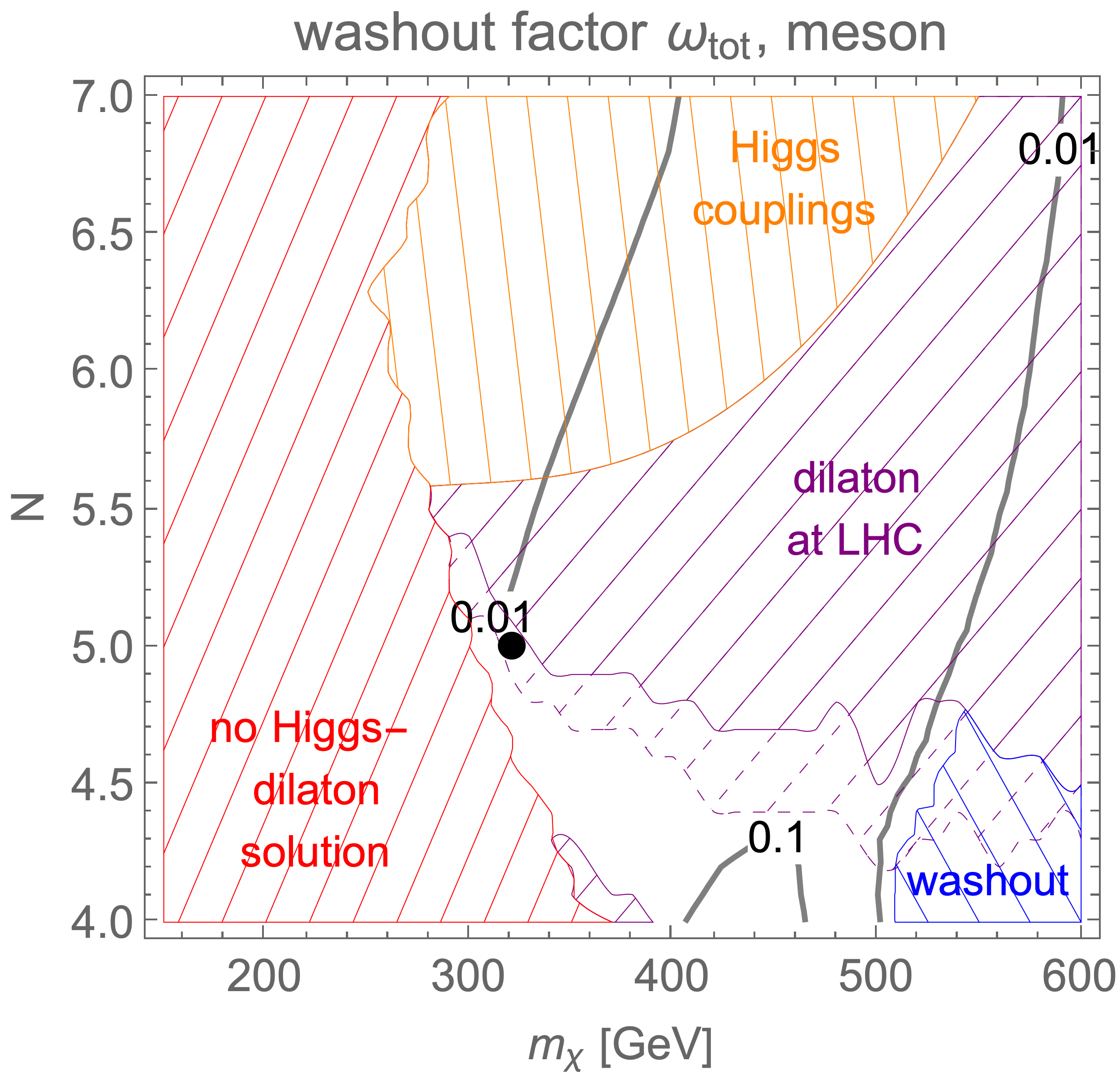}
\hspace{0.3cm}
\includegraphics[width=7.5cm]{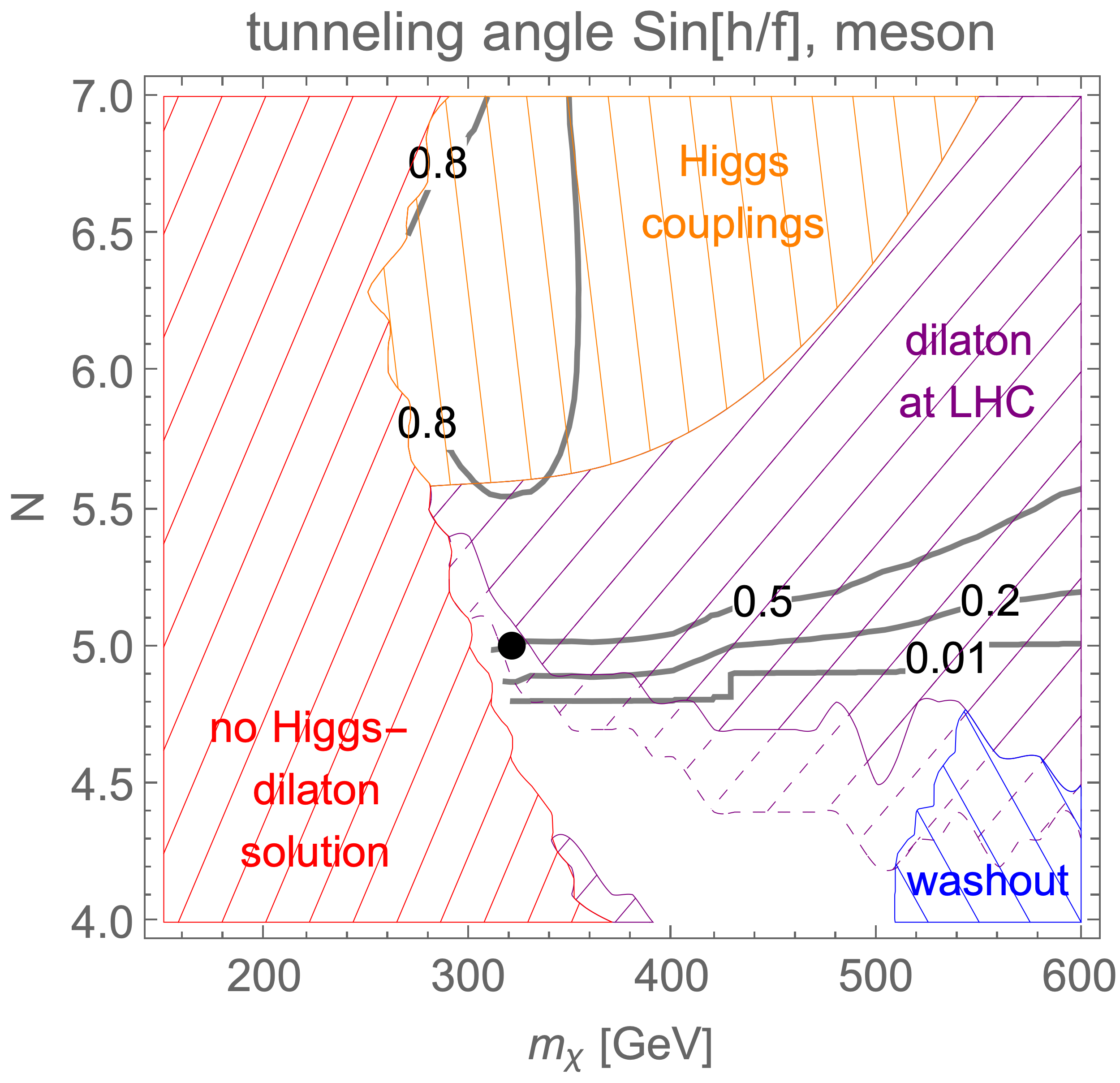} \\ 
\vspace{0.2cm}
\includegraphics[width=7.5cm]{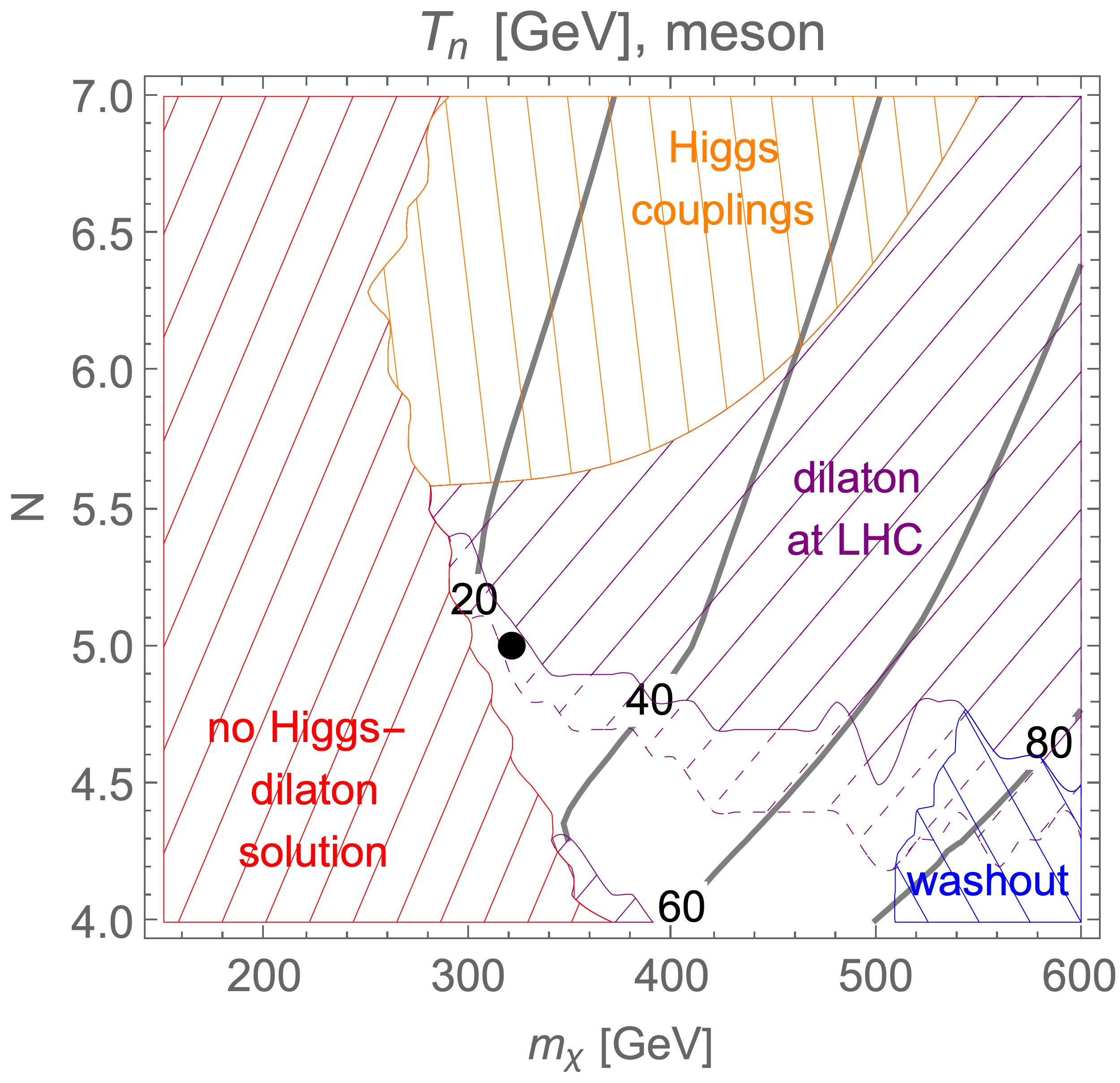}
\hspace{0.3cm}
\includegraphics[width=7.5cm]{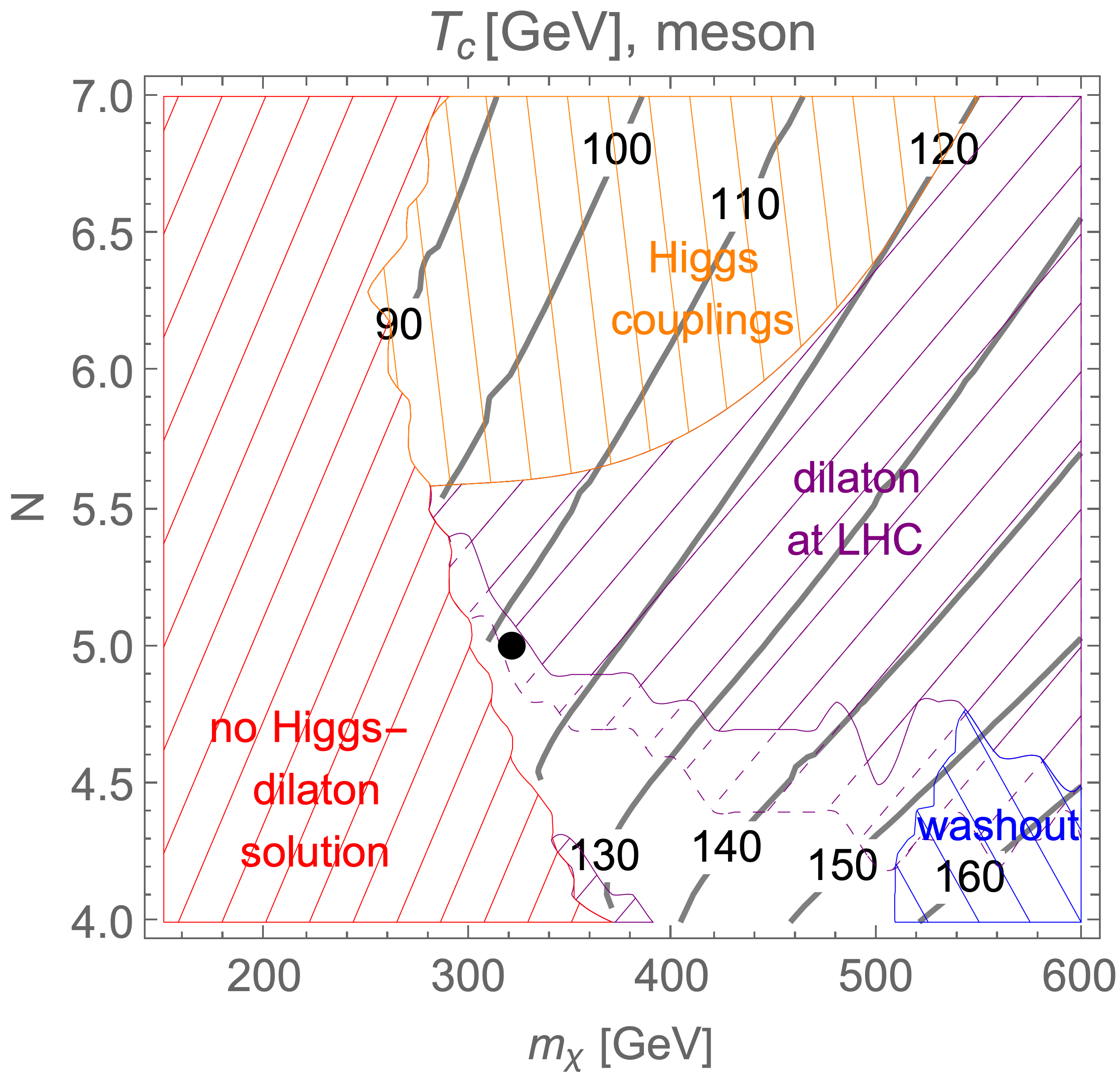}
\caption{\small \it{Results for a meson dilaton and with varying top Yukawa. The parameters that we have used are given in Table~\ref{tab:bench}. {\bf Upper left panel}: the total washout factor $\omega_{\rm tot}$ of the baryon asymmetry due to sphalerons and entropy injection. {\bf Upper right panel}:   the (sine of) the tunneling angle $\sin h/f$, which is important for the amount of CPV during the phase transition.  {\bf Lower left panel}: the nucleation temperature $T_n$ (in GeV). {\bf Lower right panel}: the critical temperature $T_c$ (in GeV). The color code for the hashed regions is the same as in Fig.~\ref{fig:MNplotsTopGlueball}. The dot at $m_{\chi}=320\,\text{GeV}, N=5$ marks the point with the largest product of $\omega_{\rm tot}$ in the upper left panel and $sin[h/f]^2$ from the upper right panel, while satisfying all constraints for $c_{gg}=0$.}}
\label{fig:MNplotsTopMeson}
\vspace{1.cm}
\end{figure}

Important quantities which describe the phase transition are the critical temperature $T_c$, where the phase transition becomes energetically possible, and the nucleation temperature $T_n$, where the phase transition completes.
We show contour lines for the critical temperature and the nucleation temperature in the lower left and right panels of Figs.~\ref{fig:MNplotsTopGlueball} and \ref{fig:MNplotsTopMeson}. In particular, the nucleation temperature should satisfy $T_n \lesssim 130 \, \text{GeV}$ in order to allow for a simultaneous confinement and electroweak phase transition. This is automatically fulfilled everywhere once $T_R < 130$~GeV.
Another interesting quantity is the ratio $T_c/T_n$ which measures the amount of supercooling of the phase transition. 
As one can see from Fig.~\ref{fig:MNplotsTopGlueball} this ratio is always greater than 1.
The value of $T_c/T_n$ is important as it enters the determination of the bubble wall velocity and
with $T_c/T_n$  larger than 1, one naively gets relativistic velocities from the perturbative estimates available in the literature, e.g.~\cite{Caprini:2019egz}. Too large bubble wall velocities would be dangerous for electroweak baryogenesis. However, we can expect the sizeable supercooling to be countered by the friction from the large number $\propto N^2$ of degrees of freedom in the plasma, and also by the strong coupling.  Unfortunately, there is no estimate in the literature for the bubble wall velocity that applies to strongly-coupled theories. It is therefore beyond the scope of our paper to conclude anything about the value of the bubble wall velocity in our framework.

As we have discussed in Section \ref{sec:EWBG}, the size of the CP-violating source depends on the squared top mass which in turn is proportional to the squared (sine of the) tunneling angle $\sin^2 h/ f$ during the phase transition. Since the size of CPV determines the amount of baryon asymmetry that is produced, we are interested in regions where $\sin h/ f$ is large. We plot contour lines of $\sin h/ f$ in the upper right panels of Figs.~\ref{fig:MNplotsTopGlueball} and \ref{fig:MNplotsTopMeson}.
As also discussed in Section \ref{sec:EWBG}, another quantity that determines the yield of EWBG in our scenario is the total washout factor $\omega_{\rm tot}$ due to sphalerons and entropy injection after reheating to the temperature $T_R$ given in Eq.~\eqref{eq:TR}. We plot contour lines of $\omega_{\rm tot}$ in the upper left panels of Figs.~\ref{fig:MNplotsTopGlueball} and \ref{fig:MNplotsTopMeson}. In order to compensate for this washout and reproduce the observed baryon asymmetry today, the latter has to be overproduced at the electroweak phase transition. We have tried different settings for the parameters that control the produced baryon asymmetry in our scenario (see \cite{Bruggisser:2018mrt} for a detailed discussion) and found that it is difficult to overproduce by much more than a factor 100. Correspondingly, $\omega_{\rm tot}$  should not be much smaller than $10^{-2}$. As can be seen in 
Figs.~\ref{fig:MNplotsTopGlueball} and \ref{fig:MNplotsTopMeson}, this requirement still leaves an open region in the $m_\chi, N$-plane.
Combining the effects of the tunneling angle and washout from sphalerons and entropy injection, we expect that our mechanism is most effective for the point in the $m_\chi, N$-plane (and some region around it) where $ \sin^2 h/f \cdot  \omega_{\rm tot}$  is maximized. Assuming $c_{gg}=0$ for the LHC constraints discussed below, this point is $m_\chi=480\, \text{GeV}, N=5.3$ ($m_\chi=320\, \text{GeV}, N=5$) for a glueball-like (meson-like) dilaton which we have marked by dots in Figs.~\ref{fig:MNplotsTopGlueball}, \ref{fig:MNplotsTopMeson} and \ref{fig:MNplotsAlphaBeta}.

\subsection{Gravitational waves}

First-order phase transitions can produce a stochastic background of gravitational waves. The two dominant sources for these gravitational waves are sound waves generated during the phase transition and collisions of bubble walls. One expects that the latter source is sizeable only if the bubble walls keep accelerating in the surrounding plasma. The calculation of the velocity of the bubble walls and whether they enter such a runaway regime is technically difficult and beyond the scope of this work. Due to the friction effects pointed out in \cite{Bodeker:2017cim}, however, one can estimate that the runaway regime only occurs for extreme amounts of supercooling (see \cite{Caprini:2019egz} for a detailed discussion). Since in the regions of the $m_{\chi},N$-plane that are allowed by constraints the phase transition is never very supercooled (cf.~the ratio $T_c/T_n$ from Figs.~\ref{fig:MNplotsTopGlueball} and \ref{fig:MNplotsTopMeson}), we will focus on sound waves as the source for the production of gravitational waves. 

\begin{figure}[t]
\centering
\vspace{.5cm}
\includegraphics[width=7.5cm]{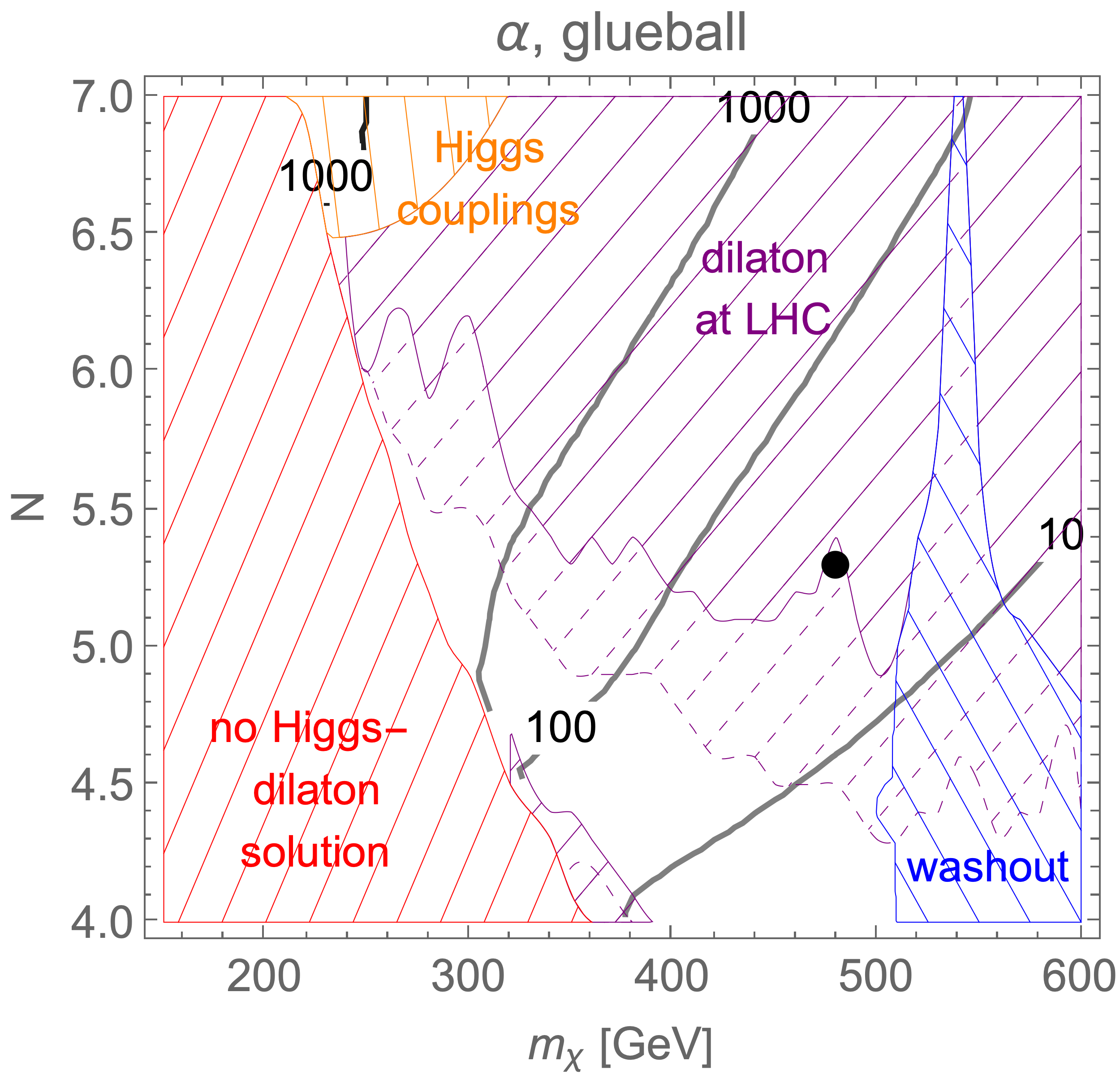}
\hspace{0.3cm}
\includegraphics[width=7.5cm]{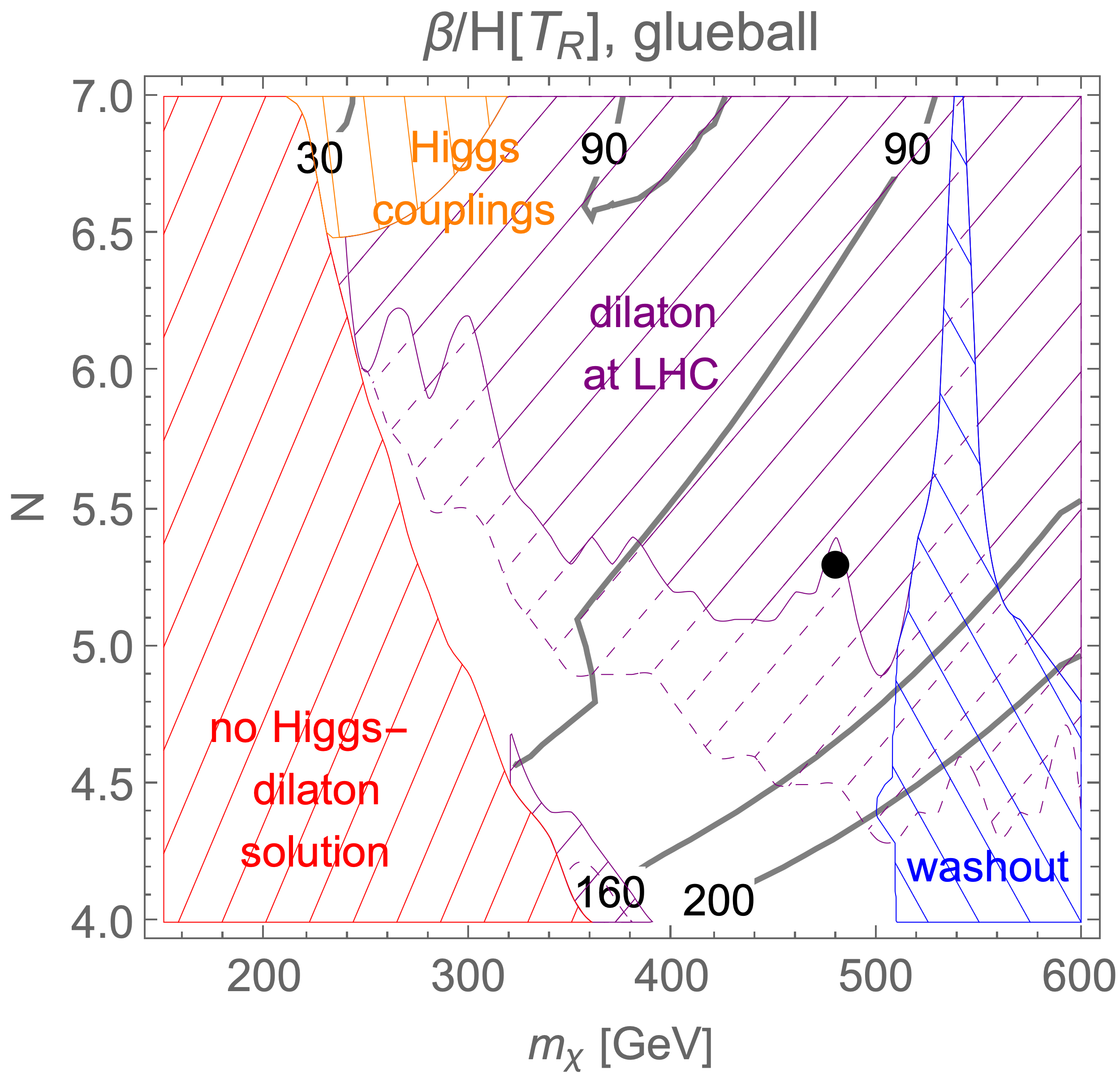} \\ 
\vspace{.5cm}
\includegraphics[width=7.5cm]{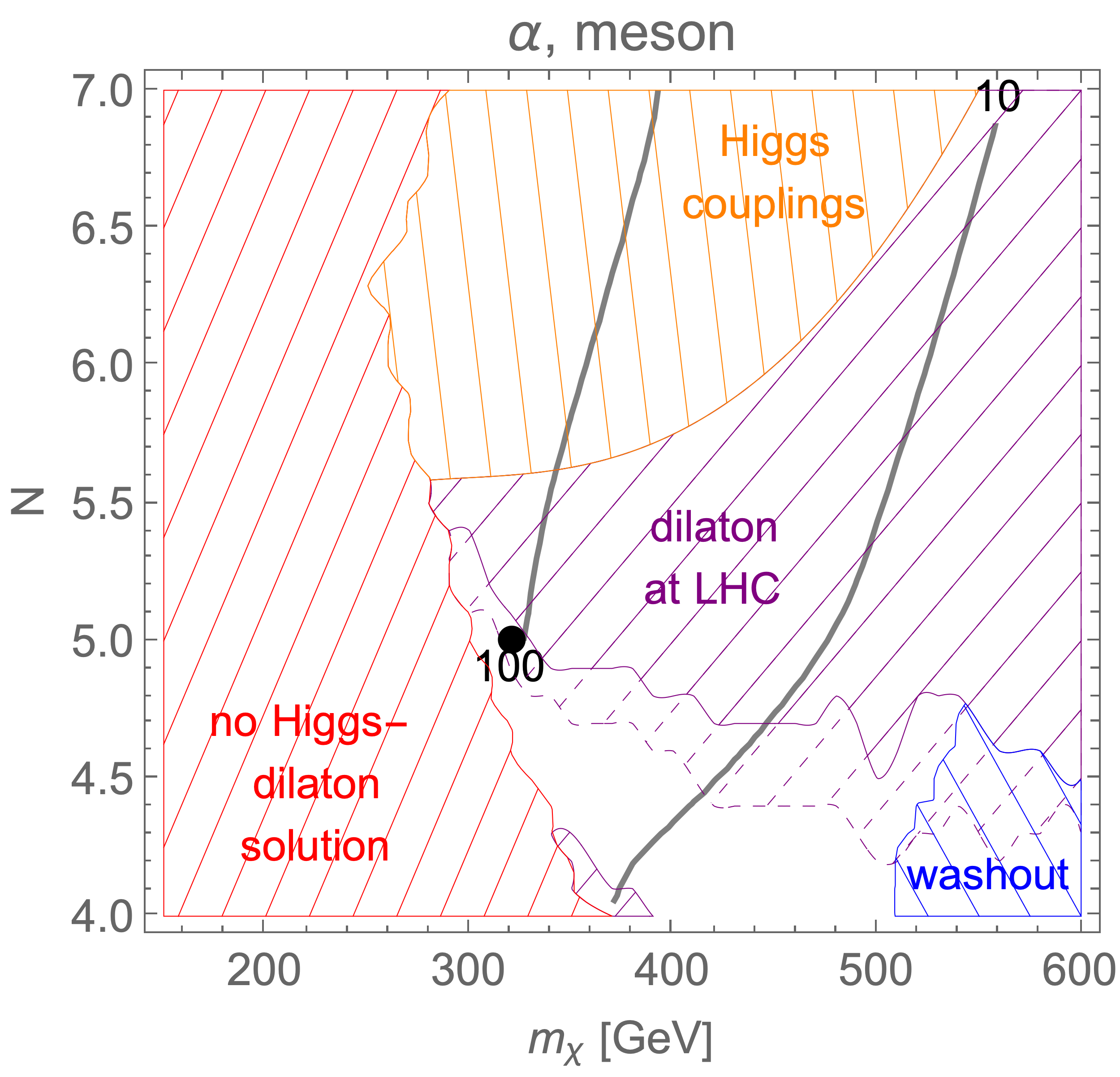}
\hspace{0.3cm}
\includegraphics[width=7.5cm]{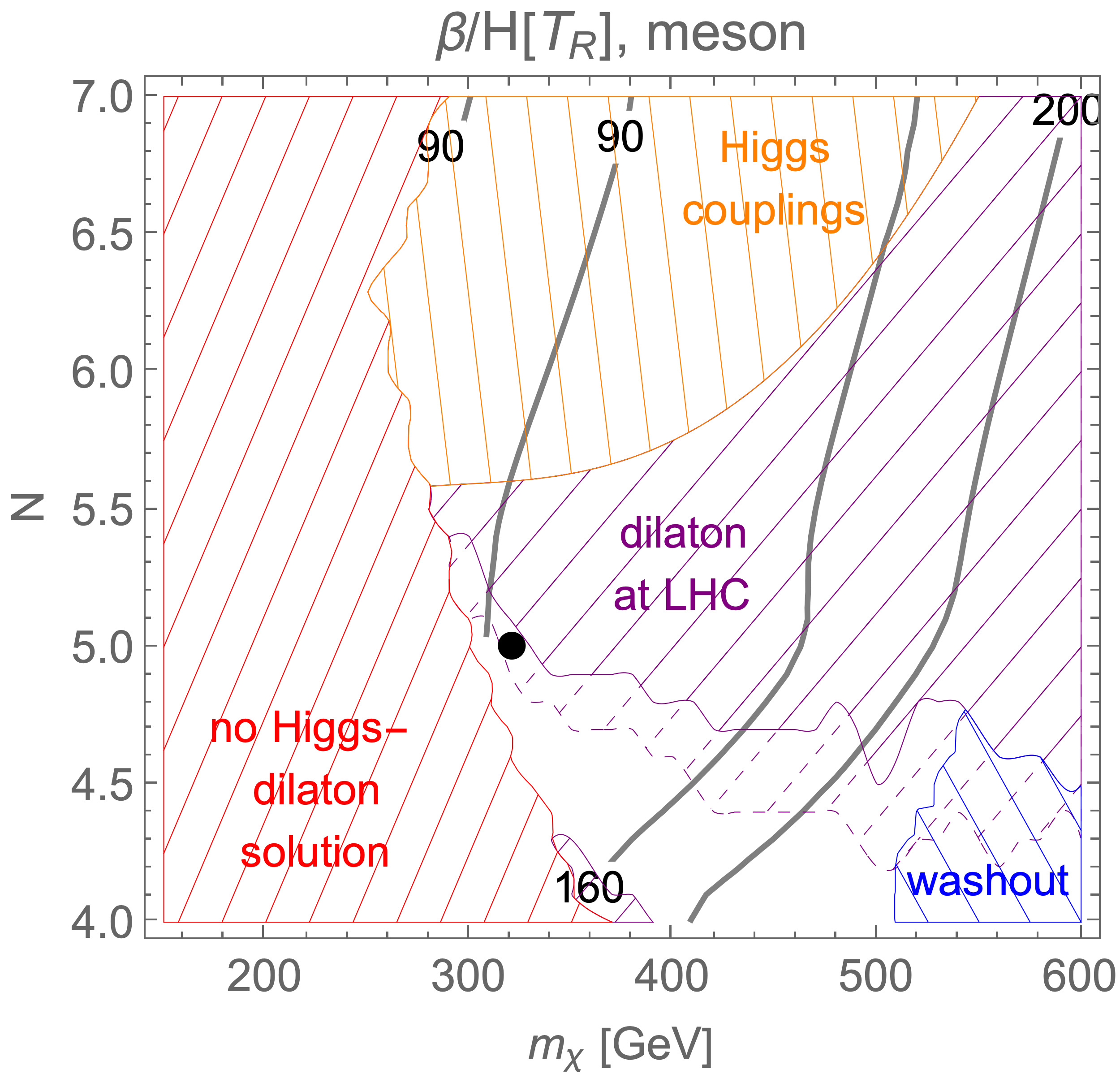}
\caption{\small \it{Results of our numerical study for a glueball dilaton ({\bf upper panels}) and a meson dilaton ({\bf lower panels}), both with varying top Yukawa. {\bf Left panels}: the strength of the phase transition $\alpha$. {\bf Right panels}: the inverse duration (over Hubble time) of the phase transition $\beta/H[T_R]$. The color code for the hashed regions is the same as in Fig.~\ref{fig:MNplotsTopGlueball}. The benchmark point for the glueball-like dilaton at $m_\chi=480\,\text{GeV}$, $N=5.3$ has $\alpha\simeq 31.3$, $\beta/H[T_R] \simeq 139$ and the one for the meson-like dilaton  at $m_\chi=320\,\text{GeV}$, $N=5$ has $\alpha\simeq 116.9$, $\beta/H[T_R] \simeq 94.5$.}}
\label{fig:MNplotsAlphaBeta} 
\vspace{1.4cm}
\end{figure}

The spectrum of these gravitational waves is mainly controlled by four parameters. The first parameter is the reheat temperature $T_R$ after the phase transition has completed, given in Eq.~\eqref{eq:TR}. Another important quantity measures the strength of the phase transition and reads
\be
\alpha \, \equiv \, \left(\frac{\Delta V}{\rho_{\rm rad}}\right)_{T_n}\simeq \, \frac{(V[0,0]-V[\chi_0,v_{\rm CH}])_{T_n}}{3 \pi^2 N^2 T_n^4/8}\, , 
\ee  
where $\Delta V$ is the latent heat released during the phase transition and $\rho_{\rm rad}$ is the energy density of the surrounding plasma at the nucleation temperature. We have plotted contour lines of $\alpha$ in the upper (lower) left panel of Fig.~\ref{fig:MNplotsAlphaBeta} for a glueball-like (meson-like) dilaton.

The spectrum also depends on $\beta \equiv [(d \Gamma/dt )/\Gamma]_{T_n}$, where $\Gamma$ is the bubble nucleation rate, which measures the inverse duration of the phase transition. Assuming fast reheating so that $H[T_n]=H[T_R]$ with $H$ being the Hubble rate, one finds
\be
\frac{\beta}{H[T_R]}\, \simeq \, \left( T \, \frac{d S_{\rm bub}}{dT}\right)_{T_n} ,
\ee
where $S_{\rm bub}$ is the bubble action.
Contour lines of $\beta/H[T_R]$ are shown in the upper (lower) right panel of Fig.~\ref{fig:MNplotsAlphaBeta} for a glueball-like (meson-like) dilaton. Finally, the fourth parameter is the bubble wall velocity $v_w$ which is the only one that we do not calculate and have to estimate.

\begin{figure}[t]
\centering
\includegraphics[width=7.3cm]{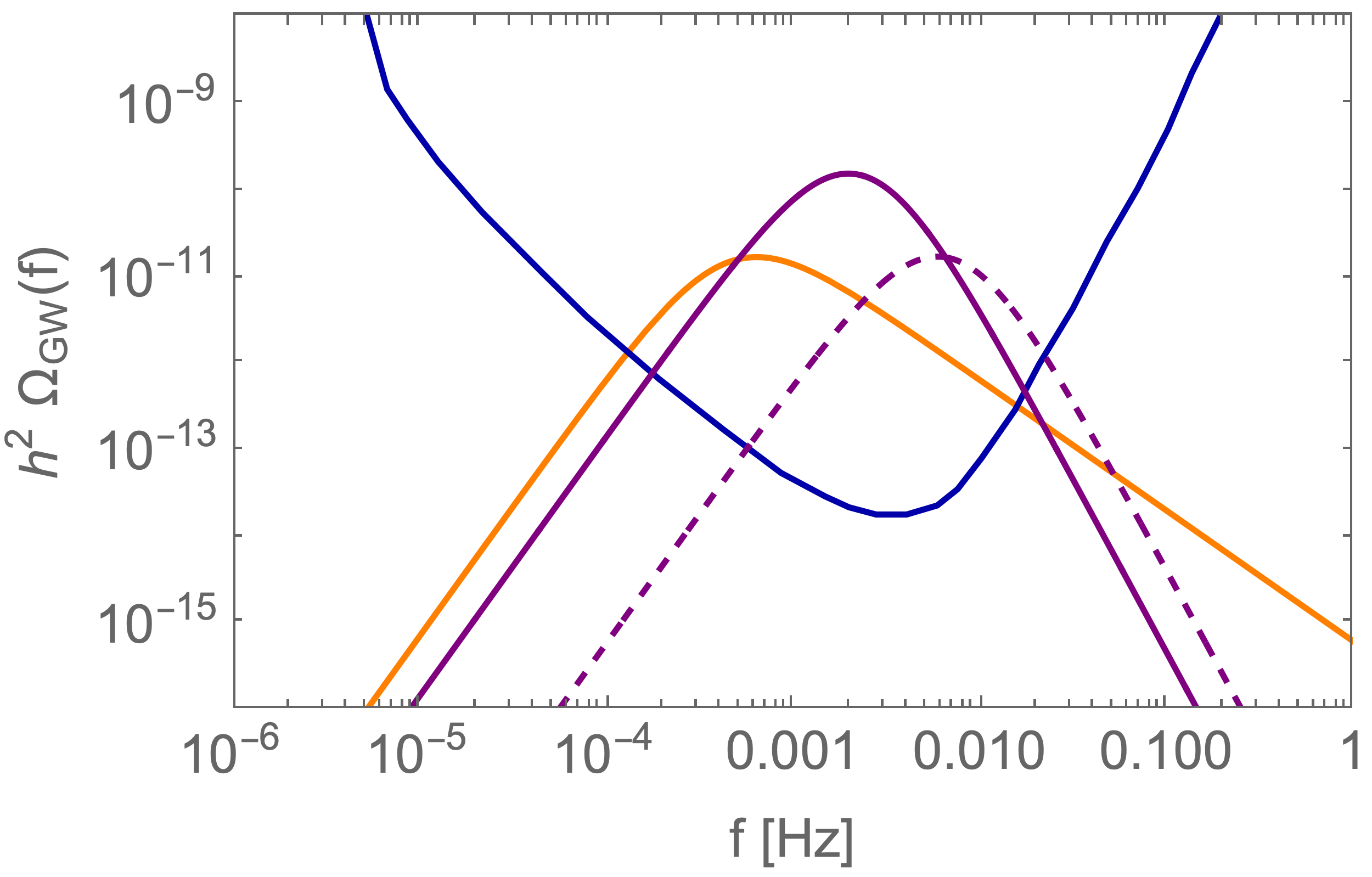} \hspace{.3cm} \includegraphics[width=7.3cm]{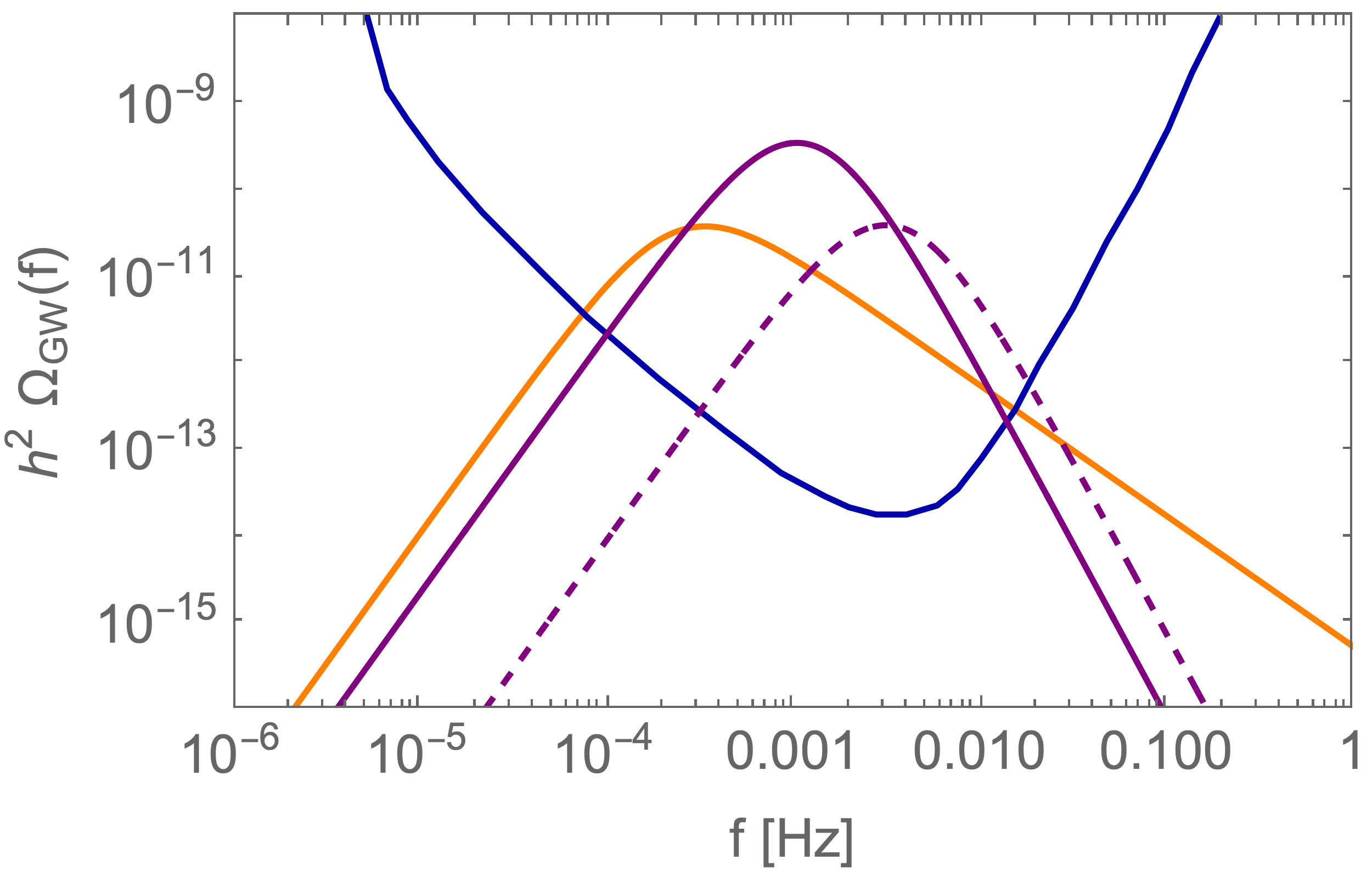}
\caption{\small \it{Gravitational-wave spectra for the benchmark points highlighted by dots in Figs.~\ref{fig:MNplotsTopGlueball}, \ref{fig:MNplotsTopMeson} and \ref{fig:MNplotsAlphaBeta}, corresponding to $m_\chi=480\,\text{GeV}$, $N=5.3$, $\alpha\simeq 31.3$, $\beta/H[T_R] \simeq 139$, $T_R=139.9 \, \text{GeV}$ for the glueball-like dilaton (left panel) and $m_\chi=320\,\text{GeV}$, $N=5$, $\alpha\simeq 116.9$, $\beta/H[T_R] \simeq 94.5$, $T_R=109.2 \, \text{GeV}$ for the meson-like dilaton (right panel). The gravitational waves are assumed to be dominantly produced by sound waves (purple lines) or bubble-wall collisions (orange lines), with the wall velocity set to $v_w=0.9$ (continous lines) and $v_w=0.3$ (dashed lines). We also show the sensitivity curve of LISA as expected for a 3-year mission (blue line).}}
\label{fig:GWspectra} 
\end{figure}

We have determined the gravitational-wave spectra for the benchmark points for the glueball and meson case which are marked by dots in Figs.~\ref{fig:MNplotsTopGlueball}, \ref{fig:MNplotsTopMeson} and \ref{fig:MNplotsAlphaBeta} (and which we estimate to have an optimal yield for the baryon asymmetry remaining at late times as discussed above). To this end, we have used the web-based tool \texttt{PTPlot} \cite{ptplot} which generates gravitational-wave spectra sourced by sound waves based on the results from \cite{Hindmarsh:2017gnf,Caprini:2019egz} and takes $T_R, \alpha, \beta/H[T_R], v_w$ as input. We have used the calculated values for the first three parameters and assumed two values for the bubble wall velocity, $v_w=0.3$ and $v_w=0.9$. The resulting spectra are shown in Fig.~\ref{fig:GWspectra}.
Note that the contribution from sound waves to the gravitational-wave spectrum has only been calculated for the case $\alpha \lesssim 0.1$. Since $\alpha$ in our case is much bigger, our prediction for the spectra from sound waves should be taken with a grain of salt. Given this uncertainty, we also show  
the spectrum obtained from bubble-wall collisions in the case $v_w=0.9$ using the results from \cite{Cutting:2018tjt}.
This is compared with the sensitivity curve of the planned gravitational-wave observatory LISA as expected for a 3-year mission. As one can see, for the assumed values of $v_w$ the gravitational waves which are produced for our benchmark points are within reach of LISA.

\subsection{Dependence on $f$ and $c^{(\chi)}_k$}

\begin{figure}[t]
\centering
\includegraphics[width=7.5cm]{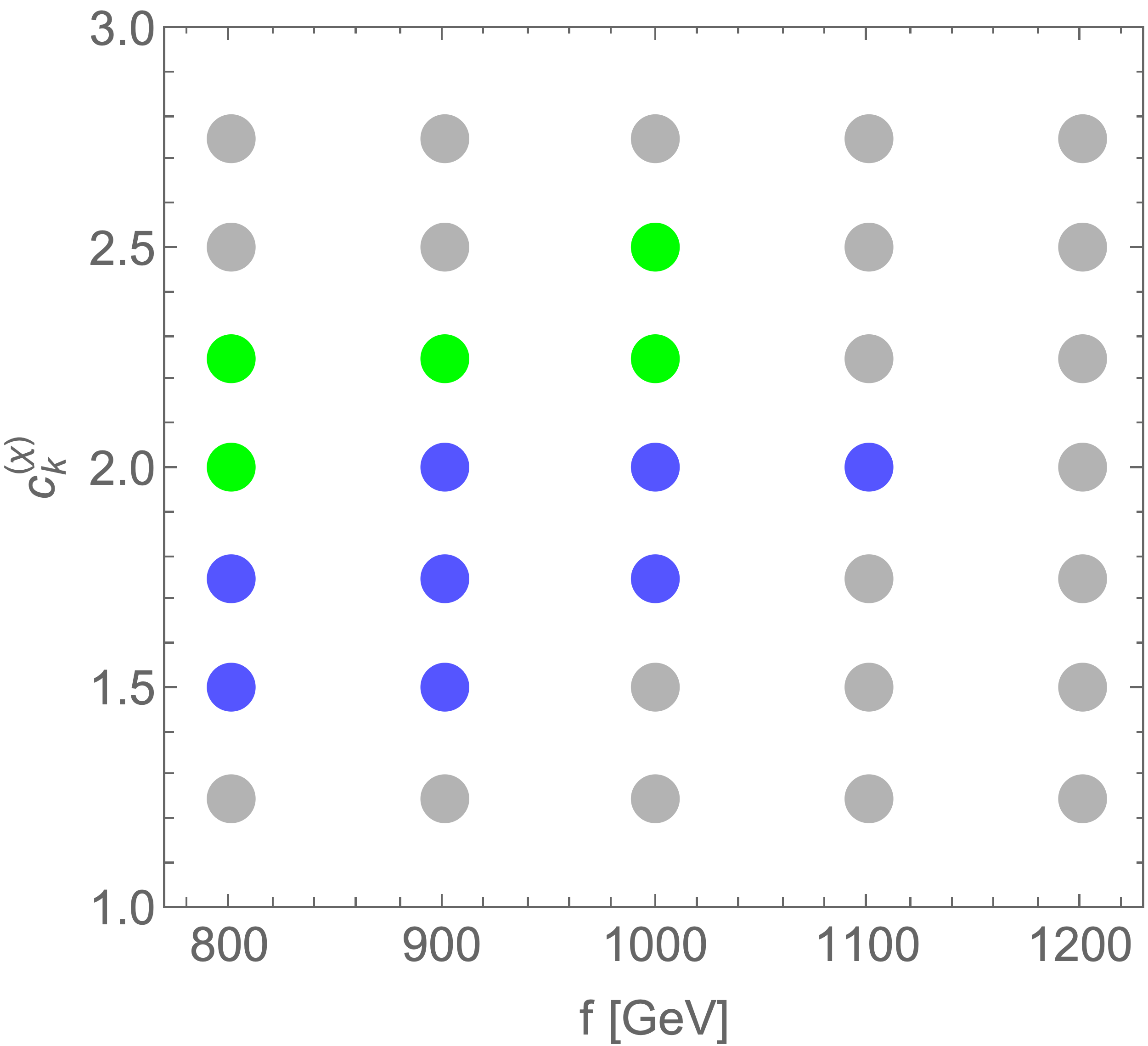}
\hspace{0.3cm}
\includegraphics[width=7.5cm]{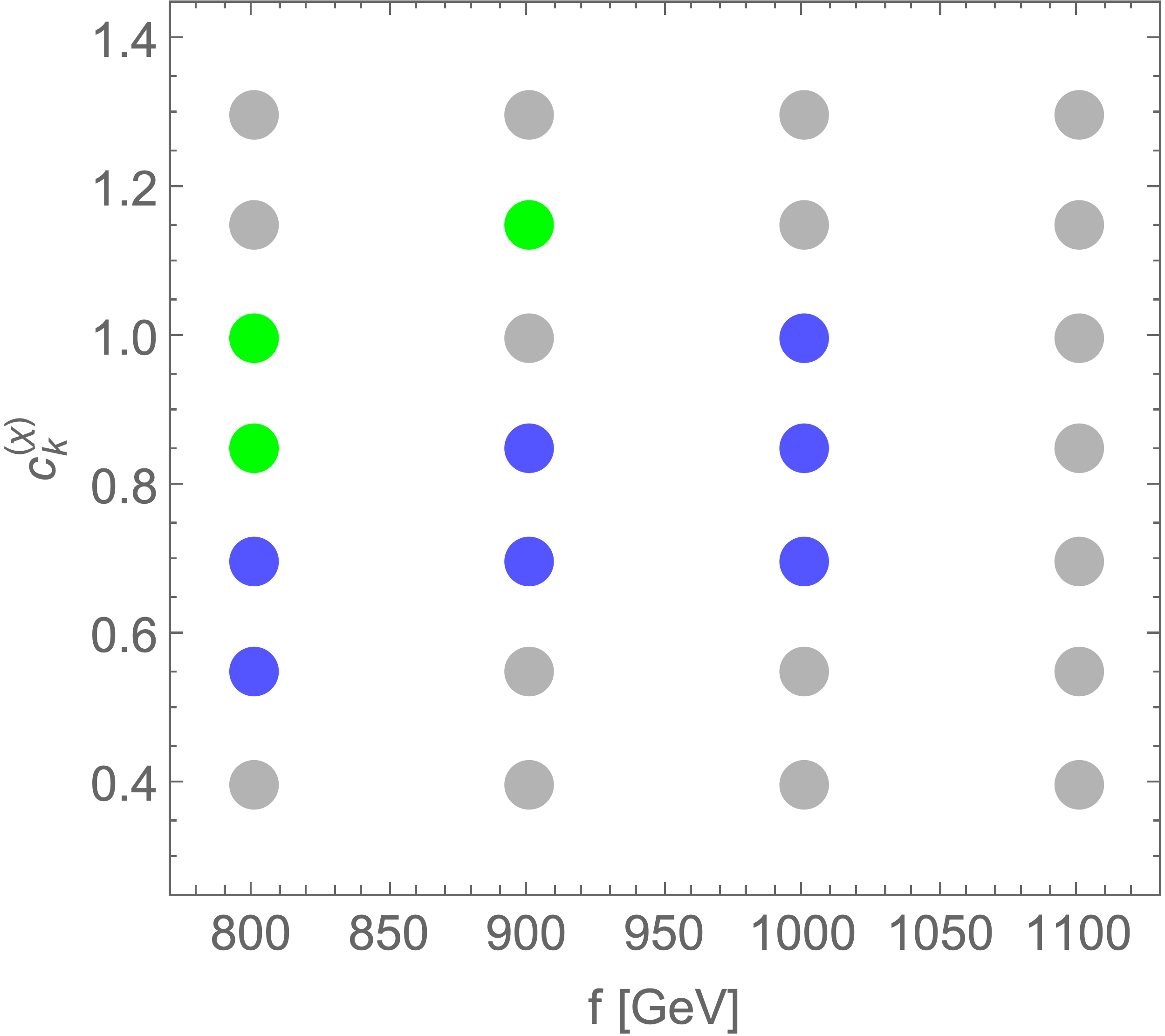} \\ 
\caption{\small \it{Results of our numerical scan over $f$ and $c_k^{(\chi)}$ for a glueball dilaton (left panel) and a meson dilaton (right panel), both with varying top Yukawa. For green points, we have found a nonvanishing region in the $m_\chi,N$-plane where $\omega_{\rm tot}$ is less than $10^{-2}$, $\sin h/f$ during the phase transition is larger than $0.1$ and collider constraints on the Higgs couplings and the dilaton are fulfilled. The corresponding values of $f$ and $c_k^{(\chi)}$ are thus expected to be suitable for EWBG. Blue points fulfill the same criteria, except the one on $\omega_{\rm tot}$ and instead $\omega_{\rm sph}$ is required to be less than $10^{-2}$. The corresponding values of $f$ and $c_k^{(\chi)}$ are expected to allow for cold electroweak baryogenesis.} The remaining grey points do not satisfy at least one of the mentioned criteria.}
\label{fig:ckfplots} 
\end{figure}

So far we have presented the results for fixed values of $c^{(\chi)}_k$ and $f=800$~GeV. Although this choice of $f$ minimizes the fine-tuning in the Higgs potential, an order-one increase of $f$ does not raise the fine-tuning drastically and therefore we will explore this possibility. Furthermore, we would like to estimate the impact of the variation of $c_{k}^{(\chi)}$ which is a free parameter of our EFT. We chose to vary this parameter as it  directly affects the relation between the Higgs-related scale $f$ and the scale $\chi_0$ which is crucial for both the phase transition properties and the collider phenomenology of the dilaton.  For reasons of computational limitations we will not vary the remaining free parameters. The result of the scan over $f$ and $c_{k}^{(\chi)}$ is presented in Fig.~\ref{fig:ckfplots}. The green points have a nonvanishing region in the $m_\chi,N$-plane where all previously mentioned constrains are satisfied,  the total washout factor $\omega_{\rm tot}$ from sphalerons and entropy injection is larger than $10^{-2}$ and the (sine of) the tunneling angle $\sin [ h/f]$ during the phase transition is greater than 0.1. The blue points feature too large a dilution of the baryon asymmetry due to entropy injection after reheating, given by the factor $(T_R/T_n)^3$. These points in parameter space however might be suitable for realizing cold electroweak baryogenesis~\cite{Krauss:1999ng,Garcia-Bellido:1999xos,Konstandin:2011ds,Servant:2014bla}. In this scenario, the baryon asymmetry is produced during the reheating of the universe after the bubble walls collide, which does not require any pre-existing density of SM particles, hence it is not sensitive to the ratio $T_R/T_n$. For these points, we therefore only require that the washout factor $\omega_{\rm sph}$ from sphalerons is larger than $10^{-2}$. Finally, the grey points in Fig.~\ref{fig:ckfplots} correspond to parameter values incompatible with both standard and cold electroweak baryogenesis. 

An important feature of the allowed parameter space shown in Fig.~\ref{fig:ckfplots} is the presence of an upper bound on the compositeness scale $f \lesssim 1000-1100$~GeV. Phenomenological constraints on the composite Higgs scenarios typically feature the opposite behaviour, predicting lower bounds on $f$, since in the limit $f\to \infty$ all the interactions become SM-like. An upper bound on $f$, on the other hand, typically comes from fine-tuning considerations.

Qualitatively, this behaviour of the parameter space favoured for EWBG can be understood as follows. The production cross-section of the dilaton goes like $1/\chi_0^2$, hence its non-observation provides a lower bound on $\chi_0 \propto f / c_k^{(\chi)}$. This results in a constraint of the type $c_k^{(\chi)} < \# f$, where $\#$ is some number, defining the upper border of the viable region in Fig.~\ref{fig:ckfplots}. According to Eq.~(\ref{eq:dilatonmass1}), the reheat temperature is related to the dilaton mass via $T_R^4 \propto  (f/c_k^{(\chi)})^2 m_\chi^2$. The upper bound on the reheat temperature from the requirement that sphalerons do not become active again after the EW phase transition  thus leads to a constraint $m_\chi^2 < \#(c_k^{(\chi)}/f)^2$. At the same time, according to Eq.~(\ref{eq:mixangle}), the Higgs-dilaton mixing goes as $\sin \theta \propto c_k^{(\chi)} f / m_\chi^2$. 
 The mass mixing is bounded from above to ensure the existence of a solution to the zero-temperature potential with the required mass and VEV for the Higgs and dilaton. In addition, it is also bounded from above due to constraints on the Higgs couplings.
The upper bound on the mixing therefore translates to a constraint $m_\chi^2 > \# c_k^{(\chi)} f$. Combining the two bounds on the dilaton mass we obtain $ (c_k^{(\chi)}/f)^2 > \#c_k^{(\chi)} f$, or $ c_k^{(\chi)} > \# f^3$. This forms the lower edge of the allowed region in Fig.~\ref{fig:ckfplots}.

\subsection{Collider and Higgs-coupling constraints}

\begin{figure}[t]
\centering
\includegraphics[width=7.5cm]{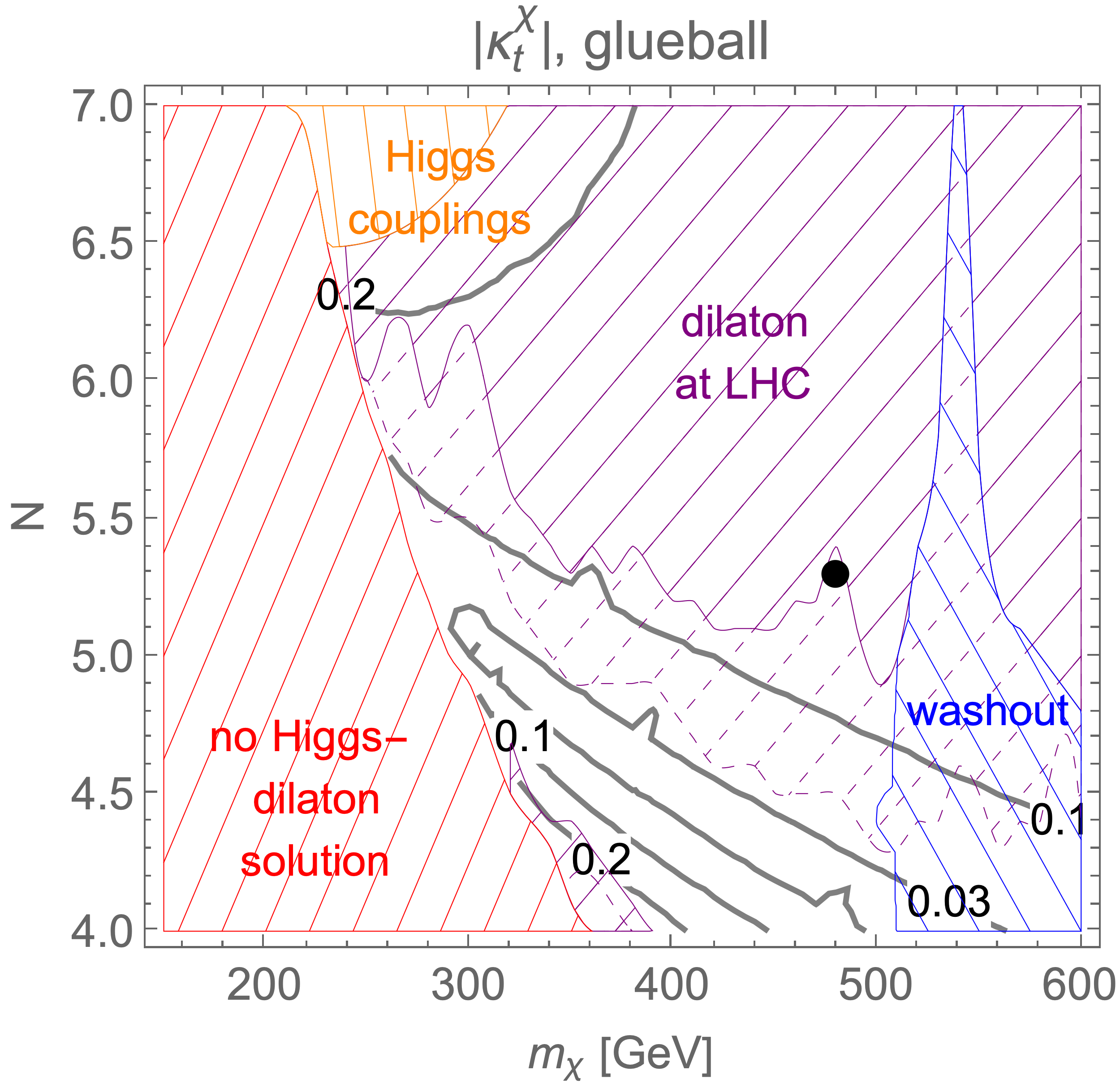}
\hspace{0.3cm}
\includegraphics[width=7.5cm]{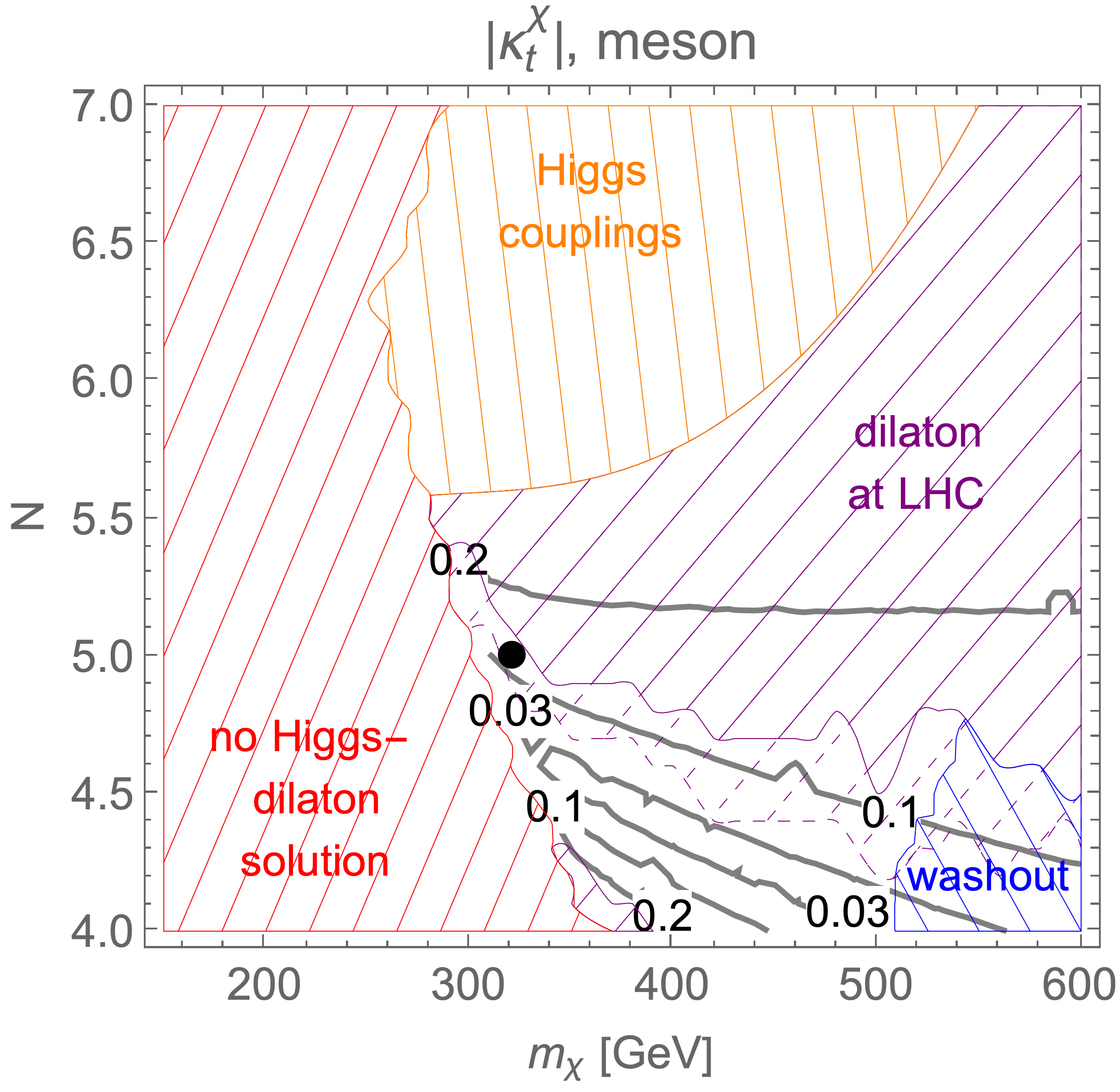} 
\caption{\small \it{Contour lines of the dilaton-top coupling $\kappa^\chi_t$ from Eq.~\eqref{eq:chitop} for a glueball dilaton (left panel) and a meson dilaton (right panel), both with varying top Yukawa. The color code for the hashed regions is the same as in Fig.~\ref{fig:MNplotsTopGlueball}.}}
\label{fig:MNplotsKappa} 
\end{figure}

We have derived the  95\%CL  LHC bounds on the dilaton using the results presented in Ref.~\cite{Bruggisser:2022ofg} which are based on the HiggsTools software~\cite{Djouadi:2018xqq,Bechtle:2020pkv,Bechtle:2020uwn,Bahl:2022igd}. The main production channel for dilaton particles is gluon fusion, and they decay dominantly into W and Z bosons. Gluon fusion production is greatly enhanced in the presence of a non-vanishing direct coupling between gluon and dilaton, which in the notation of Ref.~\cite{Bruggisser:2022ofg} reads
\be\label{eq:chigg}
{\cal L} \supset \frac {c_{gg}} 3  \frac {g_s^2}{g_*^2} \frac{c_\theta \hat\chi}{\chi_0} \,  G_{\mu \nu}^a G^{a \mu \nu}.
\ee
Here $g_s$ is the QCD coupling, $c_\theta$ is the cosine of the Higgs-dilaton mixing angle, and $\hat \chi$ is the dilaton mass eigenstate which is related to the original fields by the redefinition
\be\label{eq:massbasis}
\chi = \chi_0+ c_\theta \hat \chi - s_\theta \hat h, \quad \quad  \; h = v + c_\theta \hat h + s_\theta \hat \chi\,,
\ee 
where $\hat h$ is the Higgs mass eigenstate.
The interaction~(\ref{eq:chigg}) is generated by the strong sector and is controlled by the coefficient $c_{gg}$ whose exact value can only be inferred from a complete UV theory of the strong sector. In order to pass the current stringent experimental constraints on the dilaton we have only considered $c_{gg}=0$ and $c_{gg}=0.1$ in this section. As one can see in Fig.~\ref{fig:MNplotsTopGlueball} the allowed parameter space for $c_{gg} =  0.1$ shrinks by about 50\% compared to $c_{gg}=0$. For $c_{gg}= 1$ it almost completely vanishes. Yet, even for $c_{gg}=0$ a dilaton coupling to gluons is generated via top quark loops, proportional to the dilaton-top coupling
\bea\label{eq:dilatontop}
{\cal L}& 
\, \supset \, & \label{eq:chitop}
- \frac {\lambda_t} {\sqrt 2} \left\{s_\theta \cos \frac{v_{\rm CH}}{f}+ c_\theta (1+\gamma_{t}) \frac{v_{\text{SM}}}{\chi_0}  \right\} \bar t t \hat \chi + \text{h.c.} \, \equiv \, - \frac {\lambda_t} {\sqrt 2}  \kappa^{\chi}_t \bar t t \hat \chi + \text{h.c.} ,
\eea
where $\gamma_t = d \log \lambda_t / d \log \mu$ with $\lambda_t$ given in Eq.~\eqref{eq:topquk}.
Note that this coupling decreases if the anomalous dimension $\gamma_{t}$ or the Higgs-dilaton mixing angle $s_\theta$ are negative. In the scenario where CPV is generated by a varying top quark Yukawa coupling we indeed need $\gamma_t$ to be negative and sizeable.
This reduces the size of the second term in Eq.~\eqref{eq:chitop} and thereby the gluon-dilaton coupling. Moreover, in this case a sizeable mixing $s_\theta$ is automatically generated due to the large size of the top quark Yukawa coupling at $\chi=\chi_0$, see Eq.~(\ref{eq:mixangle}). If $s_\theta$ is negative, this results in an accidental cancellation between the two terms in Eq.~\eqref{eq:chitop} and in a further reduction of the gluon-dilaton coupling. We plot the contour lines of $\kappa^\chi_t$ in Fig.~\ref{fig:MNplotsKappa} which shows that the cancellation reduces the coupling along a valley for small $m_\chi, N$. As one can see, this produces a window in the parameter space where the LHC bounds can be satisfied (cf.~the white region in Fig.~\ref{fig:MNplotsKappa}). 
Note also that a sizeable negative $s_\theta$ can decrease the deviations of the composite Higgs couplings to massive vector bosons and quarks from their SM predictions~\cite{Bruggisser:2022ofg}. The corresponding coupling modifier with respect to the SM prediction for Higgs-W,Z couplings is given by
\be\label{eq:kapavh}
\kappa_{V}^{h} = c_\theta \cos \frac{v_{\rm CH}}{f}  -  s_\theta \frac{g_\chi}{g_*} \sin \frac{v_{\rm  CH}}{f}.
\ee
In our scans we have imposed the current $2 \sigma$ limits on the deviation of the Higgs couplings to vector bosons~\cite{CMS:2022dwd,ATLAS:2022vkf}, leading to the constraint $0.936<\kappa_{V}^{h} < 1.106$.

Furthermore, from the expression for the mixing angle~(\ref{eq:mixangle}) we find that $s_\theta \propto v_{\text{SM}}/\chi_0$. Hence both terms in the dilaton-top coupling~(\ref{eq:dilatontop}) scale as $v_{\text{SM}}/\chi_0$. Since $\chi_0 \propto (g_*/g_\chi) f$ (see Eq.~(\ref{eq:chioff})), this means that the collider constraints get relaxed for large $f$ and $c_k^{h}/c_k^{\chi}$. Additionally, for the glueball-like dilaton one finds that $\chi_0$ grows with $\sqrt N$ which also suppresses the bounds. 

The situation with the collider bounds is significantly different in the scenario with charm-induced CPV. First of all, the typical values of the Higgs-dilaton mixing angle in this case are much lower, due to the smaller charm Yukawa coupling (see Eq.~(\ref{eq:mixangle}) with $y_t \to y_c \propto \sqrt{\lambda_c}$). Moreover, $\gamma_t$ is zero in our benchmark scenario. Hence the dilaton-top coupling~(\ref{eq:dilatontop}) does not experience any accidental cancellations. This results in all the parameter space acceptable for EWBG being excluded by the collider bounds. 
In principle, the Higgs-dilaton mixing angle can be increased by raising the size of the coefficient $c_{\alpha}$. However, in this case we observe that large negative mixing angles, needed to cancel the dilaton-top coupling, are always accompanied by a sizeable detuning in the Higgs potential at $\chi < \chi_0$ in such a way that the preferred phase transition trajectory becomes $h=0$. 
In this case the CP-violating source~(\ref{eq:cpv}), which is proportional to space derivatives of the quark mass matrix, 
vanishes and no baryon asymmetry can be generated. 
Note that in the Conclusions we show a simple way to fix the problem with collider  bounds in the charm-induced CPV scenario.

\section{Constraints from the electron EDM}\label{sec:edm}

In our benchmark scenario with a varying top Yukawa, there are CP-violating couplings between the top and the Higgs and also between the top and the dilaton.  These, in turn, contribute to the electron EDM, which we will quantify in this section. Beyond the CP-violating couplings relevant for EWBG which we analyse, there are generic CPV sources in composite Higgs models that can contribute sizeably to the electron EDM~\cite{Panico:2017vlk,Panico:2018hal}, or to CP-violating flavour physics observables, leading to severe bounds on composite Higgs models. These additional CP-violating interactions are {\it a priori} independent from the interactions that are relevant for our work. 
In order to systematically suppress the unwanted contributions to the EDM induced by generic CPV sources, and also to satisfy the flavour physics constraints, one has to assume either some additional dynamical mechanisms (see e.g.~\cite{Panico:2016ull}) or symmetries (e.g.~\cite{Redi:2011zi}). However, these new ingredients can also constrain or forbid the CPV needed for EWBG. For example, the proposed $U(2)$~\cite{Redi:2012uj,Barbieri:2012uh,Barbieri:2011ci} and $U(3)$~\cite{Cacciapaglia:2007fw,DiLuzio:2019wsw,Redi:2011zi} flavour symmetries do not allow for sizeable mixing between different SM quarks at small dilaton values $\chi < \chi_0$, which is needed to generate CPV in our charm benchmark scenario.
Less constraining scenarios, like the approximately $U(1)$-symmetric and CP-conserving composite Higgs model of Ref.~\cite{Redi:2011zi},
can in principle be compatible with both our top and charm benchmark scenarios, however the allowed amount of CPV in the mixings between the elementary and composite fermions becomes constrained. A dedicated analysis would be needed to access the compatibility of this model with the current experimental bounds and EWBG.  
Note however that the assumptions about new symmetries in the composite sector can be relaxed if, for example, the ``generic'' correction to the electron EDM is accidentally suppressed by a factor of order 10. We leave a detailed analysis of this topic for future work.

\begin{figure}
\begin{center}
\includegraphics[scale=0.79]{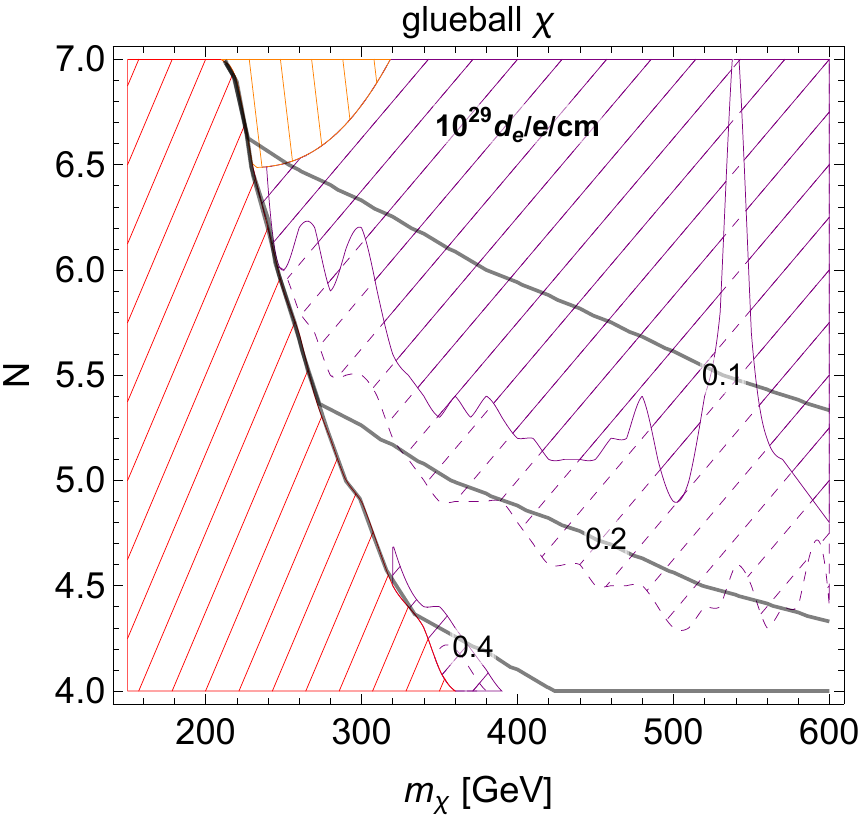}
\hspace{1cm}
\includegraphics[scale=0.785]{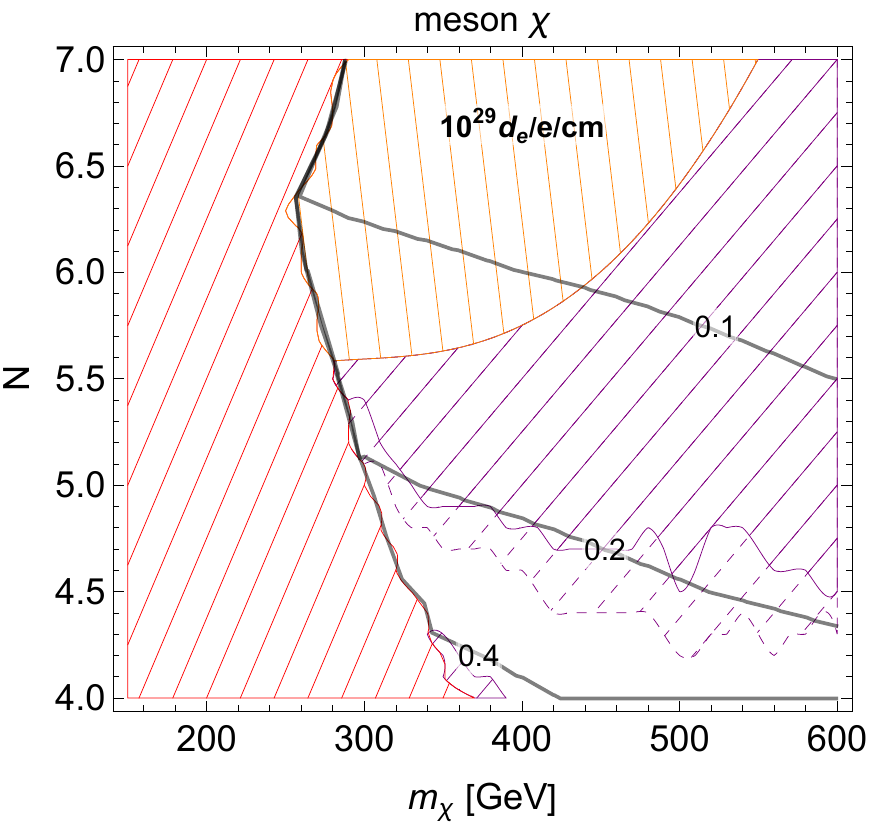}
\end{center}
\caption{{\it Electron EDM for the glueball (left panel) and meson (right panel) case, with the varying top Yukawa coupling, to be compared with the current bound $|d_e|/e <1.1 \cdot 10^{-29} \text{cm}$. The parameters are set as for the plots in Figs.~\ref{fig:MNplotsTopGlueball}, \ref{fig:MNplotsTopMeson} and \ref{fig:MNplotsAlphaBeta}, in particular $c_k^{(\chi)}=2, c_k^{(h)}=1$ ($c_k^{(\chi)}=1, c_k^{(h)}=1$) for the glueball (meson) dilaton. The complex phase $\phi$ is fixed to $0.1$. The remaining parameters are given in Table~\ref{tab:bench}.  The color code for the hashed regions is the same as in Fig.~\ref{fig:MNplotsTopGlueball}.
Note that the case with varying charm Yukawa features a suppression in the electron EDM of at least $m_c/m_t\sim 10^{-2}$, making this bound irrelevant.}}

\label{fig:edm}
\end{figure}

Let us now come back to the CP-violating interactions related to EWBG. A sizeable contribution to the electron EDM can be generated in the scenario with top-induced CPV. This contribution is sourced by the top Yukawa term
\be
{\cal L}_{\text {Yuk}} = - \frac{\lambda_t(\chi)}{\sqrt{2}} (g_\chi\chi /g_*) \sin \frac h f \bar t_L t_R + h.c. 
\supset  - \frac{1}{\sqrt{2}} \left\{\lambda_t+\frac{\partial \lambda_t}{\partial \log \chi} \frac {\chi-\chi_0}{\chi_0} \right\} v_{\text SM}  \bar t_L t_R + h.c.
\ee
The CP-violating coupling arises from the complex part of $\partial \lambda_t/\partial \log \chi$.
To evaluate it we will use the expression~(\ref{eq:topquk}) for the top Yukawa as a function of mixings 
\be\label{eq:topquk1}
\lambda_t = y_{tL} (y_t(\chi) e^{i \phi}+y_{tR}^{(2)})/g_*,
\ee
where we assume $y_{tL}, y_{tR}^{(2)}, y_t(\chi)$ to be real, with the complex phase parametrized by $\phi$. Using the  energy scaling~(\ref{eq:yukrun}) for $y_t(\mu)$, with $\partial y_t/\partial \log \mu \simeq \gamma_y y_t$, we then find
\be
\frac{\partial \lambda_t}{\partial \log \chi} = \lambda_t  \,\frac{ \gamma_y y_t(\chi_0) e^{i \phi}}{y_{tR}^{(2)}+y_t(\chi_0) e^{i \phi}} 
\equiv \lambda_t \, \gamma_t.
\ee
Finally we can write down the CPV interactions
\bea\label{eq:imtopyuk}
{\cal L}_{\text {CPV Yuk}} 
&=& 
- i \frac{|\lambda_t|}{\sqrt 2} \text{Im}[\gamma_t] \frac{\chi-\chi_0}{\chi_0} v_{\text{SM}}\bar t \gamma_5 t \\
&=&
-  i\frac{|\lambda_t|}{\sqrt 2} \text{Im}[\gamma_t]  \frac{v_{\text{SM}}}{\chi_0} \left( c_\theta \hat \chi - s_\theta \hat h  \right) \bar t \gamma_5 t 
\,\equiv \,
- i \frac{|\lambda_t|}{\sqrt 2} \left( \tilde \kappa_t^\chi \hat \chi  + \tilde \kappa_t^h \hat h\right)\bar t \gamma_5 t ,
\eea
where in the second line we switched to the mass eigenstates basis (\ref{eq:massbasis}). 

The two-loop Barr-Zee-type diagrams with one internal dilaton or Higgs propagator, one internal photon, and a top quark loop give the following contribution to the electron EDM~\cite{Barr:1990vd,Brod:2013cka}
\be\label{eq:eedm}
d_e/e = \frac{16}{3} \frac{\alpha_{\text{EM}}}{(4\pi)^3} \sqrt 2 G_F m_e \left( \kappa_e^h \, \tilde \kappa_t^h f_1\left[\frac{m_t^2}{m_h^2}\right] +\kappa_e^\chi\, \tilde \kappa_t^\chi f_1\left[\frac{m_t^2}{m_\chi^2}\right] \right),
\ee
where $G_F \simeq 1.166 \cdot 10^{-5}$ GeV$^{-2}$. For the CP-preserving electron-dilaton and electron-Higgs couplings $\kappa_e^\chi$ and $\kappa_e^h$ we use the following expressions~\cite{Bruggisser:2022ofg}  (neglecting a possible energy scaling of the electron Yukawa)
\bea\label{eq:dilatoncp}
\kappa_e^\chi  & = & s_\theta \cos \frac{v_{\rm CH}}{f} + c_\theta \frac{v_{\text{SM}}}{\chi_0}, \\
\kappa_e^h & = & c_\theta \cos \frac{v_{\rm CH}}{f} - s_\theta  \frac{v_{\text{SM}}}{\chi_0}.
\eea
Furthermore, the loop function is 
\be
f_1[x]= \frac{2x}{\sqrt{1-4x}} \left\{
\text{Li}_2 \left[ 1 - \frac{1-\sqrt{1-4x}}{2x} \right] -\text{Li}_2 \left[ 1 - \frac{1+\sqrt{1-4x}}{2x} \right] 
\right\}
\ee
with
\be
\text{Li}_2[x] = - \int_0^x d u \frac{\ln [1-u]}{u}.
\ee

Note that $f_1[m_t^2/m_h^2] \simeq 3$ while at small $x$ one has $f_1[x] \simeq x (\pi^2 + 3 \log^2 x)/3$. Using these approximations we can estimate Eq.~(\ref{eq:eedm}) as
\be\label{eq:eedmappr}
d_e/e \simeq 16 \frac{\alpha_{\text{EM}}}{(4\pi)^3} \sqrt 2 G_F m_e \text{Im}[\gamma_t]  \frac{v_{\text{SM}}}{\chi_0} \left\{ - s_\theta  + \frac{m_t^2}{m_\chi^2} \left(\frac{v_{\rm SM}}{\chi_0} +s_\theta\right) \left(1+ \frac 1 3 \log^2 \frac{m_t^2}{m_\chi^2}\right)\right\},
\ee
where we assumed $s_\theta \ll 1, v_{\rm SM}/{\chi_0} \ll 1$.
Hence, the contribution to the electron EDM decreases at large $\chi_0 \propto \sqrt N f$. Furthermore, as follows from Eq.~(\ref{eq:mixangle}), the Higgs-dilaton mixing scales as $s_\theta \propto 1/m_\chi^2$ and therefore the overall correction to $d_e/e$ also scales as $1/m_\chi^2$, leading to a suppressed EDM for large dilaton masses. However, as follows from our analysis, successful EWBG requires a relatively light dilaton and relatively low values of $N$, resulting in a non-negligible contribution to $d_e/e$.

Currently the strongest bound on the electron EDM comes from the ACME collaboration~\cite{ACME:2018yjb}. The 90\% CL upper bound reads 
\be\label{eq:imtopyukbound}
|d_e|/e \, < \, 1.1 \cdot 10^{-29} \text{cm} \, \simeq \, 5.6 \cdot 10^{-16} \text{GeV}^{-1}.
\ee
The predicted values of $d_e/e$ from Eq.~(\ref{eq:eedm}) are shown in Fig.~\ref{fig:edm}. For these plots we have set the complex phase $\phi = 0.1$ as an estimate of what would be needed to generate a sufficient amount of baryon asymmetry~\cite{Bruggisser:2018mus,Bruggisser:2018mrt}. For the points preferred by EWBG, $N\sim 4-5$ and $m_\chi \sim 300-500$~GeV (see Figs.~\ref{fig:MNplotsTopGlueball}, \ref{fig:MNplotsTopMeson}) the predicted values of $|d_e|/e$ are less than an order of magnitude away from the current limit.
Although a comprehensive analysis of the baryon asymmetry generation and its interplay with EDMs is beyond the scope of this paper, we can conclude that the next-generation EDM experiments can provide a decisive test of our  EWBG benchmark with top Yukawa-induced CPV.

\section{Discussion} \label{sec:conc}

We have presented an update of the analyses~\cite{Bruggisser:2018mus,Bruggisser:2018mrt} of EWBG in scenarios where the EW phase transition is triggered by the confinement phase transition of the new strongly-interacting sector that produces a composite Higgs boson.
To this end, we have employed an effective field theory containing the Higgs and the dilaton as pseudo-Nambu-Goldstone bosons arising from the spontaneous breaking of respectively a global flavour symmetry and conformal invariance of the strong sector.
The latter field serves as an order parameter for the confinement phase transition. A comparison with the alternative 5D approach to describing the confinement phase transition is given in Appendix~\ref{sec:45D}. 
The new elements taken into account in this work compared to Refs.~\cite{Bruggisser:2018mus,Bruggisser:2018mrt} are
\begin{itemize}
\item the complete one-loop $T=0$  corrections to the dilaton potential,
\item a more detailed computation of washout effects,
\item LHC bounds on the dilaton,
\item LHC bounds on the Higgs couplings,
\item ACME EDM bounds.
\end{itemize}
As a result, we were able to determine the remaining window of parameter space in this minimal composite Higgs realisation of EWBG. It will be fully probed by  future LHC and EDM measurements. 

The generation of the baryon asymmetry requires new sources of CPV. We have analysed two benchmark sources. CPV in the first case is produced by the top quark Yukawa coupling whose phase, being a function of the dilaton VEV, varies during the confinement phase transition. In the second case, CPV is sourced by the variation of the non-diagonal quark Yukawa matrix, which is in turn due to the variation of the charm quark mixing. While we have not performed a detailed analysis of the baryon asymmetry generation, the assumptions about the quark Yukawa variation also have paramount importance for the collider tests of the dilaton. 

The dilaton plays a crucial role for defining the properties of the phase transition. In particular, too large dilaton masses lead to reheating temperatures that are high enough for sphalerons to become active again after the phase transition, washing-out the baryon asymmetry. The relatively light dilaton which is therefore required for EWBG is effectively constrained by the LHC data~\cite{Bruggisser:2022ofg}. We then find that only the benchmark scenario with top-induced CPV passes the combination of the bounds imposed by the collider data and the requirements of EWBG. This case has a viable parameter region (see Figs.~\ref{fig:MNplotsTopGlueball}, \ref{fig:MNplotsTopMeson}) as the result of 
a suppression of the gluon-fusion production of the dilaton, easing the collider constraints. This is in turn due to 
an accidental cancellation in the dilaton coupling to top quarks.
Such a cancellation does not take place in the benchmark scenario with charm-induced CPV, and we were not able to find a viable parameter region for it. 

We have found viable parameter space for both a glueball-like and a meson-like dilaton. However, the first one features a higher efficiency of the baryon asymmetry production, which can be estimated from the product of the total washout factor $\omega_{\rm tot}$ after reheating and the squared (sine of the) tunneling angle $\sin^2 h/f$ which influences the size of CPV during the phase transition.

The viable window for the benchmark scenario with top-induced CPV is expected to be effectively covered by collider searches for the dilaton in the near future. Furthermore, in this case our estimate for the size of the electron EDM is less than an order of magnitude away from the current experimental sensitivity, and hence the corresponding experiments are also very relevant for the considered model.

Although the benchmark scenario with charm-induced CPV appears to be disfavoured by the collider bounds, there is a simple fix to it. Namely, one could envisage a scenario where the top quark mixings vary sizeably with the dilaton VEV, but do not introduce any significant CPV during the phase transition. This possibility is actually more minimal than the original assumption for the case with top-induced CPV, since it does not require one of the top chiralities to mix sizeably with two different composite operators, and one operator is sufficient. It however has the same effect on the collider phenomenology, leading to relatively large $\gamma_t$ and Higgs-dilaton mixing, which can open a window for EWBG. The CPV source could in this case instead reside in the mixing of the charm or other light quarks. Note that one then also expects that the contribution to the electron EDM is suppressed by the charm Yukawa coupling with respect to the values plotted in Fig.~\ref{fig:edm}.
In fact, a simultaneous variation of all quark Yukawa couplings (including the one of the top), with the CPV sourced by the interplay of different quark generations (as happens in our charm-induced CPV benchmark), can be achieved in a rather generic way~\cite{vonHarling:2016vhf}.

Another important test of the considered scenario are Higgs-coupling measurements. These are sensitive to the Higgs-dilaton mixing, and the overall compositeness scale $f$. The latter sets the size of the reheating temperature after the phase transition, and therefore is bounded from above.  Taking into account variations of the other parameters while ensuring successful EWBG, we find an upper bound $f \lesssim 1100$~GeV (see Fig.~\ref{fig:ckfplots}). At the same time Higgs couplings currently provide a lower bound $f \gtrsim 800$~GeV which is not far away and will be improved in the future. However, one should keep in mind that there can be an order-one variation in our bound on $f$ from the variation of the parameter $c_k^{(h)}$ which we have set to 1.

In the regions of parameter space suitable for EWBG for both the meson- and glueball-like dilaton, a gravitational wave signal is generated during the Higgs-dilaton phase transition that has the right properties to be detectable by the future observatory LISA.

We have demonstrated that the simplest benchmark scenarios, even though they allow for viable parameter space, are highly constrained by collider searches for the dilaton, and predict an electron EDM close to the current limit. It is therefore interesting to consider scenarios with  a modified thermal history of EW symmetry breaking~\cite{Meade:2018saz,Baldes:2018nel,Glioti:2018roy,Matsedonskyi:2020mlz,Matsedonskyi:2020kuy,Matsedonskyi:2021hti,Matsedonskyi:2022btb,Biekotter:2022kgf,Chang:2022psj,Bai:2021hfb} which allow to increase the overall mass and temperature scale where EWBG takes place, potentially significantly relaxing the discussed bounds. We will present an analysis of such scenarios in follow-up work~\cite{snr}. 

As a final comment we would like to emphasize that we have focused on the constraints relevant for 'vanilla-type' EWBG. For example, for the scenario of cold electroweak baryogenesis one of the most stringent constraints, limiting the dilution of the baryon asymmetry from entropy injection after reheating, does not apply. In this case, the collider bounds can be evaded by going to large $N$ provided that $c_{gg} \ll 1$ \cite{Bruggisser:2022ofg}. This motivates further analysis of alternative realizations of EWBG in the context of composite Higgs models.

\section*{Acknowledgments}

The work of SB  has been supported by the German Research Foundation (DFG) under grant no.~396021762–TRR 257.~This work is supported by the Deutsche Forschungsgemeinschaft under Germany Excellence Strategy - EXC 2121 ``Quantum Universe'' - 390833306.
OM is supported by STFC HEP Theory Consolidated grant ST/T000694/1. OM also thanks the Mainz Institute for Theoretical Physics (MITP) and ICTP-SAIFR for their hospitality and support during the completion of this work.

\appendix

\section{4D EFT vs.~5D warped models}
\label{sec:45D}

In this paper we have tried to infer the properties of the confinement phase transition from the 4D effective field theory describing the physics below the confinement scale, with the interactions derived from assumptions about the symmetries of the theory (a spontaneously-broken global symmetry with coset $SO(5)/SO(4)$ and spontaneously-broken conformal invariance, with additional small explicit breakings) and large-$N$ counting expected to hold in confining $SU(N)$ gauge theories.\footnote{See also \cite{Baldes:2020kam} for recent work on 4D modelling of confinement phase transitions.}
An alternative approach to the problem is provided by 5D models with a warped extra dimension, the Randall-Sundrum (RS) scenarios~\cite{Randall:1999ee,Creminelli:2001th,Randall:2006py,Nardini:2007me,Megias:2018dki,vonHarling:2017yew,vonHarling:2016vhf}, which can be related to consistent 4D CFTs through holography. Although more technically challenging and restrictive compared to our low-energy approach using effective field theory, these 5D theories can be more predictive, for instance providing calculable ({\it i.e.}~finite) potentials for the Higgs and the dilaton fields without {\it ad hoc} assumptions about the cutoff physics that we have to invoke for e.g.~the analysis of Coleman-Weinberg corrections in 
Appendix~\ref{sec:V1L}. Below we discuss several main differences between the 4D effective field theory approach of Ref.~\cite{Bruggisser:2018mus,Bruggisser:2018mrt} which we follow here and the description based on 5D dual models, emphasizing the sources of model dependence in the analysis of the confinement phase transition. Given the focus of this paper, we will skip an introduction to the RS scenario. After listing the main differences we will show how they affect the computation of the phase transition strength.

\subsection*{Dilaton kinetic term and quartic coupling normalization}

In order to derive the dilaton coupling $g_\chi$ we have employed large-$N$ counting. Let us take a step back here and follow its derivation. We will consider a glueball-like dilaton, to which the RS radion is a direct counterpart. The $N$-scaling of the glueball\footnote{For simplicity we refer to the fundamental constituents of the new strong dynamics as quarks and gluons and to their coupling as $g_s$, not to be confused with their SM prototypes.} two-point function (corresponding in the lowest order to a gluon loop) in the 't Hooft limit $N g_s^2/(4\pi)^2 \to 1$ is given by $N^2$, which allows to write down the kinetic term of the glueball
\be\label{eq:4dkin}
{\cal L}_{\rm kin}^{\rm CH}= \frac {N^2}{(4 \pi)^2}  \frac 1 2 (\partial_\mu \chi)^2.
\ee
The $(4 \pi)^2$ suppression factor is a naive estimate of the loop factor from the gluon loop. The two-point function however receives an infinite number of contributions from the diagrams with the same $N$-scaling and therefore this suppression can only be taken as an estimate. 
After normalizing the field $\chi$ canonically, each field insertion becomes accompanied by 
\be\label{eq:chiapp}
g_\chi = \frac{4\pi}{N},
\ee 
which allows to identify the latter with the glueball coupling strength. The $4\pi$ factor makes the theory fully strongly coupled in the limit $N\to 1$. The additional coefficient $c_k^{(\chi)}$ which we introduced in the definition of $g_\chi$ in Eq.~(\ref{eq:gchi}) compared to Eq.~(\ref{eq:chiapp}) thus can be seen as a factor reflecting possible variation in the normalization of the dilaton kinetic term.

The kinetic term of the radion field $\mu_{\text{RS}}$ in the RS setup is given by
\be\label{eq:5dkin}
{\cal L}_{\rm kin}^{\rm RS}=12 \, (M_5/k)^3 (\partial_\mu \mu_{\text{RS}})^2 =24 \, \frac {N^2}{16\pi^2}  \frac 1 2 (\partial_\mu \mu_{\text{RS}})^2,
\ee
where $M_5$ is the 5D Planck mass and $k$ is the AdS curvature scale. In the second equality we used the AdS/CFT identification
\be\label{eq:adscft}
(M_5/k)^3 = N^2/ (4\pi)^2.
\ee 

Comparing the kinetic terms (\ref{eq:4dkin}) and (\ref{eq:5dkin}) we see that there is an extra factor of $4!$ in the RS case, which means that after canonical normalization the field space becomes more ``stretched". 
This difference can be corrected by choosing $c_k^{(\chi)} = \sqrt{1/4!}$.

To make the comparison more precise we will need to also specify the scalar potentials before canonical normalization in the two cases. The Goldberger-Wise (GW)~\cite{Goldberger:1999uk} radion potential has the form~\cite{Rattazzi:2000hs} 
\be\label{eq:vrs}
V_{\text{RS}} \simeq \mu_{\text{RS}}^4 \left\{ (4+2\epsilon_{\text{RS}}) (v_1-v_0 (\mu_{\text{RS}}/\mu_{\text{RS} 0})^{\epsilon_\text{RS}})^2 - \epsilon_{\text{RS}} v_1^2 \right\},
\ee
where $\epsilon_{\text{RS}}$ corresponds to our $\gamma_\epsilon$ -- the scaling dimension of $\epsilon[\chi]$. Furthermore, $v_{0,1}$ are the VEVs of the GW field at the UV and IR brane in units of $k^{3/2}$.
For the comparison between the 4D~(\ref{eq:vchi}) and 5D~(\ref{eq:vrs}) cases
we will use the scale-independent part of the quartic dilaton coupling after canonical normalization of the kinetic term  (assuming $|\epsilon_{\text{RS}}|\ll1$):
\be
c_\chi g_\chi^2  \; \longleftrightarrow \; \frac{1}{6} v_1^2 (k/M_5)^6.
\ee
Applying the identification~(\ref{eq:adscft}) in the {\it r.h.s.}~together with $g_\chi = 4\pi/N$ in the {\it l.h.s.}, we obtain
\be\label{eq:vcomparison}
c_\chi \frac {16 \pi^2}{N^2}  \; \longleftrightarrow \; \frac{v_1^2}{144}  \frac{16 \pi^2}{N^2}.
\ee
For the typically used values of parameters, e.g.~$c_\chi\sim 0.1$ and $v_1\sim 1$, the 4D counterpart has an approximately one order of magnitude larger quartic coupling.

Note that the two dilaton potentials can not be made equal by simply choosing a different $N$ in the two descriptions to compensate for the difference of the numerical factors in Eq.~(\ref{eq:vcomparison}). The reason is that there is yet another part of the potential which is crucial for the phase transition -- the thermal dip around the dilaton origin, for which we {\it assumed} the result of ${\cal N}=4$ $SU(N)$ gauge theory, {\it i.e.}~the same as in the RS case: 
\be\label{eq:fcomparison}
F^{\rm CH}[\chi=0] = F^{\rm RS}[\mu=0] = - \frac {\pi^2 N^2}{8} T^4.
\ee
This means that any relative $N$ rescaling in the two descriptions will detune the depth of the thermal parts and change the transition properties.

\subsubsection*{Running of the dilaton quartic}

In the RS scenario with GW mechanism the VEV of the radion field is stabilized around
\be
\mu_{\text{RS}} \simeq \mu_{\text{RS}0} \left( \frac{v_0}{v_1} \right)^{-1/\epsilon_{\text{RS}}} \hspace{-.1cm},
\ee
where $\mu_{\text{RS}0}$ is of the order of the Planck mass and the desired $\mu_{\text{RS}}$ is at the TeV scale. Negative and small $\epsilon_{\text{RS}}$ 
(we chose $\epsilon_{\text{RS}}<0$ to match the negative $\gamma_\epsilon$ in our 4D scenario) allows to generate a large separation between $\mu_{\text{RS}0}$ and $\mu_{\text{RS}}$ even for a very moderate tuning  $v_0/v_1 \sim 0.1$. In this case $\epsilon_{\text{RS}}$ has to be fixed at around $-1/20$. In the 4D picture this would correspond to $\epsilon(\chi)$ having a fixed energy scaling with a constant $\gamma_\epsilon$ starting from $M_{\text{Pl}}$ down to the TeV scale. Instead, we prefer to stay more general and to not make such an assumption, treating $\gamma_\epsilon$ (and therefore the dilaton mass which depends on $\gamma_\epsilon$, see Eq.~(\ref{eq:dilatonmass})) as a free parameter. 
Note that such a situation can also be realized in less minimal 5D setups, such as that of Ref.~\cite{Agashe:2019lhy}.

\subsection*{Dilaton scale  vs.~Higgs decay constant}

In Section~\ref{sec:varf} we stated that the dilaton VEV $\chi_0$ is given by $\sqrt N f$, up to a numerical proportionality factor. This enhancement of $\chi_0$ can for example suppress the dilaton production cross-sections and the mixing with the Higgs, making it harder to track the dilaton experimentally. In the 5D dual scenario the Higgs as a pseudo-Nambu-Goldstone boson can be described by the 5th component of a 5D gauge field, and the Higgs decay constant is then given by~\cite{Agashe:2004rs}
\be
f_\text{RS}\,\simeq\, 2\mu_{\text{RS}}/g_*.
\ee 
Here $g_*$ approximately corresponds to the coupling strength of the Kaluza-Klein modes of the 5D gauge field~\cite{Contino:2003ve}, dual to meson-like composite states in the 4D picture. Taking the large-$N$ estimate for their coupling, $g_* = 4 \pi/\sqrt N$, and normalizing the dilaton field canonically, $\tilde \mu_{\text{RS}} = \sqrt{24} \mu_{\text{RS}}/g_\chi$ with $g_\chi = 4 \pi/N$, we obtain the relation
\be
f_\text{RS}\,=\, \tilde \mu_{\text{RS}} \frac {2}{\sqrt{24}} \frac {g_\chi} {g_*} \,\simeq\, \tilde \mu_{\text{RS}} \frac {0.4}{\sqrt{N}}. 
\ee 
This is very close to what we have used in Eq.~(\ref{eq:chioff}), up to an order-one factor, which can be accounted for by varying the order-one coefficients $c_i$ of our description.

\vspace{.4cm}

Having pointed out the differences in the two descriptions we can estimate their impact on the confinement phase transition. The approximation that we employ here is rather simplistic for the sake of conciseness and to avoid an overlap with the analysis of Ref.~\cite{Baldes:2021aph}, but it gives a qualitatively correct understanding of the actual physics and the numerical results that we presented before.	
We will consider the situation where the tunnelling happens from the origin to the slope of the dilaton potential far from the true minimum, which is often the case for the dilaton phase transition featuring a large amount of supercooling. In this case the quartic term of the dilaton potential is dominated by $\epsilon(\chi)$ which grows towards $\chi=0$. The relevant part of the dilaton potential can then be written as, using the high-$T$ approximation of the thermal CFT contribution~(\ref{eq:v1Tloop}),
\be\label{eq:app1}
V \simeq \frac {15}{32} N^2 T^2 g_\chi^2 \chi^2 - \epsilon[\chi] \chi^4
\equiv \frac 1 2 m_{\text{eff}}^2\chi^2 - \frac 1 4 \lambda_{\text{eff}} \chi^4,
\ee
and the corresponding tunnelling action for the $O(3)$ bounce is~\cite{Linde:1990flp}
\be
S_3/T \simeq \frac {19 \,m_{\text{eff}}}{\lambda_{\text{eff}} \,T}.
\ee
Note that by approximating the dilaton potential with quadratic and quartic terms only, we miss the Boltzmann suppression of thermal effects at $g_\chi \chi \gtrsim T$ which cuts off the quadratic term in Eq.~(\ref{eq:app1}) and decreases the barrier size.  

The size of the dilaton effective quartic coupling $\lambda_{\text{eff}}/4 \simeq \epsilon(\chi)$ is dictated by the size of the constant quartic term $c_\chi g_\chi^2$, which it has to match at $\chi=\chi_0$, and $\gamma_\epsilon$, which determines how fast $\epsilon(\chi)$ grows from that point towards smaller values of $\chi$. Before reaching an IR fixed point, $\epsilon (\chi)$ reads
\be
\epsilon (\chi) \simeq c_\chi g_\chi^2 (\chi / \chi_0)^{\gamma_\epsilon},
\ee
while the value at the IR fixed point is given by
\be
\epsilon (\chi\to0) = (\gamma_\epsilon / c_\epsilon) g_\chi^2.
\ee
As we have shown above, the constant quartic coupling $c_\chi g_\chi^2$ is set at a relatively larger value in our case, and at the same time we allow for higher $\gamma_\epsilon$. This means that in the considered parameter space of our model we typically have higher nucleation rates.

\section{Coleman-Weinberg corrections}
\label{sec:V1L}

The zero-temperature potential for the Higgs and the dilaton that we discussed in Sections~\ref{sec:Vh} and \ref{sec:Vchi} only includes the contributions generated by CFT states with mass $\sim \chi$ which can be integrated out from our description. The light degrees of freedom present in our effective field theory can also contribute to the scalar potential. Accounting for these corrections is important because they can in particular affect the Higgs-dilaton mixing. 
We will now outline our procedure for the calculation of these corrections. In comparison to the usual Coleman-Weinberg potential with dimensional regularisation, our procedure allows us to obtain the correct form of the dilaton and Higgs potential if states external to the CFT are present, with a mass independent of the dilaton field. 
While this is not necessary in the setup considered here, it is important for the case studied in the accompanying paper~\cite{snr} with high-temperature EW symmetry breaking. To consistently compare the two scenarios we will thus apply the same methods to both.

During the confinement phase transition, the mass scale of the new strong sector $m_* = g_\chi \chi$ changes from $0$ to $\sim \chi_0$. In the presence of elementary states with masses $m$ which are independent of the dilaton and have $m\ll g_* f$, the theory then transits between two very distinct regimes, with $m_* \ll m$ and $m_* \gg m$. In the former case, the one-loop correction induced by the elementary states is expected to be significantly affected by the presence of the lighter composite resonances. More precisely, the composite physics at $m_*$ is expected to cut off the momentum integrals, thus removing the sensitivity of the composite Higgs and dilaton potential to large elementary masses~\cite{Garriga:2002vf}.

Let us first consider the general form of the one-loop effective potential before regularization, in terms of Euclidean momenta, and per \emph{d.o.f.}~of a state with mass $m$,
\be\label{eq:gen1loop}
V_{\text{1L}} =   \frac {(-1)^F} 2 \int \frac {d^4 k_E}{(2\pi)^4} \ln ({k_E^2+m^2}) ,
\ee
where $F=1 (0)$ for fermions (bosons).
In order to model the effect of the new physics at the scale $m_*$, we will add a step function $\theta (m_*-k)$ to the integrand in Eq.~(\ref{eq:gen1loop}). 
This momentum cutoff leads to a finite potential of the form
\be
\begin{aligned}
V_{\text{1L}} 
&= (-1)^F  \frac 1 {2 (4 \pi)^2}  \int^{m_*^2}_0 k_E^2 d k_E^2 \ln (k_E^2+m^2)   \\
&= (-1)^F \frac{1}{64 \pi^2} \left[m_*^4 \left(\ln (m_*^2+m^2) - \frac 1  2\right) + m_*^2 m^2+ m^4 \ln \frac{m^2}{m^2+m_*^2}\right].\label{eq:loopcutoff}
\end{aligned}
\ee

\begin{figure}
\begin{center}
\includegraphics[scale=1.0]{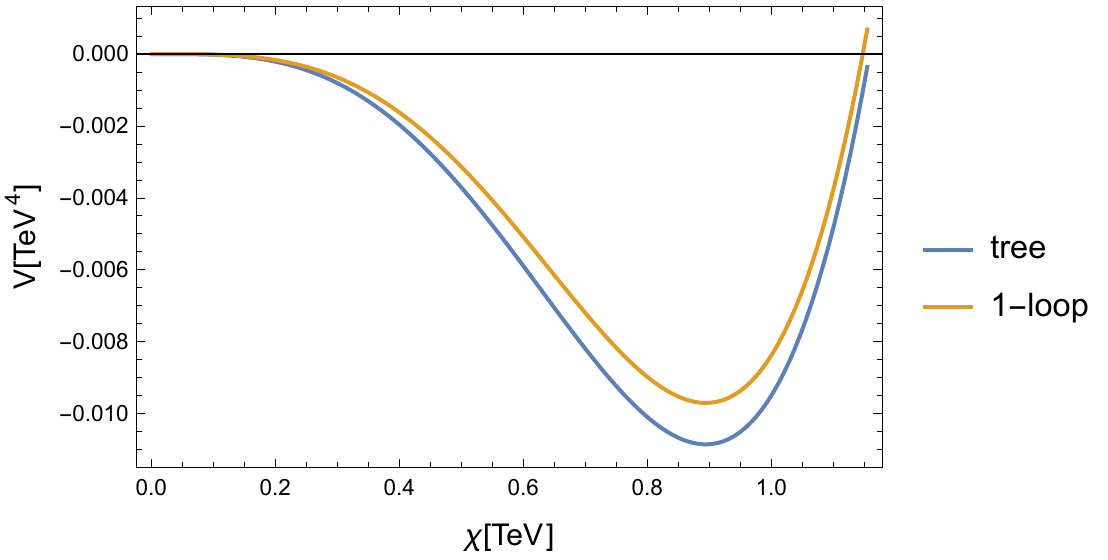}
\end{center}
\caption{{\it 
Comparison of the tree-level and one-loop potentials for a glueball dilaton with top-induced CPV, for $m_\chi = 450~\text{GeV}, c_k^{(\chi)}=2, N=5$, and the other parameters set as in Table~\ref{tab:bench}.
}}
\label{fig:Vchi}
\end{figure}

All the loop corrections to the scalar potential originating from the degrees of freedom of our effective field theory have to vanish with the elementary-composite mixings. 
To account for this we use the subtracted one-loop potential
\be\label{eq:vcwbar}
\bar V_{\text{1L}}= V_{\text{1L}}(\{g,\lambda_t,y_{L,R}\}) - V_{\text{1L}}(\{0,0,0\}).
\ee
We can rewrite the subtracted potential as
\be
\begin{aligned}
\bar V_{\text{1L}} &=
\frac {(-1)^F} 2 \int_0^{m_*} \frac {d^4 k_E}{(2\pi)^4} \ln \left[\frac{k_E^2+m^2}{k_E^2+m_0^2}\right]\\
&=
\frac {(-1)^F} 2 \int_0^{m_*} \frac {d^4 k_E}{(2\pi)^4} \ln \left[1+\frac{\delta m^2}{k_E^2+m_0^2}\right]\\
&=
\frac {(-1)^F} 2 \int_0^{m_*} \frac {d^4 k_E}{(2\pi)^4} \frac{\delta m^2}{k_E^2+m_0^2} + \dots,\label{eq:gen1loopbar}
\end{aligned}
\ee
where $\delta m^2$ is the mixing-induced correction to the square of the particle mass, and $m_0$ is the dilaton-independent part of the mass.
The expression~(\ref{eq:gen1loopbar}) indeed corresponds to the series of diagrams with increasing number of mixing insertions that we are looking for.
In the specific case that we consider here, there are no bare masses independent of the dilaton, thus $m_0 = 0$. In general, however, $m_0$ can be non-vanishing if we consider an elementary sector with a non-zero mass coupled to the CFT.

The potential~(\ref{eq:gen1loopbar}) also has a very similar form to the dilaton potential obtained using the 5D dual picture~\cite{Garriga:2002vf}, where the sharp momentum cutoff at $m_*$ is substituted with an exponential suppression by a form-factor inside of the integral.

\section{Masses of the Goldstone bosons}\label{sec:goldstones}

In the parametrization used in the main text of this paper, the four Goldstone bosons $\Pi_\alpha$, including the Higgs, can be combined into one field $\Sigma$, formally transforming as an $SO(5)$ quintuplet
\be
\Sigma=  f \left\{ \frac {\vec\Pi^T}{\Pi} \sin \frac \Pi f, \cos \frac \Pi f\right\}.
\ee
Then $|\partial_\mu \Sigma|^2$ can be used to derive the kinetic terms of the Goldstones which turn out to be non-canonically normalized. To get rid of the latter complication, we introduce an auxilliary field $\sigma$ to complete the quadruplet of Goldstones to a full linear $SO(5)$ multiplet $\hat \Sigma$ given by 
\be\label{eq:hatsigma}
\hat \Sigma = \left\{ \Sigma_1, \Sigma_2, \Sigma_3, \Sigma_4, \Sigma_5 \right\}
= \sigma \left\{ \frac {\vec\Pi^T}{\Pi} \sin \frac {\Pi} {f}, \cos \frac {\Pi} {f} \right\}.
\ee
The kinetic terms of $\hat \Sigma_i$ are now canonical and the scalar potential can be directly used to compute the mass spectrum.  The scalar potential for the linearly realized fields, including Goldstones, can be obtained from the Higgs potential Eq.~(\ref{eq:vCHtree}) using the substitution  
\bea\label{eq:goldprescr}
\sin^2 \frac \Pi f &\to& \hat \Sigma_\alpha^2/\hat \Sigma_i^2\,,\;\; \alpha=1...4,\, i=1...5.
\eea
We also assume the appropriate potential to ensure $\langle \sigma \rangle =(g_\chi/g_*) \langle \chi\rangle$, for instance 
\be
V_\sigma= - \frac 1 2 g_\chi^2 \chi^2 |\hat \Sigma|^2 + \frac 1 4 g_*^2 |\hat \Sigma|^4.
\ee
In the minimum of the potential of the heavy auxiliary field $\sigma=|\hat \Sigma|$, $V_\sigma'=0$, one then finds the Goldstones mass
\be
m_G^2 = \frac {g_*^2} {g_\chi^2 \chi^2} 2(\alpha+2 \beta \sin^2 v_{\text{CH}}/f) \cos^2 v_{\text{CH}}/f.
\ee

\bibliographystyle{JHEP}  
\bibliography{biblio}

\providecommand{\href}[2]{#2}\begingroup\raggedright\begin{thebibliography}{100}

\bibitem{Affleck:1984fy}
I.~Affleck and M.~Dine, \emph{{A New Mechanism for Baryogenesis}},
  \href{https://doi.org/10.1016/0550-3213(85)90021-5}{\emph{Nucl. Phys. B}
  {\bfseries 249} (1985) 361--380}.

\bibitem{Fukugita:1986hr}
M.~Fukugita and T.~Yanagida, \emph{{Baryogenesis Without Grand Unification}},
  \href{https://doi.org/10.1016/0370-2693(86)91126-3}{\emph{Phys. Lett. B}
  {\bfseries 174} (1986) 45--47}.

\bibitem{Shaposhnikov:1987tw}
M.~E. Shaposhnikov, \emph{{Baryon Asymmetry of the Universe in Standard
  Electroweak Theory}},
  \href{https://doi.org/10.1016/0550-3213(87)90127-1}{\emph{Nucl. Phys. B}
  {\bfseries 287} (1987) 757--775}.

\bibitem{Cohen:1990it}
A.~G. Cohen, D.~B. Kaplan and A.~E. Nelson, \emph{{Baryogenesis at the weak
  phase transition}},
  \href{https://doi.org/10.1016/0550-3213(91)90395-E}{\emph{Nucl. Phys. B}
  {\bfseries 349} (1991) 727--742}.

\bibitem{Enomoto:2022rrl}
K.~Enomoto, S.~Kanemura and Y.~Mura, \emph{{New benchmark scenarios of
  electroweak baryogenesis in aligned two Higgs double models}},
  \href{https://doi.org/10.1007/JHEP09(2022)121}{\emph{JHEP} {\bfseries 09}
  (2022) 121}, [\href{https://arxiv.org/abs/2207.00060}{{\ttfamily
  2207.00060}}].

\bibitem{Azatov:2022tii}
A.~Azatov, G.~Barni, S.~Chakraborty, M.~Vanvlasselaer and W.~Yin,
  \emph{{Ultra-relativistic bubbles from the simplest Higgs portal and their
  cosmological consequences}},
  \href{https://doi.org/10.1007/JHEP10(2022)017}{\emph{JHEP} {\bfseries 10}
  (2022) 017}, [\href{https://arxiv.org/abs/2207.02230}{{\ttfamily
  2207.02230}}].

\bibitem{Harigaya:2022ptp}
K.~Harigaya and I.~R. Wang, \emph{{First-Order Electroweak Phase Transition and
  Baryogenesis from a Naturally Light Singlet Scalar}},
  \href{https://arxiv.org/abs/2207.02867}{{\ttfamily 2207.02867}}.

\bibitem{Ellis:2022lft}
J.~Ellis, M.~Lewicki, M.~Merchand, J.~M. No and M.~Zych, \emph{{The scalar
  singlet extension of the Standard Model: gravitational waves versus
  baryogenesis}}, \href{https://doi.org/10.1007/JHEP01(2023)093}{\emph{JHEP}
  {\bfseries 01} (2023) 093},
  [\href{https://arxiv.org/abs/2210.16305}{{\ttfamily 2210.16305}}].

\bibitem{Servant:2018xcs}
G.~Servant, \emph{{The serendipity of electroweak baryogenesis}},
  \href{https://doi.org/10.1098/rsta.2017.0124}{\emph{Phil. Trans. Roy. Soc.
  Lond. A} {\bfseries 376} (2018) 20170124},
  [\href{https://arxiv.org/abs/1807.11507}{{\ttfamily 1807.11507}}].

\bibitem{vonHarling:2016vhf}
B.~von Harling and G.~Servant, \emph{{Cosmological evolution of Yukawa
  couplings: the 5D perspective}},
  \href{https://doi.org/10.1007/JHEP05(2017)077}{\emph{JHEP} {\bfseries 05}
  (2017) 077}, [\href{https://arxiv.org/abs/1612.02447}{{\ttfamily
  1612.02447}}].

\bibitem{Baldes:2016rqn}
I.~Baldes, T.~Konstandin and G.~Servant, \emph{{A first-order electroweak phase
  transition from varying Yukawas}},
  \href{https://doi.org/10.1016/j.physletb.2018.10.015}{\emph{Phys. Lett. B}
  {\bfseries 786} (2018) 373--377},
  [\href{https://arxiv.org/abs/1604.04526}{{\ttfamily 1604.04526}}].

\bibitem{Cline:2021dkf}
J.~M. Cline and B.~Laurent, \emph{{Electroweak baryogenesis from light fermion
  sources: A critical study}},
  \href{https://doi.org/10.1103/PhysRevD.104.083507}{\emph{Phys. Rev. D}
  {\bfseries 104} (2021) 083507},
  [\href{https://arxiv.org/abs/2108.04249}{{\ttfamily 2108.04249}}].

\bibitem{Postma:2022dbr}
M.~Postma, J.~van~de Vis and G.~White, \emph{{Resummation and cancellation of
  the VIA source in electroweak baryogenesis}},
  \href{https://doi.org/10.1007/JHEP12(2022)121}{\emph{JHEP} {\bfseries 12}
  (2022) 121}, [\href{https://arxiv.org/abs/2206.01120}{{\ttfamily
  2206.01120}}].

\bibitem{Cline:2021iff}
J.~M. Cline, A.~Friedlander, D.-M. He, K.~Kainulainen, B.~Laurent and
  D.~Tucker-Smith, \emph{{Baryogenesis and gravity waves from a UV-completed
  electroweak phase transition}},
  \href{https://doi.org/10.1103/PhysRevD.103.123529}{\emph{Phys. Rev. D}
  {\bfseries 103} (2021) 123529},
  [\href{https://arxiv.org/abs/2102.12490}{{\ttfamily 2102.12490}}].

\bibitem{Cline:2020jre}
J.~M. Cline and K.~Kainulainen, \emph{{Electroweak baryogenesis at high bubble
  wall velocities}},
  \href{https://doi.org/10.1103/PhysRevD.101.063525}{\emph{Phys. Rev. D}
  {\bfseries 101} (2020) 063525},
  [\href{https://arxiv.org/abs/2001.00568}{{\ttfamily 2001.00568}}].

\bibitem{Kainulainen:2021oqs}
K.~Kainulainen, \emph{{CP-violating transport theory for electroweak
  baryogenesis with thermal corrections}},
  \href{https://doi.org/10.1088/1475-7516/2021/11/042}{\emph{JCAP} {\bfseries
  11} (2021) 042}, [\href{https://arxiv.org/abs/2108.08336}{{\ttfamily
  2108.08336}}].

\bibitem{Ellis:2019flb}
S.~A.~R. Ellis, S.~Ipek and G.~White, \emph{{Electroweak Baryogenesis from
  Temperature-Varying Couplings}},
  \href{https://doi.org/10.1007/JHEP08(2019)002}{\emph{JHEP} {\bfseries 08}
  (2019) 002}, [\href{https://arxiv.org/abs/1905.11994}{{\ttfamily
  1905.11994}}].

\bibitem{Krauss:1999ng}
L.~M. Krauss and M.~Trodden, \emph{{Baryogenesis below the electroweak scale}},
  \href{https://doi.org/10.1103/PhysRevLett.83.1502}{\emph{Phys. Rev. Lett.}
  {\bfseries 83} (1999) 1502--1505},
  [\href{https://arxiv.org/abs/hep-ph/9902420}{{\ttfamily hep-ph/9902420}}].

\bibitem{Garcia-Bellido:1999xos}
J.~Garcia-Bellido, D.~Y. Grigoriev, A.~Kusenko and M.~E. Shaposhnikov,
  \emph{{Nonequilibrium electroweak baryogenesis from preheating after
  inflation}}, \href{https://doi.org/10.1103/PhysRevD.60.123504}{\emph{Phys.
  Rev. D} {\bfseries 60} (1999) 123504},
  [\href{https://arxiv.org/abs/hep-ph/9902449}{{\ttfamily hep-ph/9902449}}].

\bibitem{Konstandin:2011ds}
T.~Konstandin and G.~Servant, \emph{{Natural Cold Baryogenesis from Strongly
  Interacting Electroweak Symmetry Breaking}},
  \href{https://doi.org/10.1088/1475-7516/2011/07/024}{\emph{JCAP} {\bfseries
  07} (2011) 024}, [\href{https://arxiv.org/abs/1104.4793}{{\ttfamily
  1104.4793}}].

\bibitem{Servant:2014bla}
G.~Servant, \emph{{Baryogenesis from Strong $CP$ Violation and the QCD Axion}},
  \href{https://doi.org/10.1103/PhysRevLett.113.171803}{\emph{Phys. Rev. Lett.}
  {\bfseries 113} (2014) 171803},
  [\href{https://arxiv.org/abs/1407.0030}{{\ttfamily 1407.0030}}].

\bibitem{Hall:2019ank}
E.~Hall, T.~Konstandin, R.~McGehee, H.~Murayama and G.~Servant,
  \emph{{Baryogenesis From a Dark First-Order Phase Transition}},
  \href{https://doi.org/10.1007/JHEP04(2020)042}{\emph{JHEP} {\bfseries 04}
  (2020) 042}, [\href{https://arxiv.org/abs/1910.08068}{{\ttfamily
  1910.08068}}].

\bibitem{Carena:2022qpf}
M.~Carena, Y.-Y. Li, T.~Ou and Y.~Wang, \emph{{Anatomy of the electroweak phase
  transition for dark sector induced baryogenesis}},
  \href{https://doi.org/10.1007/JHEP02(2023)139}{\emph{JHEP} {\bfseries 02}
  (2023) 139}, [\href{https://arxiv.org/abs/2210.14352}{{\ttfamily
  2210.14352}}].

\bibitem{Flores:2022oef}
M.~M. Flores, A.~Kusenko, L.~Pearce and G.~White, \emph{{Inhomogeneous cold
  electroweak baryogenesis from early structure formation due to Yukawa
  forces}},  \href{https://arxiv.org/abs/2208.09789}{{\ttfamily 2208.09789}}.

\bibitem{Panico:2015jxa}
G.~Panico and A.~Wulzer, \emph{{The Composite Nambu-Goldstone Higgs}},
  \href{https://doi.org/10.1007/978-3-319-22617-0}{\emph{Lect. Notes Phys.}
  {\bfseries 913} (2016) pp.1--316},
  [\href{https://arxiv.org/abs/1506.01961}{{\ttfamily 1506.01961}}].

\bibitem{Espinosa:2011eu}
J.~R. Espinosa, B.~Gripaios, T.~Konstandin and F.~Riva, \emph{{Electroweak
  Baryogenesis in Non-minimal Composite Higgs Models}},
  \href{https://doi.org/10.1088/1475-7516/2012/01/012}{\emph{JCAP} {\bfseries
  01} (2012) 012}, [\href{https://arxiv.org/abs/1110.2876}{{\ttfamily
  1110.2876}}].

\bibitem{Chala:2016ykx}
M.~Chala, G.~Nardini and I.~Sobolev, \emph{{Unified explanation for dark matter
  and electroweak baryogenesis with direct detection and gravitational wave
  signatures}}, \href{https://doi.org/10.1103/PhysRevD.94.055006}{\emph{Phys.
  Rev. D} {\bfseries 94} (2016) 055006},
  [\href{https://arxiv.org/abs/1605.08663}{{\ttfamily 1605.08663}}].

\bibitem{Xie:2020bkl}
K.-P. Xie, L.~Bian and Y.~Wu, \emph{{Electroweak baryogenesis and gravitational
  waves in a composite Higgs model with high dimensional fermion
  representations}}, \href{https://doi.org/10.1007/JHEP12(2020)047}{\emph{JHEP}
  {\bfseries 12} (2020) 047},
  [\href{https://arxiv.org/abs/2005.13552}{{\ttfamily 2005.13552}}].

\bibitem{DeCurtis:2019rxl}
S.~De~Curtis, L.~Delle~Rose and G.~Panico, \emph{{Composite Dynamics in the
  Early Universe}}, \href{https://doi.org/10.1007/JHEP12(2019)149}{\emph{JHEP}
  {\bfseries 12} (2019) 149},
  [\href{https://arxiv.org/abs/1909.07894}{{\ttfamily 1909.07894}}].

\bibitem{Grojean:2004xa}
C.~Grojean, G.~Servant and J.~D. Wells, \emph{{First-order electroweak phase
  transition in the standard model with a low cutoff}},
  \href{https://doi.org/10.1103/PhysRevD.71.036001}{\emph{Phys. Rev. D}
  {\bfseries 71} (2005) 036001},
  [\href{https://arxiv.org/abs/hep-ph/0407019}{{\ttfamily hep-ph/0407019}}].

\bibitem{Bodeker:2004ws}
D.~Bodeker, L.~Fromme, S.~J. Huber and M.~Seniuch, \emph{{The Baryon asymmetry
  in the standard model with a low cut-off}},
  \href{https://doi.org/10.1088/1126-6708/2005/02/026}{\emph{JHEP} {\bfseries
  02} (2005) 026}, [\href{https://arxiv.org/abs/hep-ph/0412366}{{\ttfamily
  hep-ph/0412366}}].

\bibitem{Delaunay:2007wb}
C.~Delaunay, C.~Grojean and J.~D. Wells, \emph{{Dynamics of Non-renormalizable
  Electroweak Symmetry Breaking}},
  \href{https://doi.org/10.1088/1126-6708/2008/04/029}{\emph{JHEP} {\bfseries
  04} (2008) 029}, [\href{https://arxiv.org/abs/0711.2511}{{\ttfamily
  0711.2511}}].

\bibitem{Grinstein:2008qi}
B.~Grinstein and M.~Trott, \emph{{Electroweak Baryogenesis with a
  Pseudo-Goldstone Higgs}},
  \href{https://doi.org/10.1103/PhysRevD.78.075022}{\emph{Phys. Rev. D}
  {\bfseries 78} (2008) 075022},
  [\href{https://arxiv.org/abs/0806.1971}{{\ttfamily 0806.1971}}].

\bibitem{Bruggisser:2018mus}
S.~Bruggisser, B.~Von~Harling, O.~Matsedonskyi and G.~Servant, \emph{{Baryon
  Asymmetry from a Composite Higgs Boson}},
  \href{https://doi.org/10.1103/PhysRevLett.121.131801}{\emph{Phys. Rev. Lett.}
  {\bfseries 121} (2018) 131801},
  [\href{https://arxiv.org/abs/1803.08546}{{\ttfamily 1803.08546}}].

\bibitem{Bruggisser:2018mrt}
S.~Bruggisser, B.~Von~Harling, O.~Matsedonskyi and G.~Servant,
  \emph{{Electroweak Phase Transition and Baryogenesis in Composite Higgs
  Models}}, \href{https://doi.org/10.1007/JHEP12(2018)099}{\emph{JHEP}
  {\bfseries 12} (2018) 099},
  [\href{https://arxiv.org/abs/1804.07314}{{\ttfamily 1804.07314}}].

\bibitem{DiLuzio:2019wsw}
L.~Di~Luzio, M.~Redi, A.~Strumia and D.~Teresi, \emph{{Coset Cosmology}},
  \href{https://doi.org/10.1007/JHEP06(2019)110}{\emph{JHEP} {\bfseries 06}
  (2019) 110}, [\href{https://arxiv.org/abs/1902.05933}{{\ttfamily
  1902.05933}}].

\bibitem{ACME:2018yjb}
{\scshape ACME} collaboration, V.~Andreev et~al., \emph{{Improved limit on the
  electric dipole moment of the electron}},
  \href{https://doi.org/10.1038/s41586-018-0599-8}{\emph{Nature} {\bfseries
  562} (2018) 355--360}.

\bibitem{Kaplan:1991dc}
D.~B. Kaplan, \emph{{Flavor at SSC energies: A New mechanism for dynamically
  generated fermion masses}},
  \href{https://doi.org/10.1016/S0550-3213(05)80021-5}{\emph{Nucl. Phys. B}
  {\bfseries 365} (1991) 259--278}.

\bibitem{Agashe:2004rs}
K.~Agashe, R.~Contino and A.~Pomarol, \emph{{The Minimal composite Higgs
  model}}, \href{https://doi.org/10.1016/j.nuclphysb.2005.04.035}{\emph{Nucl.
  Phys. B} {\bfseries 719} (2005) 165--187},
  [\href{https://arxiv.org/abs/hep-ph/0412089}{{\ttfamily hep-ph/0412089}}].

\bibitem{Baldes:2016gaf}
I.~Baldes, T.~Konstandin and G.~Servant, \emph{{Flavor Cosmology: Dynamical
  Yukawas in the Froggatt-Nielsen Mechanism}},
  \href{https://doi.org/10.1007/JHEP12(2016)073}{\emph{JHEP} {\bfseries 12}
  (2016) 073}, [\href{https://arxiv.org/abs/1608.03254}{{\ttfamily
  1608.03254}}].

\bibitem{Bruggisser:2017lhc}
S.~Bruggisser, T.~Konstandin and G.~Servant, \emph{{CP-violation for
  Electroweak Baryogenesis from Dynamical CKM Matrix}},
  \href{https://doi.org/10.1088/1475-7516/2017/11/034}{\emph{JCAP} {\bfseries
  1711} (2017) 034}, [\href{https://arxiv.org/abs/1706.08534}{{\ttfamily
  1706.08534}}].

\bibitem{Bruggisser:2022ofg}
S.~Bruggisser, B.~von Harling, O.~Matsedonskyi and G.~Servant, \emph{{Dilaton
  at the LHC: complementary probe of composite Higgs}},
  \href{https://doi.org/10.1007/JHEP05(2023)080}{\emph{JHEP} {\bfseries 05}
  (2023) 080}, [\href{https://arxiv.org/abs/2212.00056}{{\ttfamily
  2212.00056}}].

\bibitem{cpr1}
R.~Contino, A.~Pomarol and R.~Rattazzi, \emph{unpublished work}, .

\bibitem{cpr2}
R.~Rattazzi, \emph{{{The naturally light dilation, talk given at Planck2010}},
  {https://indico.cern.ch/event/75810/contributions/1250635/}}, .

\bibitem{cpr3}
A.~Pomarol, \emph{{{Elementary or Composite: The particle physics dilemma, talk
  given at the XVI IFT Xmas Workshop}},
  {https://www.ift.uam-csic.es/www2/workshops/Xmas10/doc/pomarol.pdf}}, .

\bibitem{Coradeschi:2013gda}
F.~Coradeschi, P.~Lodone, D.~Pappadopulo, R.~Rattazzi and L.~Vitale, \emph{{A
  naturally light dilaton}},
  \href{https://doi.org/10.1007/JHEP11(2013)057}{\emph{JHEP} {\bfseries 11}
  (2013) 057}, [\href{https://arxiv.org/abs/1306.4601}{{\ttfamily 1306.4601}}].

\bibitem{Bellazzini:2013fga}
B.~Bellazzini, C.~Csaki, J.~Hubisz, J.~Serra and J.~Terning, \emph{{A Naturally
  Light Dilaton and a Small Cosmological Constant}},
  \href{https://doi.org/10.1140/epjc/s10052-014-2790-x}{\emph{Eur. Phys. J.}
  {\bfseries C74} (2014) 2790},
  [\href{https://arxiv.org/abs/1305.3919}{{\ttfamily 1305.3919}}].

\bibitem{Chacko:2012sy}
Z.~Chacko and R.~K. Mishra, \emph{{Effective Theory of a Light Dilaton}},
  \href{https://doi.org/10.1103/PhysRevD.87.115006}{\emph{Phys. Rev.}
  {\bfseries D87} (2013) 115006},
  [\href{https://arxiv.org/abs/1209.3022}{{\ttfamily 1209.3022}}].

\bibitem{Megias:2014iwa}
E.~Megias and O.~Pujolas, \emph{{Naturally light dilatons from nearly marginal
  deformations}}, \href{https://doi.org/10.1007/JHEP08(2014)081}{\emph{JHEP}
  {\bfseries 08} (2014) 081},
  [\href{https://arxiv.org/abs/1401.4998}{{\ttfamily 1401.4998}}].

\bibitem{Megias:2016jcw}
E.~Megias, G.~Panico, O.~Pujolas and M.~Quiros, \emph{{Light dilatons in warped
  space: Higgs boson and LHCb anomalies}},
  \href{https://doi.org/10.1016/j.nuclphysbps.2016.12.037}{\emph{Nucl. Part.
  Phys. Proc.} {\bfseries 282-284} (2017) 194--198},
  [\href{https://arxiv.org/abs/1609.01881}{{\ttfamily 1609.01881}}].

\bibitem{Pomarol:2019aae}
A.~Pomarol, O.~Pujolas and L.~Salas, \emph{{Holographic conformal transition
  and light scalars}},
  \href{https://doi.org/10.1007/JHEP10(2019)202}{\emph{JHEP} {\bfseries 10}
  (2019) 202}, [\href{https://arxiv.org/abs/1905.02653}{{\ttfamily
  1905.02653}}].

\bibitem{Grojean:2013qca}
C.~Grojean, O.~Matsedonskyi and G.~Panico, \emph{{Light top partners and
  precision physics}},
  \href{https://doi.org/10.1007/JHEP10(2013)160}{\emph{JHEP} {\bfseries 10}
  (2013) 160}, [\href{https://arxiv.org/abs/1306.4655}{{\ttfamily 1306.4655}}].

\bibitem{Matsedonskyi:2012ym}
O.~Matsedonskyi, G.~Panico and A.~Wulzer, \emph{{Light Top Partners for a Light
  Composite Higgs}}, \href{https://doi.org/10.1007/JHEP01(2013)164}{\emph{JHEP}
  {\bfseries 01} (2013) 164},
  [\href{https://arxiv.org/abs/1204.6333}{{\ttfamily 1204.6333}}].

\bibitem{Redi:2012ha}
M.~Redi and A.~Tesi, \emph{{Implications of a Light Higgs in Composite
  Models}}, \href{https://doi.org/10.1007/JHEP10(2012)166}{\emph{JHEP}
  {\bfseries 10} (2012) 166},
  [\href{https://arxiv.org/abs/1205.0232}{{\ttfamily 1205.0232}}].

\bibitem{Panico:2012uw}
G.~Panico, M.~Redi, A.~Tesi and A.~Wulzer, \emph{{On the Tuning and the Mass of
  the Composite Higgs}},
  \href{https://doi.org/10.1007/JHEP03(2013)051}{\emph{JHEP} {\bfseries 03}
  (2013) 051}, [\href{https://arxiv.org/abs/1210.7114}{{\ttfamily 1210.7114}}].

\bibitem{Witten:1979kh}
E.~Witten, \emph{{Baryons in the 1/n Expansion}},
  \href{https://doi.org/10.1016/0550-3213(79)90232-3}{\emph{Nucl. Phys.}
  {\bfseries B160} (1979) 57--115}.

\bibitem{Contino:2010rs}
R.~Contino, \emph{{The Higgs as a Composite Nambu-Goldstone Boson}},  in
  \emph{{Physics of the large and the small, TASI 09, proceedings of the
  Theoretical Advanced Study Institute in Elementary Particle Physics, Boulder,
  Colorado, USA, 1-26 June 2009}}, pp.~235--306, 2011,
  \href{https://arxiv.org/abs/1005.4269}{{\ttfamily 1005.4269}},
  \href{https://doi.org/10.1142/9789814327183_0005}{DOI}.

\bibitem{Creminelli:2001th}
P.~Creminelli, A.~Nicolis and R.~Rattazzi, \emph{{Holography and the
  electroweak phase transition}},
  \href{https://doi.org/10.1088/1126-6708/2002/03/051}{\emph{JHEP} {\bfseries
  03} (2002) 051}, [\href{https://arxiv.org/abs/hep-th/0107141}{{\ttfamily
  hep-th/0107141}}].

\bibitem{Randall:2006py}
L.~Randall and G.~Servant, \emph{{Gravitational waves from warped spacetime}},
  \href{https://doi.org/10.1088/1126-6708/2007/05/054}{\emph{JHEP} {\bfseries
  05} (2007) 054}, [\href{https://arxiv.org/abs/hep-ph/0607158}{{\ttfamily
  hep-ph/0607158}}].

\bibitem{Nardini:2007me}
G.~Nardini, M.~Quiros and A.~Wulzer, \emph{{A Confining Strong First-Order
  Electroweak Phase Transition}},
  \href{https://doi.org/10.1088/1126-6708/2007/09/077}{\emph{JHEP} {\bfseries
  09} (2007) 077}, [\href{https://arxiv.org/abs/0706.3388}{{\ttfamily
  0706.3388}}].

\bibitem{Konstandin:2010cd}
T.~Konstandin, G.~Nardini and M.~Quiros, \emph{{Gravitational Backreaction
  Effects on the Holographic Phase Transition}},
  \href{https://doi.org/10.1103/PhysRevD.82.083513}{\emph{Phys. Rev.}
  {\bfseries D82} (2010) 083513},
  [\href{https://arxiv.org/abs/1007.1468}{{\ttfamily 1007.1468}}].

\bibitem{vonHarling:2017yew}
B.~von Harling and G.~Servant, \emph{{QCD-induced Electroweak Phase
  Transition}}, \href{https://doi.org/10.1007/JHEP01(2018)159}{\emph{JHEP}
  {\bfseries 01} (2018) 159},
  [\href{https://arxiv.org/abs/1711.11554}{{\ttfamily 1711.11554}}].

\bibitem{Dillon:2017ctw}
B.~M. Dillon, B.~K. El-Menoufi, S.~J. Huber and J.~P. Manuel, \emph{{Rapid
  holographic phase transition with brane-localized curvature}},
  \href{https://doi.org/10.1103/PhysRevD.98.086005}{\emph{Phys. Rev. D}
  {\bfseries 98} (2018) 086005},
  [\href{https://arxiv.org/abs/1708.02953}{{\ttfamily 1708.02953}}].

\bibitem{Gubser:1996de}
S.~S. Gubser, I.~R. Klebanov and A.~W. Peet, \emph{{Entropy and temperature of
  black 3-branes}}, \href{https://doi.org/10.1103/PhysRevD.54.3915}{\emph{Phys.
  Rev.} {\bfseries D54} (1996) 3915--3919},
  [\href{https://arxiv.org/abs/hep-th/9602135}{{\ttfamily hep-th/9602135}}].

\bibitem{Kamada:2016cnb}
K.~Kamada and A.~J. Long, \emph{{Evolution of the Baryon Asymmetry through the
  Electroweak Crossover in the Presence of a Helical Magnetic Field}},
  \href{https://doi.org/10.1103/PhysRevD.94.123509}{\emph{Phys. Rev. D}
  {\bfseries 94} (2016) 123509},
  [\href{https://arxiv.org/abs/1610.03074}{{\ttfamily 1610.03074}}].

\bibitem{Spannowsky:2016ile}
M.~Spannowsky and C.~Tamarit, \emph{{Sphalerons in composite and non-standard
  Higgs models}}, \href{https://doi.org/10.1103/PhysRevD.95.015006}{\emph{Phys.
  Rev. D} {\bfseries 95} (2017) 015006},
  [\href{https://arxiv.org/abs/1611.05466}{{\ttfamily 1611.05466}}].

\bibitem{DOnofrio:2014rug}
M.~D'Onofrio, K.~Rummukainen and A.~Tranberg, \emph{{Sphaleron Rate in the
  Minimal Standard Model}},
  \href{https://doi.org/10.1103/PhysRevLett.113.141602}{\emph{Phys. Rev. Lett.}
  {\bfseries 113} (2014) 141602},
  [\href{https://arxiv.org/abs/1404.3565}{{\ttfamily 1404.3565}}].

\bibitem{Caprini:2019egz}
C.~Caprini et~al., \emph{{Detecting gravitational waves from cosmological phase
  transitions with LISA: an update}},
  \href{https://doi.org/10.1088/1475-7516/2020/03/024}{\emph{JCAP} {\bfseries
  03} (2020) 024}, [\href{https://arxiv.org/abs/1910.13125}{{\ttfamily
  1910.13125}}].

\bibitem{Bodeker:2017cim}
D.~Bodeker and G.~D. Moore, \emph{{Electroweak Bubble Wall Speed Limit}},
  \href{https://doi.org/10.1088/1475-7516/2017/05/025}{\emph{JCAP} {\bfseries
  05} (2017) 025}, [\href{https://arxiv.org/abs/1703.08215}{{\ttfamily
  1703.08215}}].

\bibitem{ptplot}
D.~J. Weir, ``{PTPlot: a tool for exploring the gravitational wave power
  spectrum from first-order phase transitions}.''
  \url{https://www.ptplot.org/}.
\newblock [doi:10.5281/zenodo.6949107].

\bibitem{Hindmarsh:2017gnf}
M.~Hindmarsh, S.~J. Huber, K.~Rummukainen and D.~J. Weir, \emph{{Shape of the
  acoustic gravitational wave power spectrum from a first order phase
  transition}}, \href{https://doi.org/10.1103/PhysRevD.96.103520}{\emph{Phys.
  Rev. D} {\bfseries 96} (2017) 103520},
  [\href{https://arxiv.org/abs/1704.05871}{{\ttfamily 1704.05871}}].

\bibitem{Cutting:2018tjt}
D.~Cutting, M.~Hindmarsh and D.~J. Weir, \emph{{Gravitational waves from vacuum
  first-order phase transitions: from the envelope to the lattice}},
  \href{https://doi.org/10.1103/PhysRevD.97.123513}{\emph{Phys. Rev. D}
  {\bfseries 97} (2018) 123513},
  [\href{https://arxiv.org/abs/1802.05712}{{\ttfamily 1802.05712}}].

\bibitem{Djouadi:2018xqq}
A.~Djouadi, J.~Kalinowski, M.~Muehlleitner and M.~Spira, \emph{{HDECAY:
  Twenty$_{++}$ years after}},
  \href{https://doi.org/10.1016/j.cpc.2018.12.010}{\emph{Comput. Phys. Commun.}
  {\bfseries 238} (2019) 214--231},
  [\href{https://arxiv.org/abs/1801.09506}{{\ttfamily 1801.09506}}].

\bibitem{Bechtle:2020pkv}
P.~Bechtle, D.~Dercks, S.~Heinemeyer, T.~Klingl, T.~Stefaniak, G.~Weiglein
  et~al., \emph{{HiggsBounds-5: Testing Higgs Sectors in the LHC 13 TeV Era}},
  \href{https://doi.org/10.1140/epjc/s10052-020-08557-9}{\emph{Eur. Phys. J. C}
  {\bfseries 80} (2020) 1211},
  [\href{https://arxiv.org/abs/2006.06007}{{\ttfamily 2006.06007}}].

\bibitem{Bechtle:2020uwn}
P.~Bechtle, S.~Heinemeyer, T.~Klingl, T.~Stefaniak, G.~Weiglein and
  J.~Wittbrodt, \emph{{HiggsSignals-2: Probing new physics with precision Higgs
  measurements in the LHC 13 TeV era}},
  \href{https://doi.org/10.1140/epjc/s10052-021-08942-y}{\emph{Eur. Phys. J. C}
  {\bfseries 81} (2021) 145},
  [\href{https://arxiv.org/abs/2012.09197}{{\ttfamily 2012.09197}}].

\bibitem{Bahl:2022igd}
H.~Bahl, T.~Biek\"otter, S.~Heinemeyer, C.~Li, S.~Paasch, G.~Weiglein et~al.,
  \emph{{HiggsTools: BSM scalar phenomenology with new versions of HiggsBounds
  and HiggsSignals}},
  \href{https://doi.org/10.1016/j.cpc.2023.108803}{\emph{Comput. Phys. Commun.}
  {\bfseries 291} (2023) 108803},
  [\href{https://arxiv.org/abs/2210.09332}{{\ttfamily 2210.09332}}].

\bibitem{CMS:2022dwd}
{\scshape CMS} collaboration, \emph{{A portrait of the Higgs boson by the CMS
  experiment ten years after the discovery}},
  \href{https://doi.org/10.1038/s41586-022-04892-x}{\emph{Nature} {\bfseries
  607} (2022) 60--68}, [\href{https://arxiv.org/abs/2207.00043}{{\ttfamily
  2207.00043}}].

\bibitem{ATLAS:2022vkf}
{\scshape ATLAS} collaboration, \emph{{A detailed map of Higgs boson
  interactions by the ATLAS experiment ten years after the discovery}},
  \href{https://doi.org/10.1038/s41586-022-04893-w}{\emph{Nature} {\bfseries
  607} (2022) 52--59}, [\href{https://arxiv.org/abs/2207.00092}{{\ttfamily
  2207.00092}}].

\bibitem{Panico:2017vlk}
G.~Panico, M.~Riembau and T.~Vantalon, \emph{{Probing light top partners with
  CP violation}}, \href{https://doi.org/10.1007/JHEP06(2018)056}{\emph{JHEP}
  {\bfseries 06} (2018) 056},
  [\href{https://arxiv.org/abs/1712.06337}{{\ttfamily 1712.06337}}].

\bibitem{Panico:2018hal}
G.~Panico, A.~Pomarol and M.~Riembau, \emph{{EFT approach to the electron
  Electric Dipole Moment at the two-loop level}},
  \href{https://doi.org/10.1007/JHEP04(2019)090}{\emph{JHEP} {\bfseries 04}
  (2019) 090}, [\href{https://arxiv.org/abs/1810.09413}{{\ttfamily
  1810.09413}}].

\bibitem{Panico:2016ull}
G.~Panico and A.~Pomarol, \emph{{Flavor hierarchies from dynamical scales}},
  \href{https://doi.org/10.1007/JHEP07(2016)097}{\emph{JHEP} {\bfseries 07}
  (2016) 097}, [\href{https://arxiv.org/abs/1603.06609}{{\ttfamily
  1603.06609}}].

\bibitem{Redi:2011zi}
M.~Redi and A.~Weiler, \emph{{Flavor and CP Invariant Composite Higgs Models}},
  \href{https://doi.org/10.1007/JHEP11(2011)108}{\emph{JHEP} {\bfseries 11}
  (2011) 108}, [\href{https://arxiv.org/abs/1106.6357}{{\ttfamily 1106.6357}}].

\bibitem{Redi:2012uj}
M.~Redi, \emph{{Composite MFV and Beyond}},
  \href{https://doi.org/10.1140/epjc/s10052-012-2030-1}{\emph{Eur. Phys. J. C}
  {\bfseries 72} (2012) 2030},
  [\href{https://arxiv.org/abs/1203.4220}{{\ttfamily 1203.4220}}].

\bibitem{Barbieri:2012uh}
R.~Barbieri, D.~Buttazzo, F.~Sala and D.~M. Straub, \emph{{Flavour physics from
  an approximate $U(2)^3$ symmetry}},
  \href{https://doi.org/10.1007/JHEP07(2012)181}{\emph{JHEP} {\bfseries 07}
  (2012) 181}, [\href{https://arxiv.org/abs/1203.4218}{{\ttfamily 1203.4218}}].

\bibitem{Barbieri:2011ci}
R.~Barbieri, G.~Isidori, J.~Jones-Perez, P.~Lodone and D.~M. Straub,
  \emph{{$U(2)$ and Minimal Flavour Violation in Supersymmetry}},
  \href{https://doi.org/10.1140/epjc/s10052-011-1725-z}{\emph{Eur. Phys. J. C}
  {\bfseries 71} (2011) 1725},
  [\href{https://arxiv.org/abs/1105.2296}{{\ttfamily 1105.2296}}].

\bibitem{Cacciapaglia:2007fw}
G.~Cacciapaglia, C.~Csaki, J.~Galloway, G.~Marandella, J.~Terning and
  A.~Weiler, \emph{{A GIM Mechanism from Extra Dimensions}},
  \href{https://doi.org/10.1088/1126-6708/2008/04/006}{\emph{JHEP} {\bfseries
  04} (2008) 006}, [\href{https://arxiv.org/abs/0709.1714}{{\ttfamily
  0709.1714}}].

\bibitem{Barr:1990vd}
S.~M. Barr and A.~Zee, \emph{{Electric Dipole Moment of the Electron and of the
  Neutron}}, \href{https://doi.org/10.1103/PhysRevLett.65.21}{\emph{Phys. Rev.
  Lett.} {\bfseries 65} (1990) 21--24}.

\bibitem{Brod:2013cka}
J.~Brod, U.~Haisch and J.~Zupan, \emph{{Constraints on CP-violating Higgs
  couplings to the third generation}},
  \href{https://doi.org/10.1007/JHEP11(2013)180}{\emph{JHEP} {\bfseries 11}
  (2013) 180}, [\href{https://arxiv.org/abs/1310.1385}{{\ttfamily 1310.1385}}].

\bibitem{Meade:2018saz}
P.~Meade and H.~Ramani, \emph{{Unrestored Electroweak Symmetry}},
  \href{https://doi.org/10.1103/PhysRevLett.122.041802}{\emph{Phys. Rev. Lett.}
  {\bfseries 122} (2019) 041802},
  [\href{https://arxiv.org/abs/1807.07578}{{\ttfamily 1807.07578}}].

\bibitem{Baldes:2018nel}
I.~Baldes and G.~Servant, \emph{{High scale electroweak phase transition:
  baryogenesis \textbackslash{}\& symmetry non-restoration}},
  \href{https://doi.org/10.1007/JHEP10(2018)053}{\emph{JHEP} {\bfseries 10}
  (2018) 053}, [\href{https://arxiv.org/abs/1807.08770}{{\ttfamily
  1807.08770}}].

\bibitem{Glioti:2018roy}
A.~Glioti, R.~Rattazzi and L.~Vecchi, \emph{{Electroweak Baryogenesis above the
  Electroweak Scale}},
  \href{https://doi.org/10.1007/JHEP04(2019)027}{\emph{JHEP} {\bfseries 04}
  (2019) 027}, [\href{https://arxiv.org/abs/1811.11740}{{\ttfamily
  1811.11740}}].

\bibitem{Matsedonskyi:2020mlz}
O.~Matsedonskyi and G.~Servant, \emph{{High-Temperature Electroweak Symmetry
  Non-Restoration from New Fermions and Implications for Baryogenesis}},
  \href{https://doi.org/10.1007/JHEP09(2020)012}{\emph{JHEP} {\bfseries 09}
  (2020) 012}, [\href{https://arxiv.org/abs/2002.05174}{{\ttfamily
  2002.05174}}].

\bibitem{Matsedonskyi:2020kuy}
O.~Matsedonskyi, \emph{{High-Temperature Electroweak Symmetry Breaking by SM
  Twins}}, \href{https://doi.org/10.1007/JHEP04(2021)036}{\emph{JHEP}
  {\bfseries 04} (2021) 036},
  [\href{https://arxiv.org/abs/2008.13725}{{\ttfamily 2008.13725}}].

\bibitem{Matsedonskyi:2021hti}
O.~Matsedonskyi, J.~Unwin and Q.~Wang, \emph{{Electroweak symmetry
  non-restoration from dark matter}},
  \href{https://doi.org/10.1007/JHEP12(2021)167}{\emph{JHEP} {\bfseries 12}
  (2021) 167}, [\href{https://arxiv.org/abs/2107.07560}{{\ttfamily
  2107.07560}}].

\bibitem{Matsedonskyi:2022btb}
O.~Matsedonskyi, J.~Unwin and Q.~Wang, \emph{{Towards TeV-scale supersymmetric
  electroweak baryogenesis}},
  \href{https://doi.org/10.1007/JHEP02(2023)198}{\emph{JHEP} {\bfseries 02}
  (2023) 198}, [\href{https://arxiv.org/abs/2211.09147}{{\ttfamily
  2211.09147}}].

\bibitem{Biekotter:2022kgf}
T.~Biek\"otter, S.~Heinemeyer, J.~M. No, M.~O. Olea-Romacho and G.~Weiglein,
  \emph{{The trap in the early Universe: impact on the interplay between
  gravitational waves and LHC physics in the 2HDM}},
  \href{https://doi.org/10.1088/1475-7516/2023/03/031}{\emph{JCAP} {\bfseries
  03} (2023) 031}, [\href{https://arxiv.org/abs/2208.14466}{{\ttfamily
  2208.14466}}].

\bibitem{Chang:2022psj}
J.~H. Chang, M.~O. Olea-Romacho and E.~H. Tanin, \emph{{Electroweak asymmetric
  early Universe via a scalar condensate}},
  \href{https://doi.org/10.1103/PhysRevD.106.113003}{\emph{Phys. Rev. D}
  {\bfseries 106} (2022) 113003},
  [\href{https://arxiv.org/abs/2210.05680}{{\ttfamily 2210.05680}}].

\bibitem{Bai:2021hfb}
Y.~Bai, S.~J. Lee, M.~Son and F.~Ye, \emph{{Global electroweak symmetric
  vacuum}}, \href{https://doi.org/10.1007/JHEP07(2021)225}{\emph{JHEP}
  {\bfseries 07} (2021) 225},
  [\href{https://arxiv.org/abs/2103.09819}{{\ttfamily 2103.09819}}].

\bibitem{snr}
B.~von Harling, O.~Matsedonskyi and G.~Servant, \emph{{High-Temperature
  Electroweak Baryogenesis with Composite Higgs}},
  \href{https://arxiv.org/abs/2307.14426}{{\ttfamily 2307.14426}}.

\bibitem{Baldes:2020kam}
I.~Baldes, Y.~Gouttenoire and F.~Sala, \emph{{String Fragmentation in
  Supercooled Confinement and Implications for Dark Matter}},
  \href{https://doi.org/10.1007/JHEP04(2021)278}{\emph{JHEP} {\bfseries 04}
  (2021) 278}, [\href{https://arxiv.org/abs/2007.08440}{{\ttfamily
  2007.08440}}].

\bibitem{Randall:1999ee}
L.~Randall and R.~Sundrum, \emph{{A Large mass hierarchy from a small extra
  dimension}}, \href{https://doi.org/10.1103/PhysRevLett.83.3370}{\emph{Phys.
  Rev. Lett.} {\bfseries 83} (1999) 3370--3373},
  [\href{https://arxiv.org/abs/hep-ph/9905221}{{\ttfamily hep-ph/9905221}}].

\bibitem{Megias:2018dki}
E.~Meg\'\i{}as, G.~Nardini and M.~Quir\'os, \emph{{Gravitational waves and
  collider signatures from holographic phase transitions in soft walls}},
  \href{https://doi.org/10.22323/1.336.0227}{\emph{PoS} {\bfseries
  Confinement2018} (2018) 227},
  [\href{https://arxiv.org/abs/1811.10891}{{\ttfamily 1811.10891}}].

\bibitem{Goldberger:1999uk}
W.~D. Goldberger and M.~B. Wise, \emph{{Modulus stabilization with bulk
  fields}}, \href{https://doi.org/10.1103/PhysRevLett.83.4922}{\emph{Phys. Rev.
  Lett.} {\bfseries 83} (1999) 4922--4925},
  [\href{https://arxiv.org/abs/hep-ph/9907447}{{\ttfamily hep-ph/9907447}}].

\bibitem{Rattazzi:2000hs}
R.~Rattazzi and A.~Zaffaroni, \emph{{Comments on the holographic picture of the
  Randall-Sundrum model}},
  \href{https://doi.org/10.1088/1126-6708/2001/04/021}{\emph{JHEP} {\bfseries
  04} (2001) 021}, [\href{https://arxiv.org/abs/hep-th/0012248}{{\ttfamily
  hep-th/0012248}}].

\bibitem{Agashe:2019lhy}
K.~Agashe, P.~Du, M.~Ekhterachian, S.~Kumar and R.~Sundrum, \emph{{Cosmological
  Phase Transition of Spontaneous Confinement}},
  \href{https://doi.org/10.1007/JHEP05(2020)086}{\emph{JHEP} {\bfseries 05}
  (2020) 086}, [\href{https://arxiv.org/abs/1910.06238}{{\ttfamily
  1910.06238}}].

\bibitem{Contino:2003ve}
R.~Contino, Y.~Nomura and A.~Pomarol, \emph{{Higgs as a holographic
  pseudoGoldstone boson}},
  \href{https://doi.org/10.1016/j.nuclphysb.2003.08.027}{\emph{Nucl. Phys. B}
  {\bfseries 671} (2003) 148--174},
  [\href{https://arxiv.org/abs/hep-ph/0306259}{{\ttfamily hep-ph/0306259}}].

\bibitem{Baldes:2021aph}
I.~Baldes, Y.~Gouttenoire, F.~Sala and G.~Servant, \emph{{Supercool composite
  Dark Matter beyond 100 TeV}},
  \href{https://doi.org/10.1007/JHEP07(2022)084}{\emph{JHEP} {\bfseries 07}
  (2022) 084}, [\href{https://arxiv.org/abs/2110.13926}{{\ttfamily
  2110.13926}}].

\bibitem{Linde:1990flp}
A.~D. Linde, \emph{{Particle physics and inflationary cosmology}},
  {\emph{Contemp. Concepts Phys.} {\bfseries 5} (1990) 1--362},
  [\href{https://arxiv.org/abs/hep-th/0503203}{{\ttfamily hep-th/0503203}}].

\bibitem{Garriga:2002vf}
J.~Garriga and A.~Pomarol, \emph{{A Stable hierarchy from Casimir forces and
  the holographic interpretation}},
  \href{https://doi.org/10.1016/S0370-2693(03)00301-0}{\emph{Phys. Lett. B}
  {\bfseries 560} (2003) 91--97},
  [\href{https://arxiv.org/abs/hep-th/0212227}{{\ttfamily hep-th/0212227}}].

\end{thebibliography}\endgroup

\end{document}